\documentclass[twocolumn,prd,aps,superscriptaddress]{revtex4}
\usepackage{graphicx}
\usepackage[usenames]{color}
\usepackage{amsmath,amsfonts,amssymb}
\usepackage[colorlinks,bookmarks]{hyperref}
\definecolor{linkblue}{rgb}{0,0,0.8}
\definecolor{linkgreen}{rgb}{0,0.5,0}
\hypersetup{pdfpagemode=None, pdfstartview=FitH, linkcolor=linkblue, %
             citecolor=linkgreen, urlcolor=linkblue}

\usepackage{color}
\usepackage{natbib}
\usepackage{url}
\usepackage{multirow}
\usepackage{epsf}
\usepackage{epsfig}
\usepackage[letterpaper,margin=0.75in]{geometry}

\usepackage{ifthen}

\usepackage[T1]{fontenc}
\usepackage{lmodern}
\usepackage{ifxetex,ifluatex}

\usepackage{scrextend}

\usepackage{morefloats}

\def\setsymbol#1#2{\expandafter\def\csname #1\endcsname{#2}}
\def\getsymbol#1{\csname #1\endcsname}

\def\Planck{\textit{Planck}}

\newbox\tablebox    \newdimen\tablewidth
\def\leaderfil{\leaders\hbox to 5pt{\hss.\hss}\hfil}
\def\tablenote#1 #2\par{\begingroup \parindent=0.8em
    \abovedisplayshortskip=0pt\belowdisplayshortskip=0pt
    \noindent
    $$\hss\vbox{\hsize\tablewidth \hangindent=\parindent \hangafter=1 \noindent
    \hbox to \parindent{$^#1$\hss}\strut#2\strut\par}\hss$$
    \endgroup}

\def\L2{\ifmmode L_2\else $L_2$\fi}

\def\DeltaT{\ifmmode \Delta T\else $\Delta T$\fi}
\def\deltat{\ifmmode \Delta t\else $\Delta t$\fi}
\def\fknee{\ifmmode f_{\rm knee}\else $f_{\rm knee}$\fi}
\def\Fmax{\ifmmode F_{\rm max}\else $F_{\rm max}$\fi}
\def\solar{\ifmmode{\rm M}_{\mathord\odot}\else${\rm M}_{\mathord\odot}$\fi}
\def\Msolar{\ifmmode{\rm M}_{\mathord\odot}\else${\rm M}_{\mathord\odot}$\fi}
\def\Lsolar{\ifmmode{\rm L}_{\mathord\odot}\else${\rm L}_{\mathord\odot}$\fi}
\def\inv{\ifmmode^{-1}\else$^{-1}$\fi}
\def\mo{\ifmmode^{-1}\else$^{-1}$\fi}
\def\sup#1{\ifmmode ^{\rm #1}\else $^{\rm #1}$\fi}
\def\expo#1{\ifmmode \times 10^{#1}\else $\times 10^{#1}$\fi}
\def\,{\thinspace}
\def\lsim{\mathrel{\raise .4ex\hbox{\rlap{$<$}\lower 1.2ex\hbox{$\sim$}}}}
\def\gsim{\mathrel{\raise .4ex\hbox{\rlap{$>$}\lower 1.2ex\hbox{$\sim$}}}}

\def\simprop{\mathrel{\raise .4ex\hbox{\rlap{$\propto$}\lower 1.2ex\hbox{$\sim$}}}}
\def\deg{\ifmmode^\circ\else$^\circ$\fi}
\def\pdeg{\ifmmode $\setbox0=\hbox{$^{\circ}$}\rlap{\hskip.11\wd0 .}$^{\circ}
          \else \setbox0=\hbox{$^{\circ}$}\rlap{\hskip.11\wd0 .}$^{\circ}$\fi}
\def\arcs{\ifmmode {^{\scriptstyle\prime\prime}}
          \else $^{\scriptstyle\prime\prime}$\fi}
\def\arcm{\ifmmode {^{\scriptstyle\prime}}
          \else $^{\scriptstyle\prime}$\fi}
\newdimen\sa  \newdimen\sb
\def\parcs{\sa=.07em \sb=.03em
     \ifmmode \hbox{\rlap{.}}^{\scriptstyle\prime\kern -\sb\prime}\hbox{\kern -\sa}
     \else \rlap{.}$^{\scriptstyle\prime\kern -\sb\prime}$\kern -\sa\fi}
\def\parcm{\sa=.08em \sb=.03em
     \ifmmode \hbox{\rlap{.}\kern\sa}^{\scriptstyle\prime}\hbox{\kern-\sb}
     \else \rlap{.}\kern\sa$^{\scriptstyle\prime}$\kern-\sb\fi}
\def\ra[#1 #2 #3.#4]{#1\sup{h}#2\sup{m}#3\sup{s}\llap.#4}
\def\dec[#1 #2 #3.#4]{#1\deg#2\arcm#3\arcs\llap.#4}
\def\deco[#1 #2 #3]{#1\deg#2\arcm#3\arcs}
\def\rra[#1 #2]{#1\sup{h}#2\sup{m}}

\def\dots{\relax\ifmmode \ldots\else $\ldots$\fi}
\def\WHzsr{\ifmmode $W\,Hz\mo\,sr\mo$\else W\,Hz\mo\,sr\mo\fi}
\def\mHz{\ifmmode $\,mHz$\else \,mHz\fi}
\def\GHz{\ifmmode $\,GHz$\else \,GHz\fi}
\def\mKs{\ifmmode $\,mK\,s$^{1/2}\else \,mK\,s$^{1/2}$\fi}
\def\muKs{\ifmmode \,\mu$K\,s$^{1/2}\else \,$\mu$K\,s$^{1/2}$\fi}
\def\muKRJs{\ifmmode \,\mu$K$_{\rm RJ}$\,s$^{1/2}\else \,$\mu$K$_{\rm RJ}$\,s$^{1/2}$\fi}
\def\muKHz{\ifmmode \,\mu$K\,Hz$^{-1/2}\else \,$\mu$K\,Hz$^{-1/2}$\fi}
\def\MJysr{\ifmmode \,$MJy\,sr\mo$\else \,MJy\,sr\mo\fi}
\def\MJysrmK{\ifmmode \,$MJy\,sr\mo$\,mK$_{\rm CMB}\mo\else \,MJy\,sr\mo\,mK$_{\rm CMB}\mo$\fi}
\def\microns{\ifmmode \,\mu$m$\else \,$\mu$m\fi}

\def\muK{\ifmmode \,\mu$K$\else \,$\mu$\hbox{K}\fi}
\def\microK{\ifmmode \,\mu$K$\else \,$\mu$\hbox{K}\fi}
\def\muW{\ifmmode \,\mu$W$\else \,$\mu$\hbox{W}\fi}
\def\kms{\ifmmode $\,km\,s$^{-1}\else \,km\,s$^{-1}$\fi}
\def\kmsMpc{\ifmmode $\,\kms\,Mpc\mo$\else \,\kms\,Mpc\mo\fi}
\providecommand{\sorthelp}[1]{}

\def\eqref#1{(\ref{#1})}

\def\smica{{\tt SMICA}}

\def\LCDM{$\Lambda$CDM}

\def\eq{\begin{eqnarray}}
\def\qe{\end{eqnarray}}
\def\beq{\begin{equation}}
\def\eeq{\end{equation}}

\def\curl{\mathcal}
\def\({\left(}
\def\){\right)}

\def\and{\quad \mbox{and} \quad}
\def\barQ{\kern2pt\overline{\kern-2pt\curl{Q}}}

\def\barR{\kern2pt\overline{\kern-2pt\curl{R}}}

\def\nn{\nonumber}
\def\bargamma{\kern2pt\overline{\kern-2pt\gamma}}

\def\leaderfil{\leaders\hbox to 5pt{\hss.\hss}\hfil}

\newcommand{\be}{\begin{equation}}
\newcommand{\ee}{\end{equation}}
\newcommand{\bea}{\begin{eqnarray}}
\newcommand{\eea}{\end{eqnarray}}

\renewcommand{\L}[0]{\mathbf{L}}

\def\inv{^{-1}}

\newcommand{\spinup}{\;\raise1.0pt\hbox{$'$}\hskip-6pt\partial\;}
\newcommand{\spindown}{\;\overline{\raise1.0pt\hbox{$'$}\hskip-6pt\partial}\;}

\newcommand{\Tr}[1]{{\mathrm{Tr}\left[ #1 \right]}}

\def\bra{\langle}
\def\ket{\rangle}
\def\ld{\left}
\def\rd{\right}
\def\wtl{\widetilde}
\def\fr{\frac}
\def\dhat{{\hat{\mathbf{d}}}}
\def\nhat{{\hat{\mathbf{n}}}}

\def\vk{{\boldsymbol{k}}}
\def\vx{{\boldsymbol{x}}}
\def\Ph{{{\cal P}^{\rm hi}}}
\def\Pl{{{\cal P}^{\rm lo}}}

\def\del{\delta}
\def\oo{\frac{1}}
\def\rls{r_{\rm LS}}
\def\alme{a_{\ell m}^E}
\def\almb{a_{\ell m}^B}
\def\almei{a_{\ell m}^{E,i}}
\def\almbi{a_{\ell m}^{B,i}}
\def\almpei{a_{\ell'm'}^{E,i}}
\def\almpbi{a_{\ell'm'}^{B,i}}
\def\2Ylm{{_{\pm2}Y_{\ell m}}}
\def\p2Ylm{{_{\pm2}Y_{\ell'm'}}}

\begin{document}

\title{Testing physical models for dipolar asymmetry with CMB polarization}

\author{D. Contreras} \email{dagocont@phas.ubc.ca}
\affiliation{Department of Physics \& Astronomy\\
University of British Columbia, Vancouver, BC, V6T 1Z1  Canada}

\author{J. P. Zibin} \email{zibin@phas.ubc.ca}
\affiliation{Department of Physics \& Astronomy\\
University of British Columbia, Vancouver, BC, V6T 1Z1  Canada}

\author{D. Scott} \email{dscott@phas.ubc.ca}
\affiliation{Department of Physics \& Astronomy\\
University of British Columbia, Vancouver, BC, V6T 1Z1  Canada}

\author{A. J. Banday} \email{anthony.banday@irap.omp.eu}
\affiliation{Universit\'e de Toulouse\\
UPS-OMP, IRAP, F-31028 Toulouse cedex 4, France}
\affiliation{CNRS, IRAP\\
9 Av.\ colonel Roche, BP 44346, F-31-28 Toulouse cedex 4, France}

\author{K. M. G\'orski} \email{krzysztof.m.gorski@jpl.nasa.gov}
\affiliation{Jet Propulsion Laboratory, California Institute of Technology\\
4800 Oak Grove Drive, Pasadena, California, U.S.A}
\affiliation{Warsaw University Observatory\\
Aleje Ujazdowskie 4, 00-478 Warszawa, Poland}

\date{\today}

\begin{abstract}
The cosmic microwave background (CMB) temperature anisotropies exhibit a
large-scale dipolar power asymmetry.  To determine whether this is due to a
real, physical modulation or is simply a large statistical fluctuation requires
the measurement of new modes.  Here we forecast how well CMB polarization data
from \Planck\ and future experiments will be able to confirm or constrain
physical models for modulation.  Fitting several such models to the \Planck\
temperature data allows us to provide predictions for polarization asymmetry.
While for some models and parameters \Planck\ polarization will decrease error
bars on the modulation amplitude by only a small percentage, we show,
importantly, that cosmic-variance-limited (and in some cases even \Planck)
polarization data can decrease the errors by considerably better than the
expectation of $\sqrt 2$ based on simple $\ell$-space arguments. We project
that if the primordial fluctuations are truly modulated (with parameters as
indicated by \Planck\ temperature data) then \Planck\ will be able to make a
2$\sigma$ detection of the modulation model with 20--75\% probability,
increasing to 45--99\% when cosmic-variance-limited polarization is considered.
We stress that these results are quite model dependent.  Cosmic variance in
temperature is important: combining statistically isotropic polarization with
temperature data will spuriously increase the significance of the temperature
signal with 30\% probability for \Planck.
\end{abstract}

\maketitle

\section{Introduction}
\label{sec:Intro}

The largest scales of our cosmic microwave background (CMB) temperature sky
exhibit a dipolar asymmetry with an amplitude at the percent
level~\cite{Eriksen2004, Hanson2009, Bennett2011, planck2013-p09,
planck2014-a18}. While the cosmological origin of this signal is not in dispute
(existing in both WMAP and \Planck\ data with very different systematic effects
and frequency coverage), its statistical significance is debated (see
\cite{Bennett2011,planck2014-a18} and references therein).  Without correction
for the effects of {\em a posteriori} selection, this large-scale signal has a
significance at approximately the $3\sigma$ level and an amplitude of roughly
6--7\%, when restricted to a multipole range of $\ell \lesssim 65$.  Applying
look-elsewhere correction on the maximum multipole reduces the corresponding
$p$-value to of order 10\%~\cite{Bennett2011, planck2014-a18}.  However, the
asymmetry is present on scales that are roughly super-Hubble at last
scattering, suggesting a possible physical origin related to
very-early-Universe physics.  If this were the case, the former significance
estimate might be more relevant given a model that predicted it.

The dipolar asymmetry in temperature is now characterized as well as it ever
will be, because the measurements on these scales are limited by cosmic
variance and hence essentially all the cosmological information has already
been extracted.  The only way to unambiguously determine if this signal is
due to a physical modulation of fluctuations or is just a random statistical
fluctuation is to acquire new
information (i.e., measure new and independent modes \cite{Scott2016}). This
can be achieved with the addition of large-scale polarization data.
Polarization data are available from the \Planck\ satellite; however, residual
systematics (particularly at large angular scales) have so far prevented a full
investigation of the polarization dipole asymmetry signal.  Other probes of new
modes that have been examined in this context include large-scale structure
\cite{Hirata09,fvpbmb14,yhgks14,bafn14}, CMB lensing~\cite{Zibin2015,hbf16},
$21\,$cm measurements~\cite{smkr16}, and spectral distortions~\cite{pesky}.

The use of polarization in the investigation of dipole asymmetry has been
examined previously in Refs.~\cite{dph08, Paci2010, Chang2013, Paci2013, pesky,
Ghosh2015, Kothari2015, Kothari2015a, Namjoo2015, Bunn2016}.  In particular, it
was appreciated~\cite{dph08, Namjoo2015} that a modulation model must be
constructed in position (or $k$) space and propagated to map (or spherical
harmonic) space in order to consistently describe both temperature and
polarization modes. In addition, it was found that the predicted signature in
polarization is quite dependent on the assumed model for the $k$-space
modulation~\cite{Namjoo2015}.  Here we will use the temperature data to predict
the polarization signature given a $k$-space modulation model, following the
formalism developed in Ref.~\cite{Zibin2015}.

Naively, one might expect that the temperature signal, together with the
assumption of a physical modulation of the large-scale three-dimensional
fluctuation field, would predict a roughly 6--7\% asymmetry in polarization in
the $\ell$ range of 2 to 65. However, this is not necessarily the case, since
the mapping (transfer functions) from $k$ to $\ell$ space differs between
temperature and polarization. Therefore, in the absence of a detailed physical
modulation model we cannot use polarization to test for such a physical origin.
In addition, as stressed in Ref.~\cite{Zibin2015}, the signal in the
temperature data alone is not strong enough to pick out a well-defined
modulation scale dependence. In this paper we will explore how these issues are
modified with the inclusion of polarization data in more detail.

One well-known concern when considering polarization is the correlation between
temperature and gradient- (or $E$-) mode polarization.  More specifically,
modes we can measure from polarization are not completely independent of
temperature.  This correlation may alter a possible polarization asymmetry
signal or mimic such a signal in its absence. We deal with the correlation by
calibrating our estimator on \Planck\ Full Focal Plane (FFP8) temperature and
polarization simulations \cite{planck2014-a14} that are appropriately
correlated. A further concern is the spurious enhancement of a modulation
signal when including polarization. That is, simply due to cosmic variance and
noise, adding polarization to temperature might sometimes increase the
significance of a signal even when there is no true underlying modulation. This
is especially true when the original temperature signal is of low to moderate
significance, as is the case with our real CMB sky.  We address this concern by
quantifying the expected effect of adding polarization both with and without an
underlying physical modulation.

Our main goals are to determine how likely it is that a physical origin
for a modulation could be confirmed or refuted, and to quantify the expected improvement
in constraints on the $k$-space modulation model parameters, with the addition
of polarization data.  We will present projections for \Planck\ as well as a
cosmic-variance-limited polarization measurement.  We will \emph{not} perform a
blind multipole-space dipole asymmetry search as has been done with
temperature, though this can be done with the estimator we employ here and will
be important to perform once the data are available.

For this study we use the FFP8 cosmological parameters, with Hubble parameter
$H_0 = 100h\,{\rm km\,s}^{-1}{\rm Mpc}^{-1}$, where $h = 0.6712$, baryon
density $\Omega_bh^2 = 0.0222$, cold dark matter (CDM) density $\Omega_ch^2 =
0.1203$, neutrino density $\Omega_\nu h^2 = 0.00064$, cosmological constant
density parameter $\Omega_\Lambda = 0.6823$, primordial comoving curvature
perturbation power spectrum amplitude $A_{\rm s} = 2.09\times10^{-9}$ (at pivot
scale $k_0 = 0.05\,{\rm Mpc}^{-1}$) and tilt $n_{\rm s} = 0.96$, and optical
depth to reionization $\tau = 0.065$.

Note that at high $\ell$ our results would be biased due to the effect of
aberration \cite{Aghanim:2013suk}, which is not present in the FFP8 simulations
\cite{planck2014-a14}. In Appendix~\ref{sec:aberration} we describe how we
detect aberration in the temperature data and remove it so as not to bias our
results.

\section{Modulation approach}
\label{sec:models}

\subsection{Formalism}
\label{sec:formn}

Our goal is to construct position- or $k$-space models that generate
scale-dependent dipolar asymmetry, while remaining agnostic as to the detailed
origin (presumably inflationary) of the modulation.  Based on the temperature
signal, we would like to modulate the largest scales while maintaining
consistency with the usual isotropic $\Lambda$CDM power spectra. We consider
three different types of model: the first is a modulated adiabatic mode which
comprises a part of the total primordial power spectrum~\footnote{One
particularly motivated model of this form would be a modulated integrated
Sachs-Wolfe component, which will be examined in~\cite{short}.}; the second is
a modulated CDM isocurvature mode; and the third is a modulated tensor
mode~\footnote{We note that a modulation of the optical depth to
reionization~\cite{pesky} is likely extremely disfavoured since it would
imply temperature asymmetry to the smallest scales, contrary to the
observations~\cite{hansenetal09,Hanson2009,fh13,planck2014-a18,qn15,awkkf15}.}.
For the adiabatic scalar case, we must modulate a large-scale part of the
spectrum.  The contributions from CDM isocurvature and tensor modes, however,
are naturally restricted to scales $\ell \lesssim 100$ (at least for
near-scale-invariant spectra).  Therefore in these cases we only need to apply
a scale-invariant modulation to the tensor or isocurvature component.  We will
analyze each of these models with a slight generalization of the approach in
\cite{Zibin2015}, which considered only the adiabatic scalar case, and we refer
to that reference for full details.  Our approach will be readily applicable to
other modulation models.

We begin by decomposing the primordial fluctuations into two components.  The
first, $\wtl Q^{\rm lo}(\vx)$, is spatially linearly modulated, and hence
its intersection with our last-scattering surface will be dipole modulated.
It takes the form
\beq
\wtl Q^{\rm lo}(\vx) = Q^{\rm lo}(\vx)\ld(1 + A\fr{\vx\cdot\dhat}{r_{\rm LS}}\rd),
\label{modflucts}
\eeq
where $Q^{\rm lo}(\vx)$ is statistically isotropic with power spectrum
$\Pl(k)$, $A$ is the modulation amplitude, $\dhat$ is the direction of
modulation, and $r_{\rm LS}$ is the comoving distance to last scattering.  The
second, unmodulated component, $Q^{\rm hi}(\vx)$, is statistically isotropic
with power spectrum $\Ph(k)$.  The two fields are uncorrelated, i.e.,
\beq
\ld\bra Q^{\rm lo}(\vk)Q^{\rm hi*}(\vk')\rd\ket = 0.
\eeq
The field $\wtl Q^{\rm lo}(\vx)$ will correspond to the isocurvature, tensor,
or modulated adiabatic component, while $Q^{\rm hi}(\vx)$ will be the
remaining, unmodulated adiabatic component.  The superscripts ``lo'' and ``hi''
refer to the fact that generally these components will dominate at low and high
$k$, respectively.  Strictly, we should consider only amplitudes $A \le 1$,
since for larger $A$ the fluctuations in Eq.~(\ref{modflucts}) will vanish
somewhere inside the last scattering surface and the details in this case may
depend on the specific (presumably inflationary) realization of the model.

As shown in Ref.~\cite{Zibin2015}, the total temperature anisotropies will be
given to very good approximation by the sum of the uncorrelated contributions
from the modulated and unmodulated fluctuations, i.e.,
\beq
  \frac{\delta T(\nhat)}{T_0} = \frac{\delta T^{\rm lo}(\nhat)}{T_0}\ld(1 +
  A\nhat\cdot\dhat\rd) + \frac{\delta T^{\rm hi}(\nhat)}{T_0}.
\label{eq:totTanis}
\eeq
The anisotropies $\delta T^{\rm lo}/T_0$, with power spectrum $C_\ell^{T,\rm
lo}$, are generated by the perturbations with power spectrum $\Pl(k)$, while
$\delta T^{\rm hi}/T_0$, with spectrum $C_\ell^{T,\rm hi}$, are generated by
the uncorrelated perturbations with power spectrum $\Ph(k)$.  The form of
Eq.~(\ref{eq:totTanis}) is easy to understand in the limit where the
anisotropies are much smaller than the length scale of modulation (i.e.,
$r_{\rm LS}$).  The large-scale case is less obvious, but
Eq.~(\ref{eq:totTanis}) still holds to very good
approximation~\cite{Zibin2015}.  We have ignored the ISW effect here, since
it would introduce a negligible modification \cite{Zibin2015}.

   For the $E$-mode polarization anisotropies, we similarly use
\beq
  \frac{\delta E(\nhat)}{T_0} = \frac{\delta E^{\rm lo}(\nhat)}{T_0}\ld(1 +
  A\nhat\cdot\dhat\rd) + \frac{\delta E^{\rm hi}(\nhat)}{T_0}.
\label{eq:Emod}
\eeq
Due to the effects of reionization and the non-local definition of $E$ modes,
Eq.~(\ref{eq:Emod}) becomes inaccurate on the very largest scales, $\ell
\lesssim 10$.  In Appendix~\ref{app:EBmod}, we derive the effect of a
spatially linear modulation on the $E$ (and $B$) modes, taking the
non-locality into effect.  We find that omiting this correction results in
a bias to our recovered modulation
amplitudes of roughly $3\%$ in the worst case (the most red-tilted power
law model we consider).  Hence we do not apply this correction here (but
plan to implement it in future work).  In Appendix~\ref{app:EBmod}, we also
show that the correct treatment results in novel couplings between
$B$ modes and $E$ or $T$ modes for a linear modulation.  The very-large-scale
nature of the corrections implies weak statistical weight, but
these unusual couplings could in principle be used to constrain large-scale
signals.

In terms of spherical harmonic coefficients, the total fluctuations in
Eqs.~(\ref{eq:totTanis}) or (\ref{eq:Emod}) can be written as
\beq
a_{\ell m} = a_{\ell m}^{\rm lo} + a_{\ell m}^{\rm hi}
   + \sum_M \Delta X_M \sum_{\ell'm'}a_{\ell'm'}^{\rm lo}\xi^M_{\ell m\ell'm'},
\label{eq:modulatedmodes}
\eeq
where $a_{\ell m}^{\rm lo}$ are the statistically isotropic modes, and the
$\Delta X_M$ are the spherical harmonic decomposition of $A\nhat\cdot\dhat$
(the dipolar nature ensures that $M = -1, 0, 1$). The $\xi^M_{\ell m\ell'm'}$
are coupling coefficients defined by
\beq
\xi^M_{\ell m\ell'm'} \equiv \sqrt{\fr{4\pi}{3}}
   \int Y_{\ell'm'}(\nhat)Y_{1M}(\nhat)Y_{\ell m}^*(\nhat)d\Omega_\nhat,
\label{eq:xidef}
\eeq
and given explicitly by
\begin{align}
 \xi^0_{\ell m\ell'm'} &= \delta_{m'm}\left(\delta_{\ell'\ell-1}A_{\ell-1\,m}
+ \delta_{\ell'\ell+1}A_{\ell m}\right), \\
 \xi^{\pm 1}_{\ell m\ell'm'} &= \delta_{m'm\mp 1}\left(
\delta_{\ell'\ell-1}B_{\ell-1\,\pm m-1} - \delta_{\ell'\ell+1}B_{\ell\,\mp m}
\right),
\label{eq:couplingcoeffs}
\end{align}
where
\begin{align}
 A_{\ell m} &= \sqrt{\frac{(\ell+1)^2 - m^2}{(2\ell+1)(2\ell+3)}},
 \label{eq:Acoeff} \\
 B_{\ell m} &= \sqrt{\frac{(\ell+m+1)(\ell+m+2)}{2(2\ell+1)(2\ell+3)}}.
 \label{eq:Bcoeff}
\end{align}

From Eq.~\eqref{eq:modulatedmodes} we can find the covariance of the total
temperature or polarization anisotropy multipoles to first order in the
modulation amplitude $\Delta X \equiv \sqrt{\sum_M|\Delta X_M|^2} = A$:
\begin{align}
C_{\ell m\ell'm'} &\equiv \bra a_{\ell m}a_{\ell'm'}^*\ket\\
   &= C_\ell\delta_{\ell\ell'}\delta_{mm'} +
      \frac{\delta C_{\ell\ell'}}{2}\sum_M \Delta X_M\xi^M_{\ell m\ell'm'},
\label{eq:cmbcovariance}
\end{align}
where $\delta C_{\ell\ell'} \equiv 2(C_\ell^{\rm lo} + C_{\ell'}^{\rm lo})$.
The above equation explicitly shows that a dipole modulation will lead to
coupling of $\ell$ to $\ell\pm1$ modes in the multipole
covariance~\cite{pubb05}. In Eq.~(\ref{eq:cmbcovariance}) $C_\ell$ is the the
total isotropic power spectrum, which, since the two fluctuation components are
uncorrelated, is given to linear order in the asymmetry by
\beq
C_\ell^T = C_\ell^{T,\rm lo} + C_\ell^{T,\rm hi},
\label{eq:Cltot}
\eeq
and similarly for polarization.  Clearly, $C_\ell^T$ must be consistent with
measurements of the isotropic power.  For the adiabatic cases, we take
\begin{equation}
\mathcal{P}^{\rm lo}_\mathcal{R}(k) + \mathcal{P}^{\rm hi}_\mathcal{R}(k)
   = \mathcal{P}^{\Lambda{\rm CDM}}_\mathcal{R}(k),
\label{lohiLCDM}
\end{equation}
where
\beq
\mathcal{P}^{\Lambda{\rm CDM}}_\mathcal{R}(k)
   = A_{\rm s}\left(\frac{k}{k_0}\right)^{n_{\rm s} - 1}
\label{LCDMPS}
\eeq
is the comoving curvature power spectrum for $\Lambda$CDM, so that the
isotropic power constraints are automatically satisfied.  However, constraints
on isocurvature and tensor isotropic contributions~\cite{planck2014-a24,
Ade:2015tva,BKP16} should be incorporated in addition to the constraints from
the temperature asymmetry in order to obtain the tightest constraints for those
models~\cite{short}.

\subsection{Adiabatic modulation}
\label{sec:adiabatic_model}
\subsubsection{$\tanh$ spectrum}
\label{sec:tanh}

For scalar adiabatic modes~\cite{Zibin2015} the fluctuation fields $Q^{\rm
lo}(\vx)$ and $Q^{\rm hi}(\vx)$ both correspond to the comoving curvature
perturbation, $\cal{R}$.  Our first specific model is intended to capture a
large-scale modulation with a small number of parameters. We choose a modulated
component spectrum of the form
\begin{align}
  \mathcal{P}^{\rm lo}_{\mathcal{R}}(k)
  &= \frac{1}{2} \mathcal{P}^{\Lambda{\rm CDM}}_\mathcal{R}(k)
     \left[1 - \tanh{\left(\frac{\ln k - \ln k_{\rm c}}{\Delta\ln k}\right)}\right].
  \label{eq:power_low}
\end{align}
This smooth step function in $k$ ensures that mainly the largest scales, $k
\lesssim k_{\rm c}$, are modulated.  The quantity $\Delta\ln k$ determines the
sharpness of the transition from modulated to unmodulated scales.  The other
parameters of the model are the amplitude of modulation ($A = A_\mathcal{R}$)
and its direction ($l,\,b$) in Galactic coordinates.  The unmodulated
contribution is fixed by Eq.~(\ref{lohiLCDM}).  It must be noted that the
temperature data alone are not strong enough to constrain all five modulation
parameters ($k_{\rm c},\,\Delta \ln k,\,A_\mathcal{R},\,l,\,b$); see
Ref.~\cite{Zibin2015}.  The shapes of the best-fit asymmetry spectra for the
$\tanh$ model (and all of the others) will be illustrated in
Sec.~\ref{sec:tempresults}.

As shown in detail in~\cite{Zibin2015}, the fact that we have split the
adiabatic fluctuations into two uncorrelated parts does not restrict the
modulation mechanism (presumably inflationary) in any way. Instead this is a
convenient way to describe an adiabatic spectrum modulated with arbitrary scale
dependence.

\subsubsection{Power-law spectrum}

The $\tanh$ modulation model in Eq.~(\ref{eq:power_low}) is not explicitly
motivated by any early-Universe model.  Perhaps better motivated would be a
simple power-law modulation, i.e.,
\beq
\mathcal{P}^{\rm lo}_{\mathcal{R}}(k)
   = \mathcal{P}^{\Lambda{\rm CDM}}_\mathcal{R}(k_0^{\rm lo})
     \left(\frac{k}{k_0^{\rm lo}}\right)^{n^{\rm lo}_{\rm s} - 1},
   \label{eq:powerlaw}
\eeq
where $n^{\rm lo}_{\rm s}$ is the tilt and $k_0^{\rm lo}$ is the pivot scale of
the modulated component of fluctuations.  This model will be abbreviated
``ad.-PL''.  We again impose Eq.~(\ref{lohiLCDM}) in order to define the
unmodulated $\mathcal{P}^{\rm hi}_\mathcal{R}(k)$. We consider only red tilts
with $n^{\rm lo}_{\rm s} \le n_{\rm s}$, and choose $k_0^{\rm lo} =
1.5\times10^{-4}\,{\rm Mpc}^{-1}$, which corresponds roughly to quadrupolar
angular scales.  Larger $k_0^{\rm lo}$ would contradict the positivity of
$\mathcal{P}^{\rm hi}_\mathcal{R}(k)$ on the largest observable scales, while
smaller $k_0^{\rm lo}$ would be degenerate with the modulation amplitude.
Again we modulate $\mathcal{P}^{\rm lo}_{\mathcal{R}}(k)$ with amplitude
$A_\mathcal{R}$, so the total modulation fraction approaches $A_\mathcal{R}$ on
large angular scales. This model also should be taken with some caution, since
for large departures from scale invariance (i.e., $1 - n^{\rm lo}_{\rm s}
\hskip0.4mm\not\hskip-0.4mm\ll 1$), we might expect higher-order terms such as
running, running of running, etc.\ in the modulation spectrum.

\subsubsection{Modulated scalar spectral index}
\label{sec:nsmodel}

Next we consider a single-component adiabatic model with a linear gradient in
the tilt, $n_{\rm s}$, of the primordial power spectrum~\cite{Moss2011}.  We
abbreviate this
model ``$n_{\rm s}$-grad''.  In this case we do not strictly follow the
two-component formalism of Sec.~\ref{sec:formn}, but instead can directly write
the asymmetry spectrum as~\cite{Moss2011}
\beq
C_\ell^{\rm lo} = -\fr{1}{2}\fr{dC_\ell^{\Lambda{\rm CDM}}}{dn_{\rm s}}.
\label{tiltCllo}
\eeq
Here we have used a linear approximation for the effect of the gradient, which
will be well justified by our results.  We allow for free modulation amplitude,
$\Delta n_{\rm s}$, and let the tilt pivot scale, $k_*$, vary.  Note that this
treatment is degenerate with fixed pivot $k_*$ and additional modulation of the
primordial amplitude, $A_{\rm s}$.  Since a modulation of tilt produces extra
power on large scales in the $-\dhat$ direction, we have included a minus sign
in Eq.~(\ref{tiltCllo}) so that the best-fit modulation directions will be
directly comparable to those of the other models.

\subsection{Tensor modulation}
\label{sec:tensmodel}

The possibility that a modulated tensor component is present is particularly
well motivated observationally \cite{pesky,Scott:2014qea, Chluba2014}, since the
contribution of (near-scale-invariant or red-tilted) tensors is negligible at
small scales.  Tight constraints on the tensor-to-scalar ratio $r$ will make it
difficult to achieve sufficient modulation, however, via the isotropic power
constraint~\cite{Chluba2014,short}.

In this case we take the unmodulated component $Q^{\rm hi}(\vx)$ to be
adiabatic fluctuations with the standard $\Lambda$CDM form, Eq.~\eqref{LCDMPS}.
The modulated component $Q^{\rm lo}(\vx)$ will be a scale-invariant tensor
contribution, i.e.,
\begin{align}
  \mathcal{P}_{\rm t}(k) &=
  r_{0.05}\mathcal{P}^{\Lambda\rm{CDM}}_\mathcal{R}(k_0),
  \label{eq:tensormodel}
\end{align}
with $r_{0.05} = 0.07$ (this is the 95\% upper limit from~\cite{BKP16}) at
pivot scale $k_0 = 0.05\,$Mpc$^{-1}$.  The tensor modes
are uncorrelated with the adiabatic fluctuations (note that even in the
presence of such correlations, the scalar and tensor anisotropy power spectra
would still be additive~\cite{zibin14}) and modulated in a
\emph{scale-invariant} way with amplitude $A = A_{\rm t}$.  As before, the
tensors produce anisotropy power $C^{T,\rm lo}_\ell$ (and similarly for $E$).

\subsection{Isocurvature modulation}
\label{sec:isomodel}

Also well motivated as a modulation model is the CDM isocurvature spectrum,
since it naturally contributes mainly at large angular scales for
near-scale-invariant (or red-tilted) isocurvature modes~\cite{ehk09}.  In
this case we
assume that the unmodulated $Q^{\rm hi}(\vx)$ is an adiabatic contribution
which takes the standard, $\Lambda$CDM form, Eq.~\eqref{LCDMPS}. The modulated
part $Q^{\rm lo}(\vx)$ will be the isocurvature component, which we take to be
scale-invariant, i.e.,
\beq
\mathcal{P}_\mathcal{I}(k)
 = \frac{\alpha}{1 - \alpha}\mathcal{P}^{\Lambda{\rm CDM}}_\mathcal{R}(k_0),
\label{eq:isomodel}
\eeq
where we use the same pivot scale $k_0$ as for the tensors. The isocurvature
modes are taken to be uncorrelated with the adiabatic fluctuations and
modulated in a {\em scale-invariant} way with amplitude $A = A_{\mathcal{I}}$.
This isocurvature spectrum then determines the modulated component of the CMB
fluctuations, $C^{\rm lo}_{\ell}$. The isocurvature fraction, $\alpha$, should
properly be constrained by the isotropic \Planck\ likelihood~\cite{short}, but
here we simply choose the \Planck\ upper limit for uncorrelated, {\em
scale-invariant} isocurvature, $\alpha = 0.04$ \cite{planck2014-a24}.

\section{Dipole asymmetry estimator}

In this section we describe the estimator that we use to extract modulation
parameters from data or simulations given a modulation model. This estimator is
applied in harmonic space, exploiting the fact that (to leading order in the
anisotropy) dipole modulation is equivalent to the coupling of $\ell$ with
$\ell\pm1$ modes, as we saw in Sec.~\ref{sec:formn}.

\subsection{Connection to previous approaches}

The estimator that we use was originally developed for temperature
data~\cite{Moss2011,planck2014-a18, Zibin2015}, but can equally be applied to
polarization (subject to the caveats discussed in Sec.~\ref{sec:formn}).
Our current implementation includes new improvements to the
treatment compared with~\cite{planck2014-a18, Zibin2015} in order to account
for the expectation of noisy polarization data. Differences in the
implementation of the estimator between this work and Ref.~\cite{Zibin2015} are
outlined in Appendix~\ref{sec:filter}, while the consequent differences in
temperature results can be seen by comparing Table~\ref{tab:bestfitparams} with
table~I of Ref.~\cite{Zibin2015} (though the differences are not significant).
Here we present a condensed description of our estimator (the full details are
in Appendix C of \cite{planck2014-a18}).

We note that our estimator is essentially identical to that of
Ref.~\cite{Hanson2009} and can also be rewritten in terms of BiPoSH
coefficients \cite{Hajian2003, Hajian2006}; however, the physical motivation
for this approach comes from Ref.~\cite{Moss2011}, where general cosmological
parameter modulations were explored.  Our implementation to deal with masking
and noise uses an inverse-variance filter on the spherical harmonic
coefficients that optimally account for the masking, as used in
Refs.~\cite{planck2013-p12, planck2014-a17, planck2014-a18}.

\subsection{Full-sky, noise-free case}
\label{sec:blindderiv}

From Eq.~\eqref{eq:cmbcovariance} it is clear that the multipole covariance can
be decomposed into an isotropic part and a small anisotropic part proportional
to $\Delta X_M$. In other words, we can make the identification $C_{\ell
m\ell'm'} = C_I + C_A$, with $C_I$ being the first term on the right-hand side
of Eq.~\eqref{eq:cmbcovariance} and $C_A$ being the second term. The inverse
covariance matrix can then be written as $C^{-1}_I - C^{-1}_I C_A C^{-1}_I$ (to
linear order in the anisotropy).  The best-fit $\Delta X_M$ values are given by
\begin{align}
 \Delta X_M &= \frac{d^\dagger C^{-1}_I \partial C_A/\partial \Delta X_M\,
 C^{-1}_I d}
 {\Tr{C^{-1}_I \partial C_A/\partial\Delta X_M\,C^{-1}_I \partial
 C_A/\partial \Delta X_M}},
 \label{eq:genest}
\end{align}
for multipole vector $d$.  These can be written more explicitly as
\begin{align}
 \Delta X_0 &= \frac{6\sum_{\ell m}\delta
C_{\ell\ell+1}C_{\ell}^{-1}C_{\ell+1}^{-1}A_{\ell m}a^*_{\ell
m}a_{\ell+1\,m}}{\sum_{\ell}\delta C_{\ell\ell+1}^2C_\ell^{-1}
C_{\ell+1}^{-1}(\ell + 1)},
 \label{eq:mossest0}\\
 \Delta X_{+1} &= \frac{6\sum_{\ell m}\delta
C_{\ell\ell+1}C_{\ell}^{-1}C_{\ell+1}^{-1}B_{\ell m}a^*_{\ell
m}a_{\ell+1\,m+1}}{\sum_{\ell}\delta C_{\ell\ell+1}^2C_\ell^{-1}
C_{\ell+1}^{-1}(\ell + 1)},
 \label{eq:mossest1}
\end{align}
and $\Delta X_{-1} = -\Delta X^*_{+1}$. We note that Eq.~\eqref{eq:genest} is
completely general and can be used to examine modulation of any kind (i.e.,
beyond simple dipolar modulation). The cosmic variance of the modulation
amplitude estimator is given by
\beq
\sigma_X^2 \equiv \ld\bra|\Delta X_M|^2\rd\ket = 12\ld(\sum_\ell(\ell + 1)
             \fr{\delta C_{\ell\ell + 1}^2}{C_\ell C_{\ell + 1}}\rd)^{-1}.
\label{eq:sigmax}
\eeq

As we mentioned in Sec.~\ref{sec:formn}, we strictly only consider values of
the modulation amplitude ($A_\mathcal{R}$, $\Delta n_{\rm s}$, $A_{\rm t}$, or
$A_\mathcal{I}$) less than unity.  Note, however, that the $\mathcal{O}(\Delta
X^2)$ terms which were dropped in the multipole covariance,
Eq.~(\ref{eq:cmbcovariance}), couple $\ell$ with $\ell$ and $\ell \pm 2$.
Therefore the estimator, Eqs.~(\ref{eq:mossest0}) and (\ref{eq:mossest1}), will
be insensitive to them.  In other words, our approach will recover modulation
amplitudes as large as unity without bias: we have no need to restrict to
small $\Delta X$.  However, the results will not be optimal, and could be
improved for large $\Delta X$ by incorporating these extra couplings into
the estimator.

\subsection{Realistic skies}

The estimators presented in Sec.~\ref{sec:blindderiv} are only adequate for
full sky coverage with no noise. In this subsection we show how we include the
effects of masking and noise in the data.

The combination of $C^{-1}_I d$ in Eq.~(\ref{eq:genest}) is suggestive that we
should apply an inverse-covariance filter to the data.  This is exactly what we
do. The effects of masking are readily dealt with by employing inverse-variance
filtering to the data, as described in Refs.~\cite{planck2013-p12,
planck2014-a17}. Pixels within the mask are given infinite variance and thus
are given zero weight, which optimally accounts for masking effects. The
effects of inhomogeneous noise could also readily be handled by including its
variance contribution in the inverse-variance filter.  However, we have not
included these effects, which means we will have a slightly suboptimal though
still \emph{unbiased} estimate.  Residual effects not captured by our approach
(e.g. inhomogeneous noise) are handled by subtracting a mean field term,
derived from simulations with the same foreground and noise properties as the
data. We note that this same approach has been used for lensing estimators in
Refs.~\cite{planck2013-p12, planck2014-a17} and specifically for temperature
dipole modulation in Refs.~\cite{planck2014-a18,Zibin2015}.  Cut-sky
aberration effects~\cite{jcdkw14} are taken into account by removing
aberration from the data as described in Appendix~\ref{sec:aberration}.

The estimator, Eqs.~(\ref{eq:mossest0})--(\ref{eq:mossest1}), then becomes
\begin{align}
 \tilde{X}^{WZ}_0 &= \frac{6\sum_{\ell m} \delta C^{WZ}_{\ell
 \ell+1}A_{\ell m} S^{(WZ)}_{\ell m\,\ell+1\, m+M}}
 {\sum_{\ell}\ld(\delta C^{WZ}_{\ell \ell +1}\rd)^2(\ell +
 1)F^{(W}_{\ell}F^{Z)}_{\ell+1}},
 \label{eq:est0}\\
 \tilde{X}^{WZ}_{+1} &= \frac{6\sum_{\ell m} \delta C^{WZ}_{\ell
 \ell+1}B_{\ell m} S^{(WZ)}_{\ell m\,\ell+1\, m+M}}
 {\sum_{\ell}\ld(\delta C^{WZ}_{\ell \ell +1}\rd)^2(\ell +
 1)F^{(W}_{\ell}F^{Z)}_{\ell+1}},
\label{eq:est1}
\end{align}
with
\begin{align}
  S^{WZ}_{\ell m\ell'm'} &\equiv W^*_{\ell m} Z_{\ell'm'} - \left<W^*_{\ell
  m} Z_{\ell'm'}\right>.
  \label{eq:transformed_mfc}
\end{align}
Here, $WZ = {TT, TE, EE}$; $W_{\ell m}$ and $Z_{\ell m}$ are inverse-covariance
filtered data; $F^{W}_\ell \simeq \left< W_{\ell m} W^*_{\ell m} \right>$; and
the last term on the right-hand-side of Eq.~\eqref{eq:transformed_mfc} denotes
the mean-field correction (details including the precise form of $F^{W}_\ell$
can be found in Appendix A.1 of Ref.~\cite{planck2014-a17} and
Appendix~\ref{sec:filter} of this paper). The parentheses in the superscripts
indicate symmetrization over the enclosed variables.

The estimators of Eqs.~\eqref{eq:est0}--\eqref{eq:est1} can be combined with
inverse-variance weighting over all data combinations ($TT$, $TE$, $EE$) to
obtain a combined minimum-variance estimator, given by
\begin{align}
\Delta \tilde{X}_M &=
  \frac{\sum_{WZ} \Delta\tilde{X}^{WZ}_M\left(\sigma_X^{WZ}\right)^{-2}}
       {\sum_{WZ}\left(\sigma_X^{WZ}\right)^{-2}}.
  \label{eq:fullest}
\end{align}
Here we calculate the variance from the scatter of simulations, although they
agree closely with the Fisher errors given in Ref.~\cite{planck2014-a18}.

\section{Results}
\label{sec:results}

\subsection{Temperature only}
\label{sec:tempresults}

In this section we present constraints using \Planck\ temperature data only.
Specifically, we use the \smica\ 2015 temperature solution, one of four
\Planck\ component-separation methods~\cite{planck2014-a11}, all of which
produce very similar results~\cite{planck2014-a18}.  The temperature data are
evaluated up to a maximum multipole $\ell_{\max} = 1000$ (no significant
difference was found when extending to higher multipoles), with the exception
of the modulated $n_{\rm s}$ model, which uses $\ell_{\max} = 2000$.

\begin{figure*}
\begin{center}
\includegraphics[width=0.45\hsize]{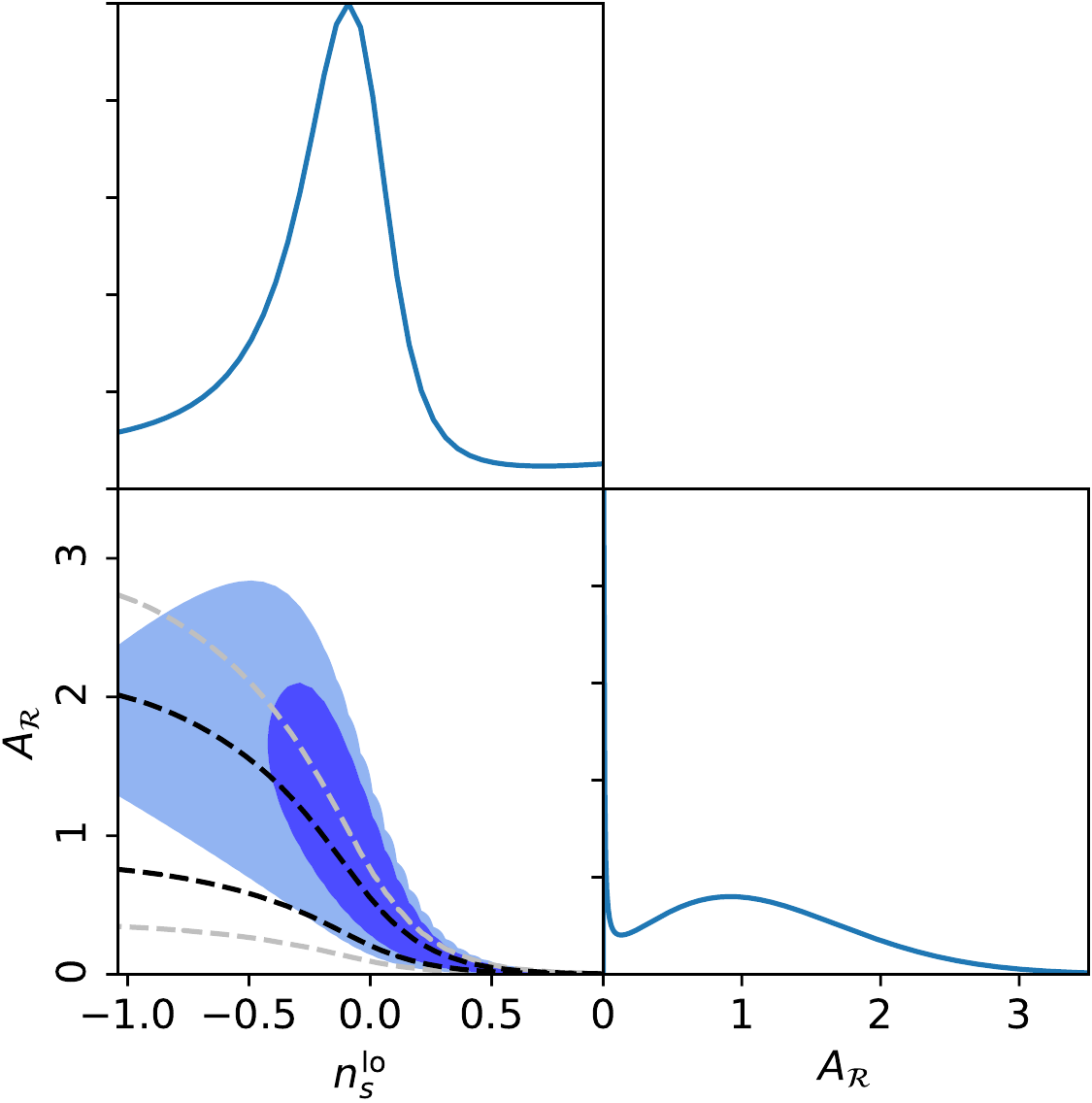}
\includegraphics[width=0.45\hsize]{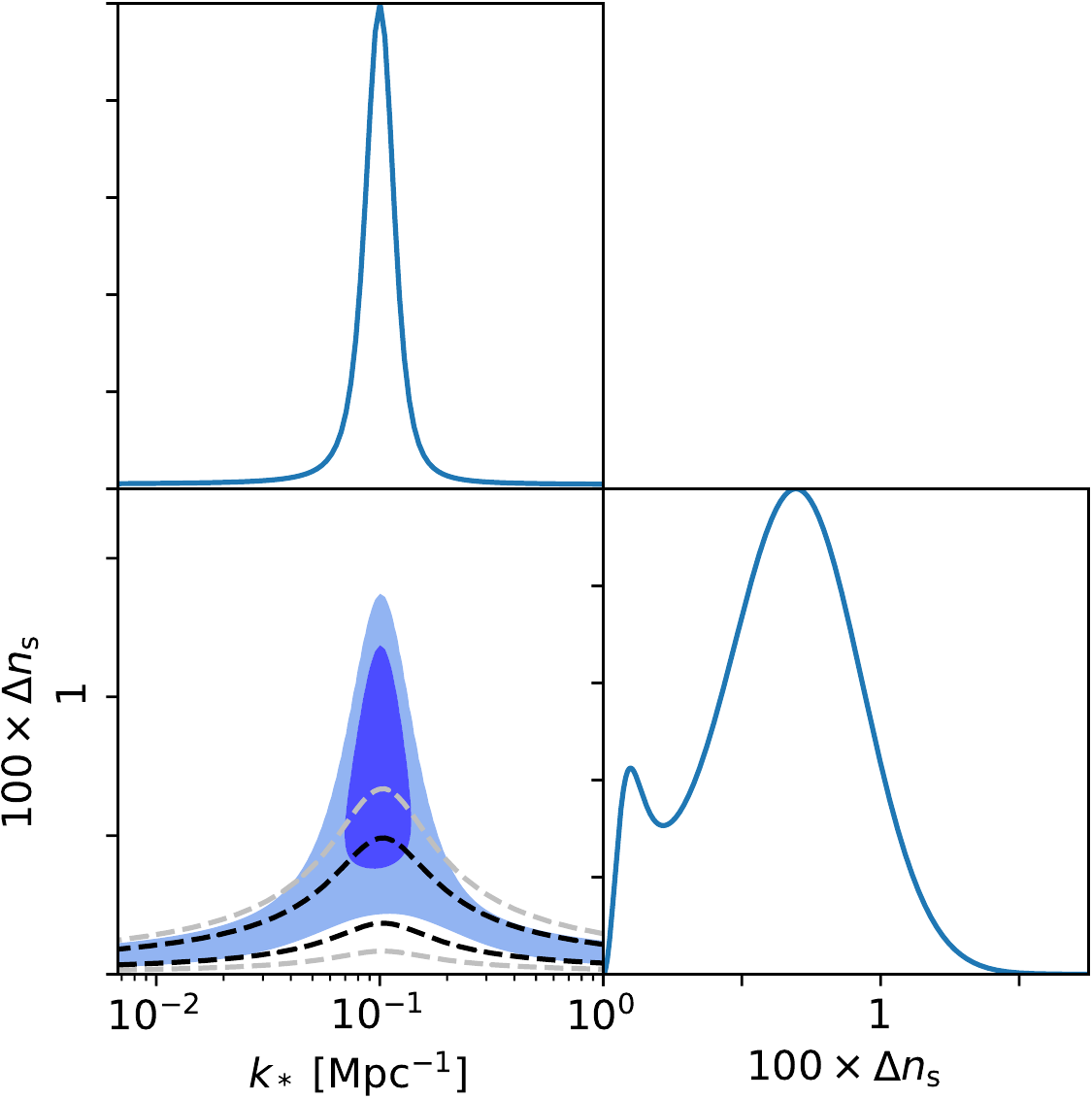}
\includegraphics[width=0.5\hsize]{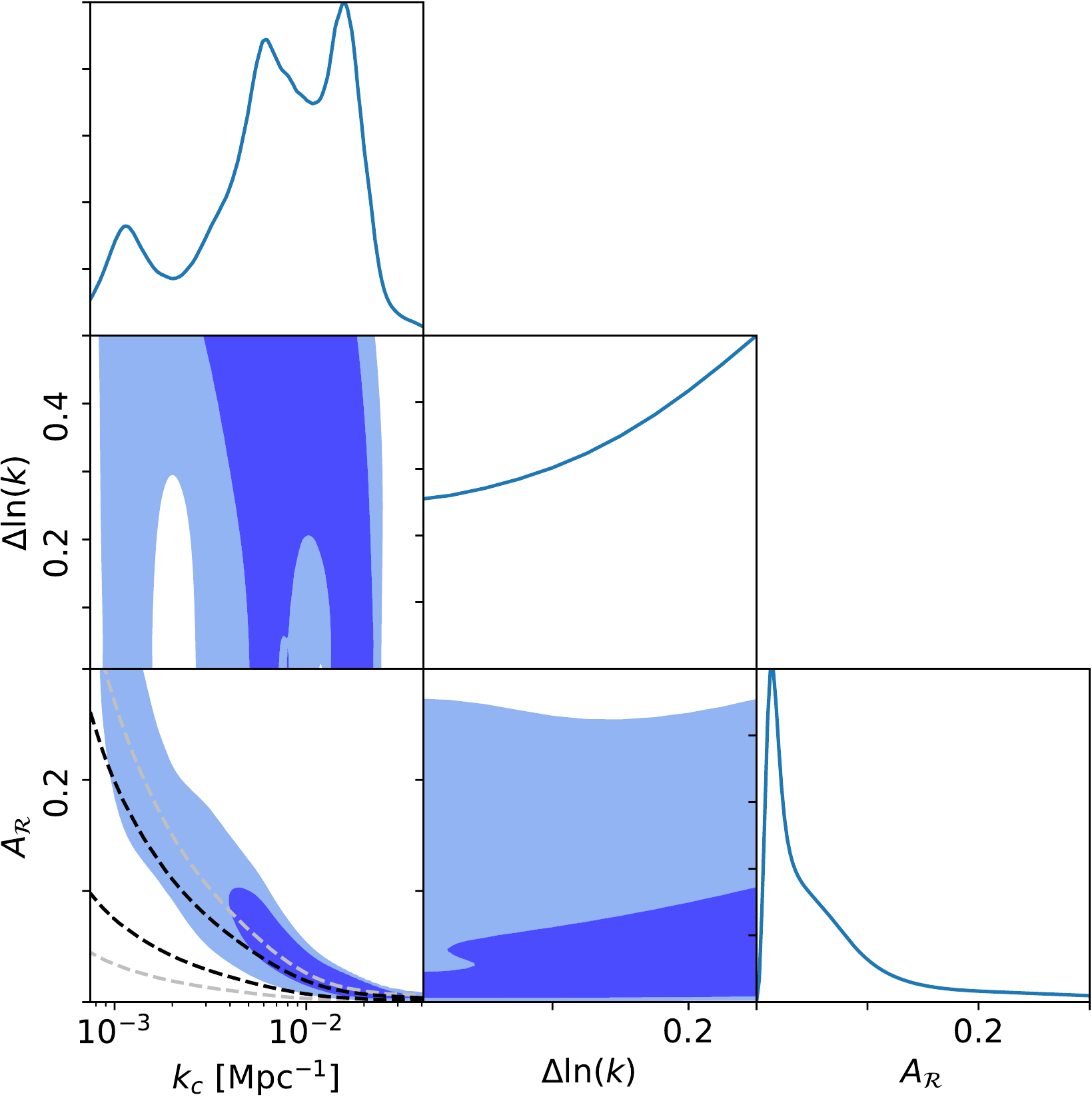}
\end{center}
\caption{Marginalized posteriors for the adiabatic power-law (top left triangle
plot), $n_{\rm s}$ gradient (top right), and $\tanh$ (bottom) models, using
\Planck\ temperature data only.  Dark and light blue regions enclose 68\% and
95\% of the likelihood, respectively. The black dashed curves represent the
theoretical distributions of the parameters coming solely from cosmic variance
in statistically isotropic skies.}
\label{fig:powerlawtriangle}
\end{figure*}

Fig.~\ref{fig:powerlawtriangle} shows the posteriors of the modulation
parameters for our adiabatic models.  Results for the $\tanh$ model have
already been commented on in \cite{Zibin2015}; here we simply reiterate that
the model is (unsurprisingly) not constrained well with temperature alone. The
pile-up of the posterior at low modulation amplitudes (present for all models,
though most notably for the adiabatic power-law model) is a volume effect that
arises due to our choice of prior, which expects uniform posteriors for the
$k$-space parameters ($\ln k_{\rm c}, \Delta \ln k, \ln k_*, n^{\rm lo}_s$)
for statistically isotropic data.

The best-fit modulation parameters for all models are presented in
Table~\ref{tab:bestfitparams}. We see that the modulation amplitudes required
for the tensor and isocurvature models exceed unity, i.e., the best fits have
$A_{{\rm t}},\, A_{\mathcal{I}} > 1$.  This suggests that, for the case of
tensor and isocurvature isotropic contributions at the current upper limits
($r_{0.05} = 0.07$ and $\alpha = 0.04$), maximal modulation (i.e., modulation
amplitude unity) is insufficient to produce the observed temperature asymmetry.
For this reason we do not discuss these models further here. This result will
be addressed more quantitatively in~\cite{short}.  Note that this conclusion
for a specific isocurvature model was originally made in
\cite{planck2014-a24} and the difficulty of providing sufficient modulation
for tensor models was discussed in~\cite{pesky,Scott:2014qea,Chluba2014}.
Furthermore, the best-fit value of $\Delta n_{\rm s}
= 0.014$ for the $n_{\rm s}$-grad model is tighter than the corresponding
value of $\Delta n_{\rm s} = 0.03$ in~\cite{pesky}.  This illustrates the
constraining power of the small-scale anisotropies in our analysis.

\begin{table}[!htp]
\begin{tabular}{lccccc}
\hline
\hline
Parameter                           & $\tanh$  & ad.-PL   & $n_{\rm s}$-grad & tensors  & isocurvature \\
\hline
$10^3 k_{\rm c}\, [{\rm Mpc}^{-1}]$ & $7.45$   & $\cdots$ & $\cdots$         & $\cdots$ & $\cdots$     \\
$\Delta \ln k$                      & $0.5$    & $\cdots$ & $\cdots$         & $\cdots$ & $\cdots$     \\
$n_{\rm s}^{\rm lo}$                & $\cdots$ & $-0.09$  & $\cdots$         & $\cdots$ & $\cdots$     \\
$k_*\, [{\rm Mpc}^{-1}]$            & $\cdots$ & $\cdots$ & $0.10$           & $\cdots$ & $\cdots$     \\
$\Delta X$                          & $-0.065$ & $-0.457$ & $-0.011$         & $-2.2$   & $-1.2$       \\
$\Delta Y$                          & $-0.044$ & $-0.566$ & $-0.008$         & $-2.0$   & $-1.0$       \\
$\Delta Z$                          & $-0.035$ & $-0.499$ & $-0.004$         & $-1.3$   & $-0.8$       \\
$A_{\mathcal{R}}$                   & $0.086$  & $0.882$  & $\cdots$         & $\cdots$ & $\cdots$     \\
$\Delta n_{\rm s}$                  & $\cdots$ & $\cdots$ & $0.014$          & $\cdots$ & $\cdots$     \\
$A_{{\rm t}}$                       & $\cdots$ & $\cdots$ & $\cdots$         & $3.3$    & $\cdots$     \\
$A_{\mathcal{I}}$                   & $\cdots$ & $\cdots$ & $\cdots$         & $\cdots$ & $1.8$        \\
$l\,[^\circ]$                       & $214$    & $231$    & $214$            & $221$    & $221$        \\
$b\,[^\circ]$                       & $-24$    & $-34$    & $-17$            & $-24$    & $-28$        \\
\hline
\end{tabular}
\caption{Best-fit modulation parameters for the \Planck\ temperature data,
given the models described in Sec.~\ref{sec:models}.}
\label{tab:bestfitparams}
\end{table}

\begin{figure}[!ht]
\centerline{\includegraphics[width=\hsize]{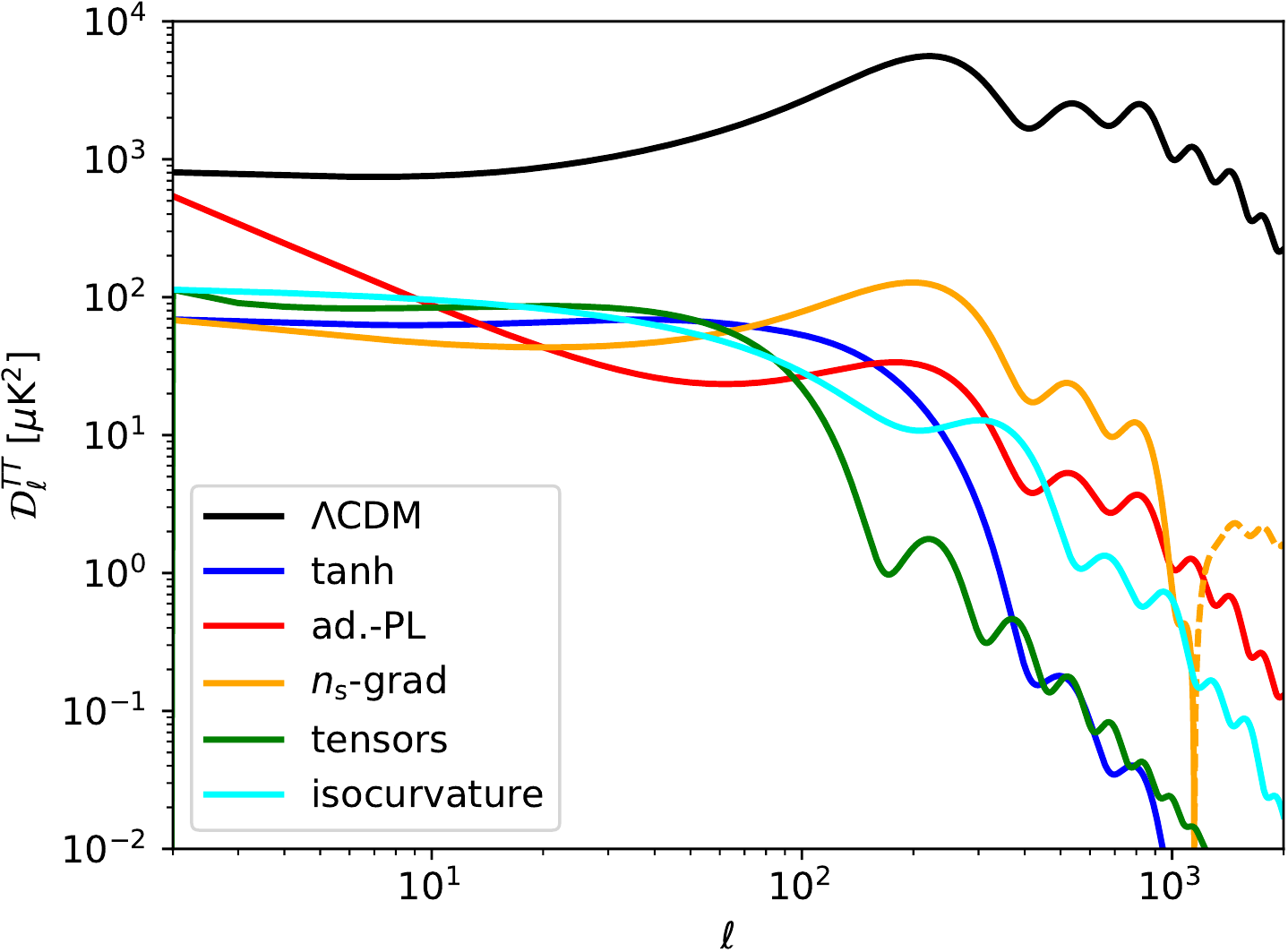}}
\centerline{\includegraphics[width=\hsize]{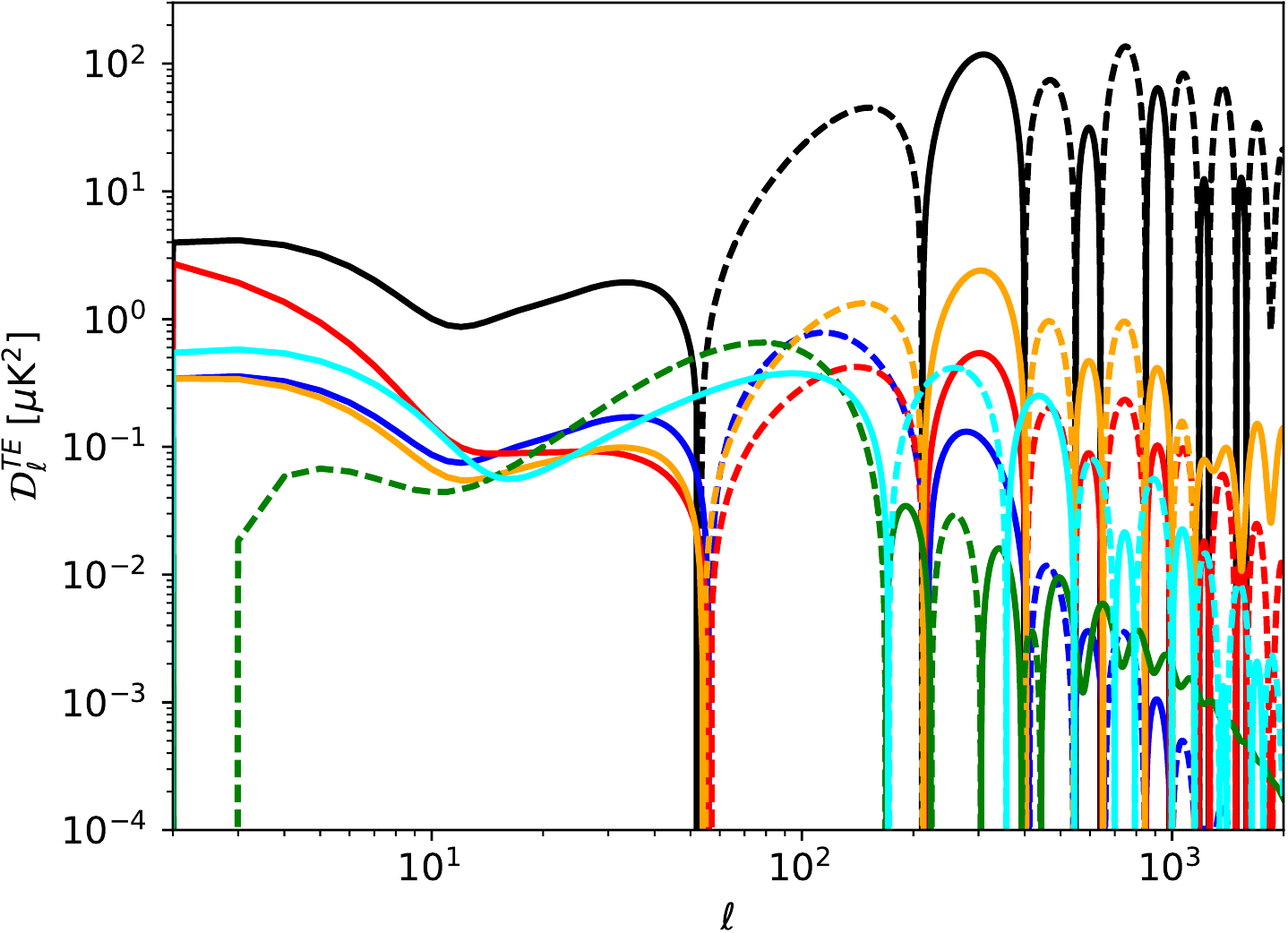}}
\centerline{\includegraphics[width=\hsize]{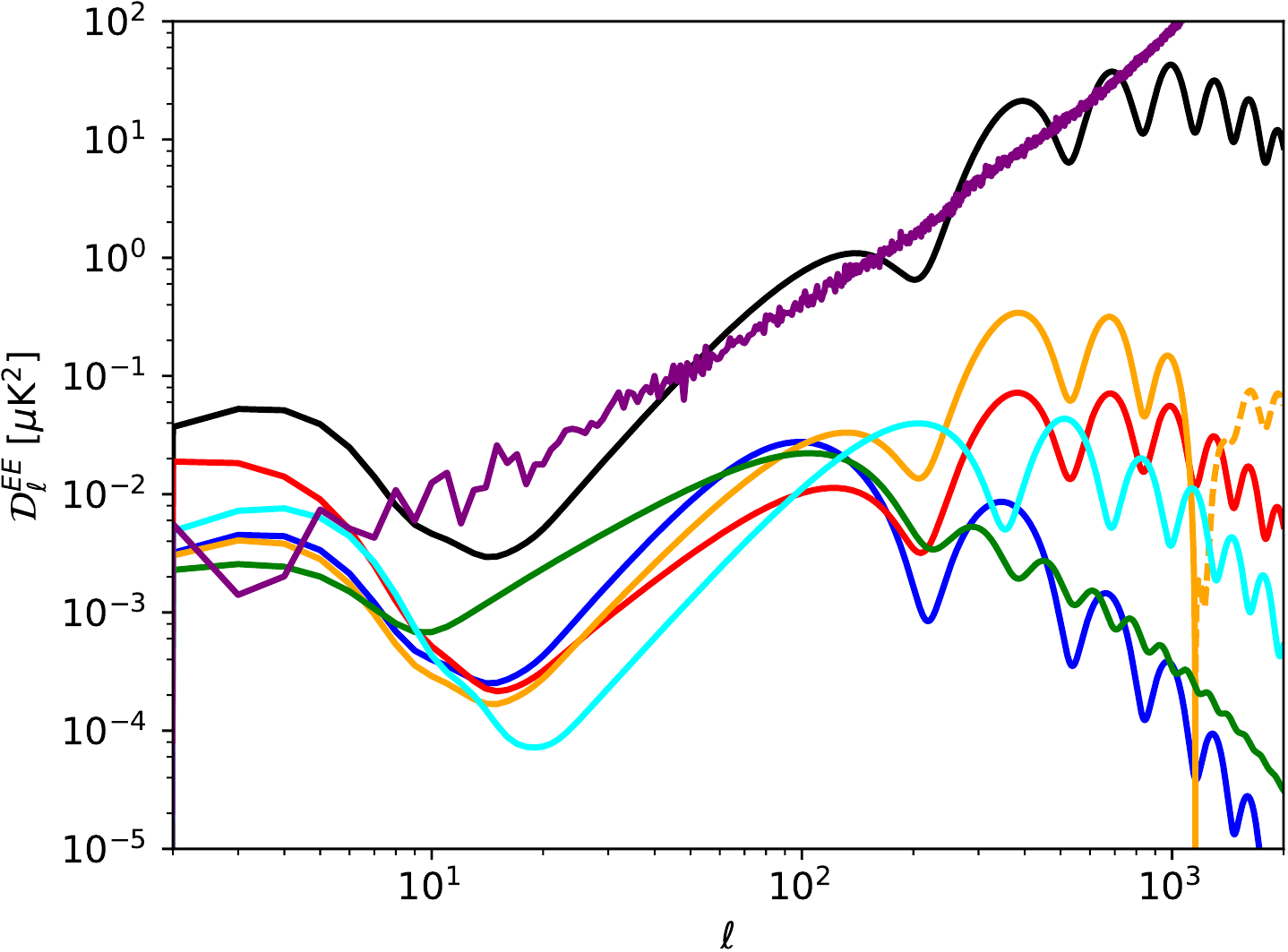}}
\caption{$\Lambda$CDM power spectra for $TT$, $TE$, and $EE$ (top to bottom
panels) compared to the best-fit asymmetry spectra, $AC^{\rm lo}_\ell$, to the
\Planck\ temperature data (see Table~\ref{tab:bestfitparams}), for the various
models.  The purple curve in the bottom panel is the noise power spectrum for a
single FFP8 noise realization. The best-fit $TT$ asymmetry spectra give
several-percent-level asymmetry for $\ell \lesssim 100$, as expected.  Here
${\cal D}_\ell \equiv \ell(\ell + 1)C_\ell/(2\pi)$.}
\label{fig:modpol}
\end{figure}

In Fig.~\ref{fig:modpol} we compare the $\Lambda$CDM power spectra to the
asymmetry spectra, $AC^{\rm lo}_\ell$, for the temperature best-fit modulation
parameters given in Table~\ref{tab:bestfitparams}, for each of our models.
Note that the best-fit $TT$ asymmetry spectra correspond very roughly to $5\%$
modulation in amplitude out to $\ell \simeq 60$, as expected from the previous
$\ell$-space analyses of the asymmetry.

It is worth reiterating that none of these models are currently favoured over
base $\Lambda$CDM: the goodness of fit of these models to the asymmetry (and
isotropic) data is discussed in detail in~\cite{short}.  Hence the interest
here in pursuing polarization data to improve constraints and test for a
physical origin to the asymmetry.

\subsection{Including polarization}
\label{sec:paramfitting}
\subsubsection{$E$ vs.\ $T$ asymmetry spectra}

\begin{figure}[!ht]
\centerline{\includegraphics[width=\hsize]
{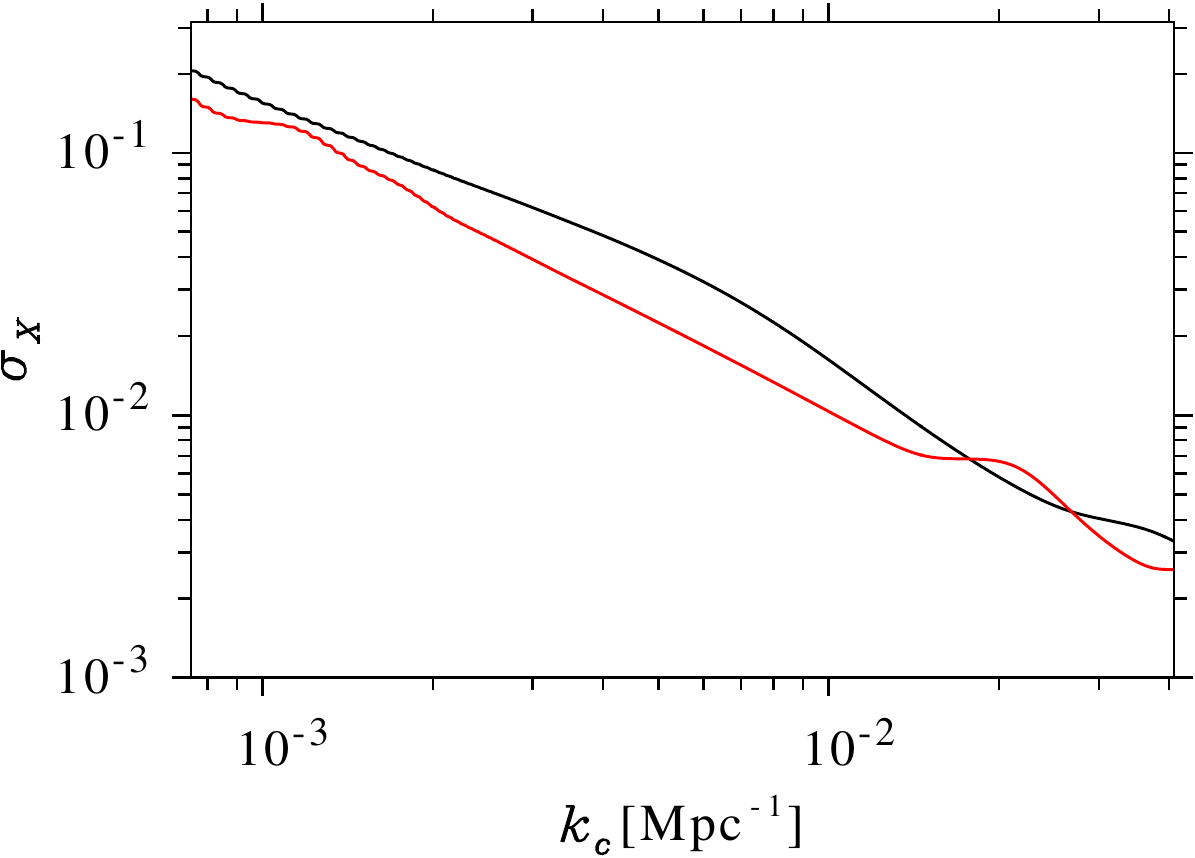}}
\caption{Cosmic variance for a measurement of the amplitude of modulation for
the $\tanh$ model (with $\Delta\ln k = 0.01$) for $TT$ (black curve) and $EE$
(red).  Polarization does considerably better than the naive $\ell$-space
expectation of identical cosmic variance for $TT$ and $EE$.}
\label{fig:sigT_E_kc}
\end{figure}

We claimed in Sec.~\ref{sec:Intro} that it was not reasonable to take the
observed multipole-space temperature asymmetry, e.g., $6\%$ asymmetry to $\ell
= 65$, and predict a $6\%$ asymmetry in $E$ to $\ell = 65$.  We demonstrate
this explicitly in this subsection, using the $\tanh$ model as an example.
First, in Fig.~\ref{fig:sigT_E_kc} we illustrate the cosmic variance
[calculated via Eq.~(\ref{eq:sigmax})] for a measurement of the amplitude of
$\tanh$ modulation versus cutoff scale $k_{\rm c}$, for $\Delta\ln k = 0.01$ (which
corresponds closely to a step function in $k$-space, with modulation only for
$k < k_{\rm c}$).  It is clear that for most values of $k_{\rm c}$, the cosmic variance is
considerably smaller for an $EE$ measurement than for a $TT$ measurement.  In
particular, this applies around the best-fit value, $k_{\rm c} = 7.45\times10^{-3}$
Mpc$^{-1}$, where the $EE$ standard deviation is smaller by a factor of nearly
two than the $TT$ value.  This is in stark contrast to the naive $\ell$-space
expectation, where identical modulation cutoffs for $TT$ and $EE$ leads to
identical cosmic variance. However, we can also see that there are some values
of $k_{\rm c}$ for which the $EE$ cosmic variance is comparable to or even worse than
the $TT$ value.

\begin{figure}[!ht]
\centerline{\includegraphics[width=\hsize]
{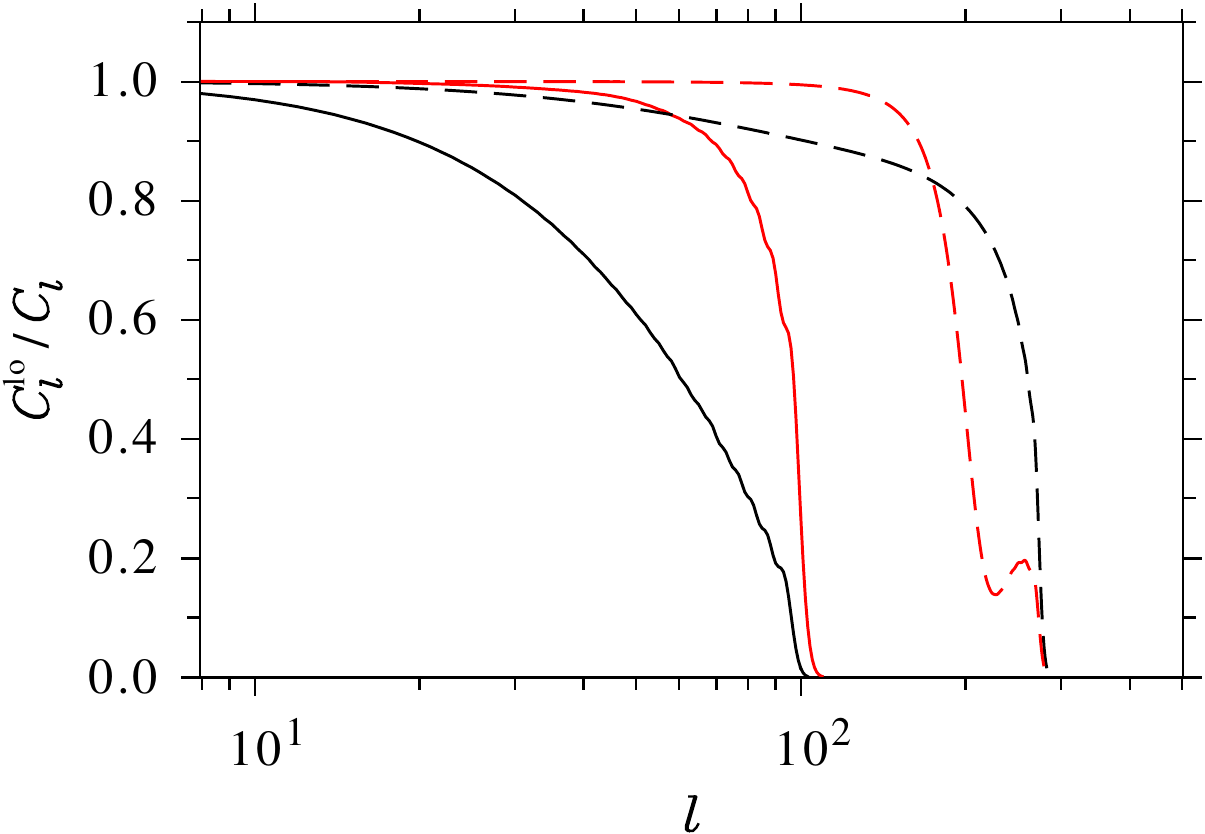}}
\caption{Asymmetry spectra $C_\ell^{\rm lo}$ for temperature (black curves) and
$E$-mode polarization (red) for the $\tanh$ model of Sec.~\ref{sec:tanh}, with
$\Delta\ln k = 0.01$ and $k_{\rm c} = 7.45\times10^{-3}\,{\rm Mpc}^{-1}$ (solid
curves) and $k_{\rm c} = 2\times10^{-2}\,{\rm Mpc}^{-1}$ (dashed).  The same physical
$k$-space modulation produces substantially more modulation of $E$ than of $T$
for the lower $k_{\rm c}$ value, and conversely for the higher $k_{\rm c}$ value.}
\label{fig:TEdiff}
\end{figure}

To help understand these differences between $TT$ and $EE$, we plot in
Fig.~\ref{fig:TEdiff} asymmetry spectra $C_\ell^{T,\rm lo}$ and $C_\ell^{E,\rm
lo}$ for the $\tanh$ model with $\Delta\ln k = 0.01$. For the case $k_{\rm c} =
7.45\times10^{-3}\,{\rm Mpc}^{-1}$ (i.e., the best-fit value), we can see that
the asymmetry spectra differ substantially between $T$ and $E$ (note that this
effect is also visible in figure 1 of~\cite{Namjoo2015}).  In particular, we
predict substantially larger modulation, and hence lower cosmic variance, in
$E$ than in $T$, since $C_\ell^{E,\rm lo} > C_\ell^{T,\rm lo}$ for all $\ell$.
The reason is that the transfer functions from $k$- to $\ell$-space
are narrower for $E$ than for $T$, and hence the step in $k$-space is better
resolved in $E$.  On the other hand, for the case $k_{\rm c} =
2\times10^{-2}\,{\rm Mpc}^{-1}$
some fine $k$-space structure (a dip in this case) is resolved by polarization
and leads to lower asymmetry power for $EE$ and hence larger cosmic variance.
This illustrates the necessity of working in $k$-space rather than
$\ell$-space when testing physical models for modulation.

\begin{figure}
\centerline{\includegraphics[width=\hsize]{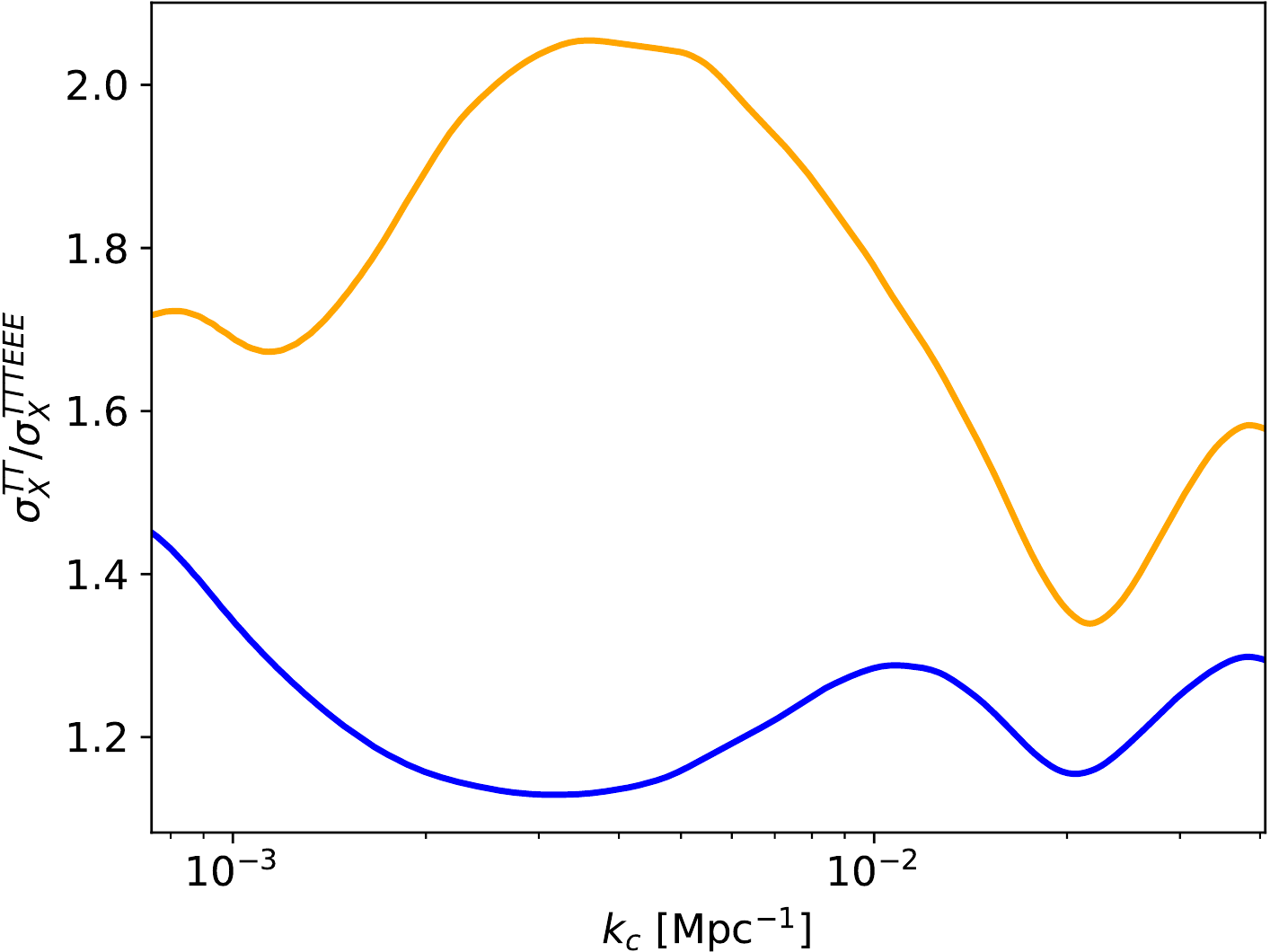}}
\centerline{\includegraphics[width=\hsize]{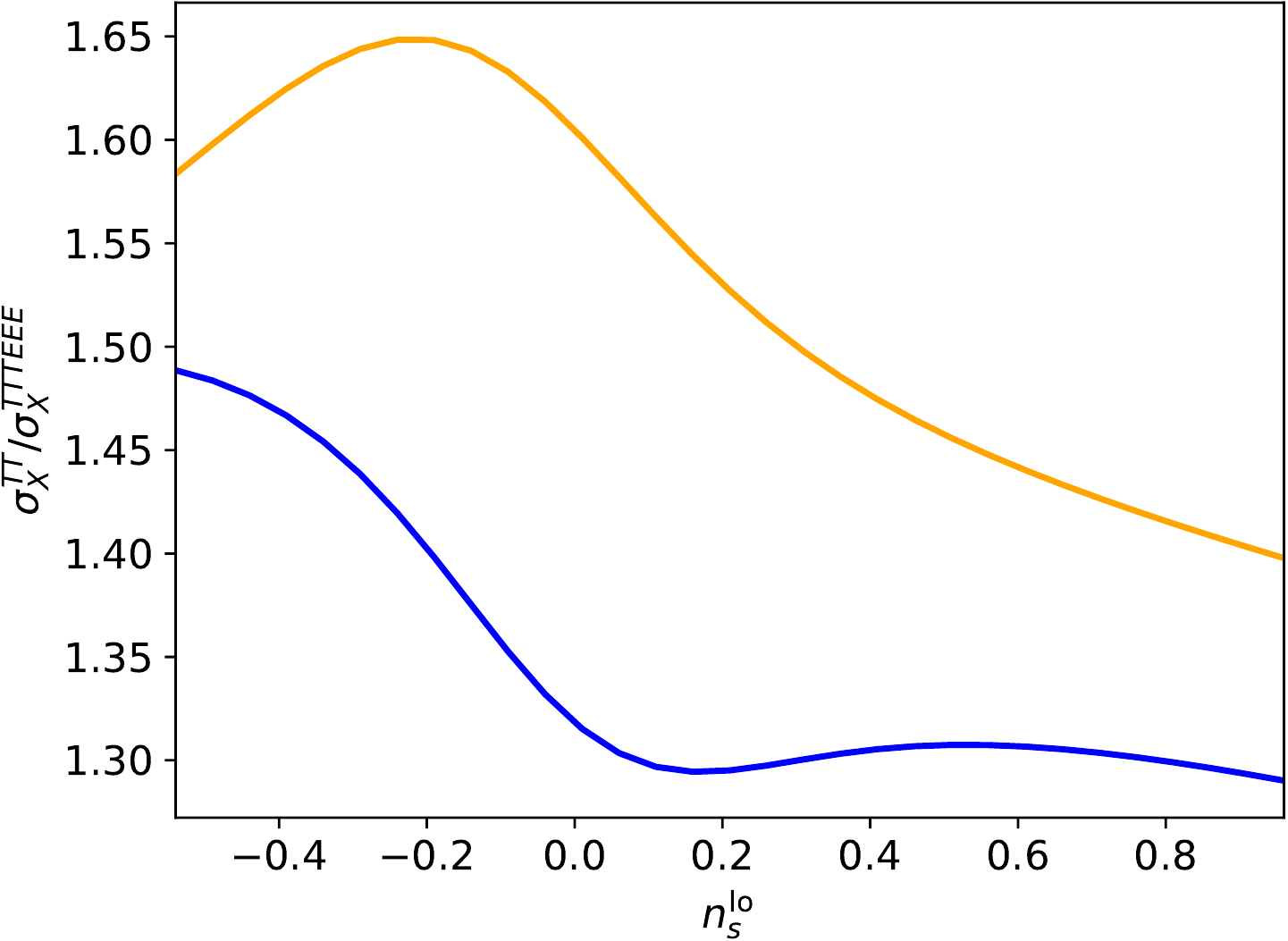}}
\centerline{\includegraphics[width=\hsize]{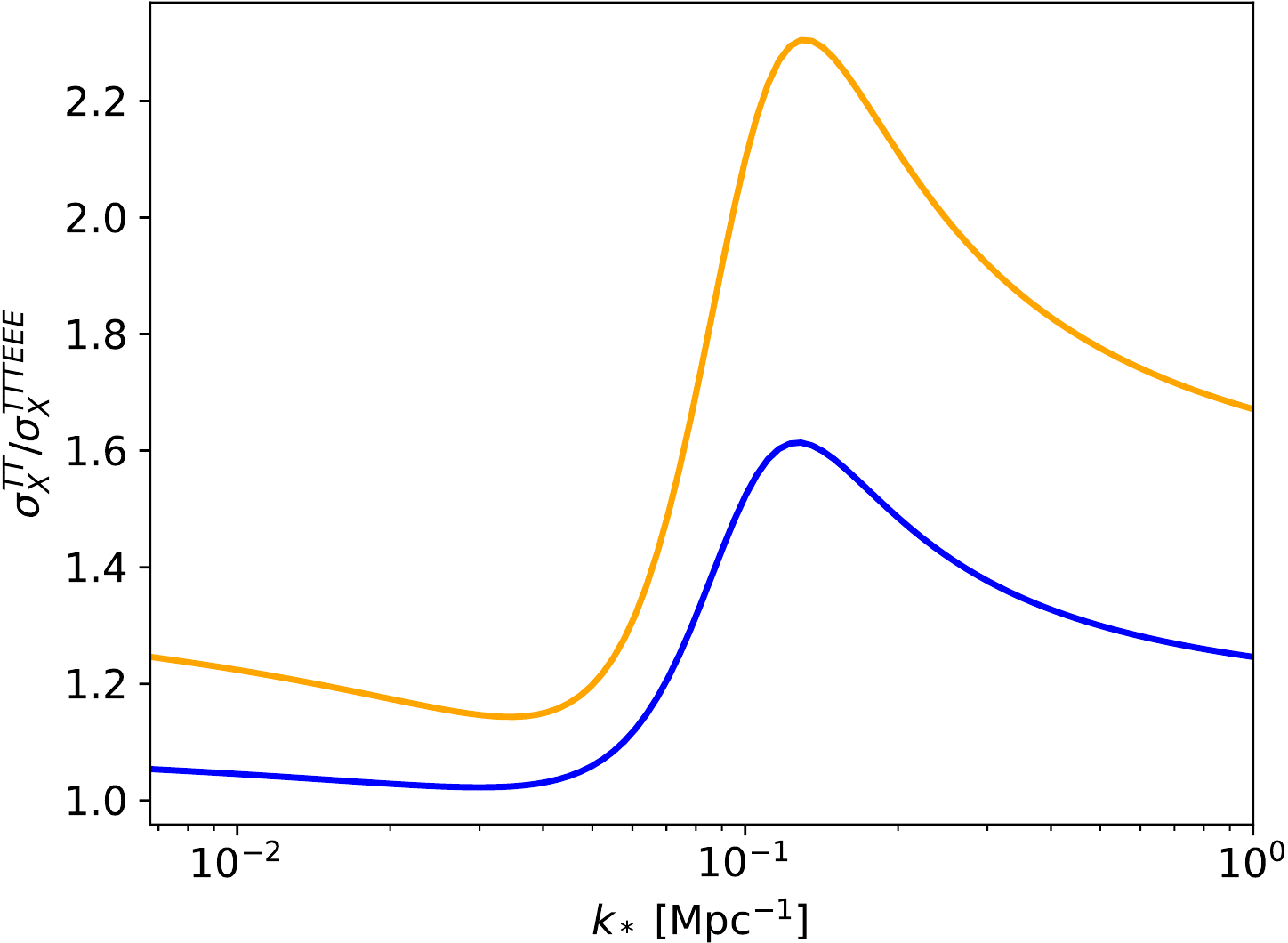}}
\caption{Improvement in the error bar for a measurement of the amplitude of
modulation for the $\tanh$, adiabatic power-law, and $n_{\rm s}$ gradient
models (top to bottom panels), assuming that the modulation direction is known,
when \Planck\ (blue curves) or cosmic-variance-limited (orange) simulated
polarization data (to $\ell_{\max} = 1000$) are added to \Planck\ temperature.
The dependence on $k_{\rm c}$ for the \Planck\ case $\tanh$ model follows the peak
structure of the $EE$ power spectra relative to the noise (see
Fig.~\ref{fig:modpol}, bottom) and can exceed the naive expectation of
$\sqrt{2}$.}
\label{fig:tanh_sigimprovement_1D}
\end{figure}

Importantly, these results imply that an ideal polarization measurement can
improve on the $TT$ modulation amplitude measurement error considerably better
than the naive $\ell$-space expectation of $\sqrt{2}$.  In
Fig.~\ref{fig:tanh_sigimprovement_1D} we show the expected improvement to the
error bar on the amplitude of modulation in a known direction ($\sigma_X$) when
adding \Planck\ or cosmic-variance-limited polarization to \Planck\
temperature, as a function of the modulation parameters.  (Recall that for
temperature we use $\ell_{\max} = 2000$ for the $n_{\rm s}$ gradient model and
$\ell_{\max} = 1000$ for all others.) For the $\tanh$ model the dependence on
$\Delta \ln k$ is quite weak and so we have averaged over it and only show the
dependence on $k_{\rm c}$. We can see that even with \Planck\ polarization (blue
curves), there are parameter values for which the addition of polarization
decreases the error bar by more than the naive expectation of $\sqrt 2$. This,
again, is due to the difference in the $k$-to-$\ell$ transfer functions between
polarization and temperature.  That is, for the same $\mathcal{P}^{\rm lo}(k)$
modulation, polarization modes are more strongly modulated (on many scales)
than temperature, as we saw in Fig.~\ref{fig:TEdiff}.  It is also worth noting
that the variations in improvement follow the peak structure of the $EE$ power
spectra, i.e., the first three peaks that are above the noise level in
Fig.~\ref{fig:modpol}. This is most clearly evident with the $\tanh$ model. The
expected improvement when adding cosmic-variance-limited polarization exceeds a
factor of two for some models and parameter ranges, substantially exceeding the
naive value of $\sqrt{2}$.

\subsubsection{Constraints}

We cross-check and validate our method through the use of FFP8
component-separated CMB + noise simulations \cite{planck2014-a14}, masked
appropriately using the \Planck\ 2015 common mask for polarization, with
unmasked fraction
$f_{\rm sky} = 0.75$.  For our cosmic-variance-limited results we remove the
noise in the polarization simulations and apply \emph{no} mask. For all models
we examine our polarization simulations up to a maximum multipole $\ell_{\max}
= 1000$. Throughout we will consider adding statistically isotropic or
anisotropic polarization to \Planck\ temperature data. In both scenarios we
will use the same set of FFP8 simulations, modified to either include the
appropriate temperature-polarization correlation with the \emph{given}
temperature data (for the case of statistically isotropic polarization, see
Appendix~\ref{sec:isoestimates}), or to include the appropriate modulation for
the specific model and parameters considered (see Appendix~\ref{sec:modsims}).

\begin{figure}
\centerline{\includegraphics[width=\hsize]{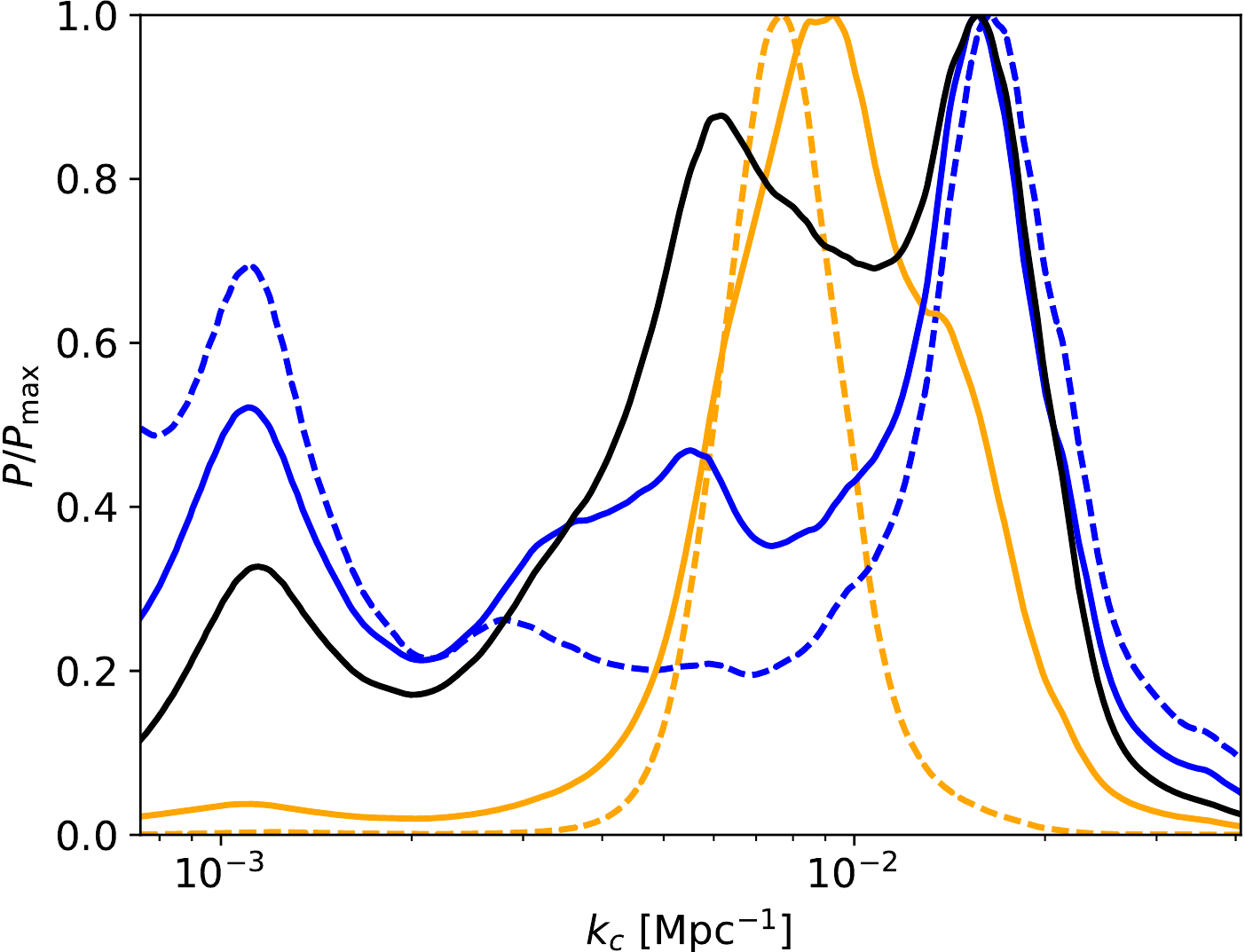}}
\centerline{\includegraphics[width=\hsize]{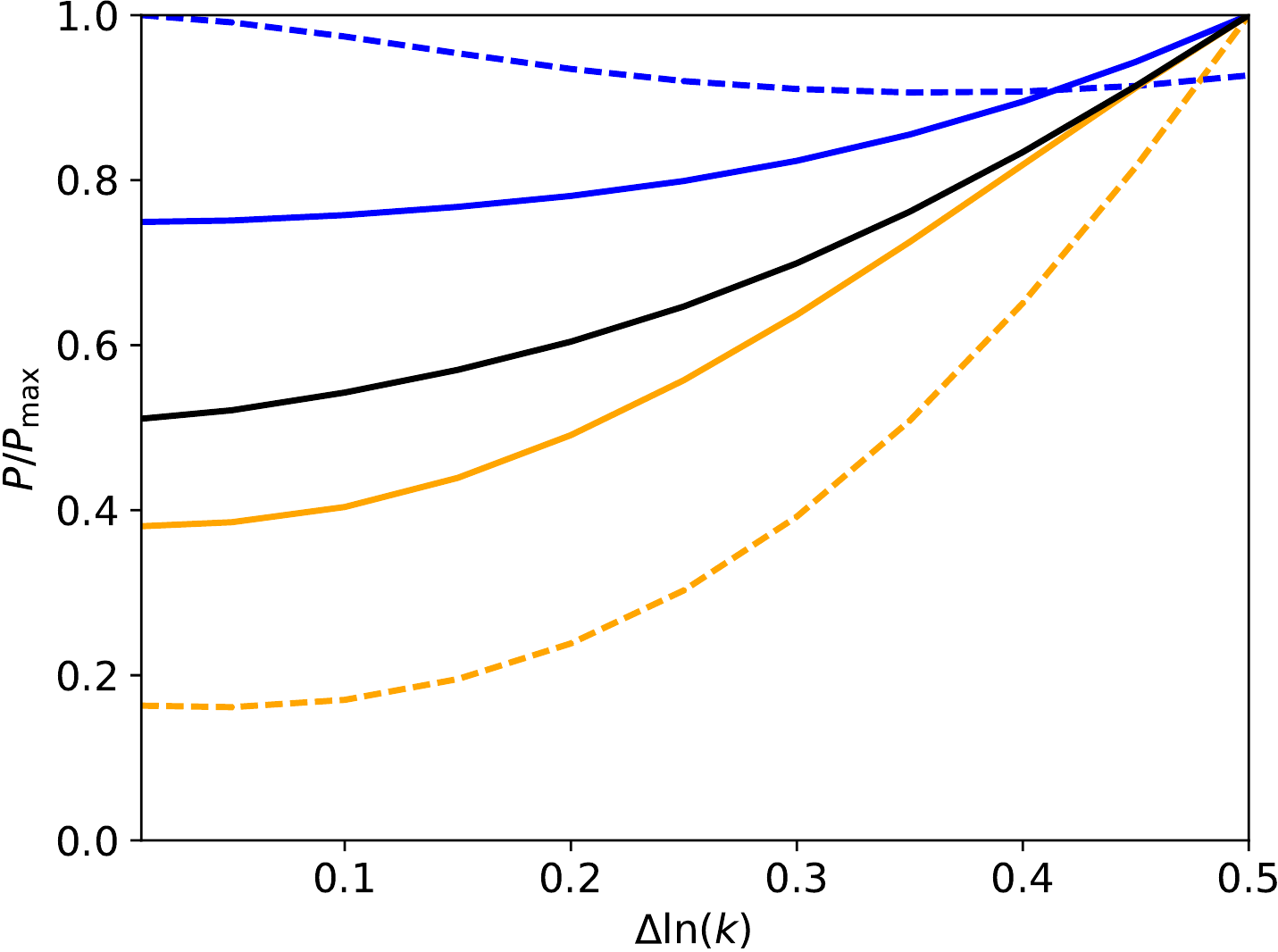}}
\centerline{\includegraphics[width=\hsize]{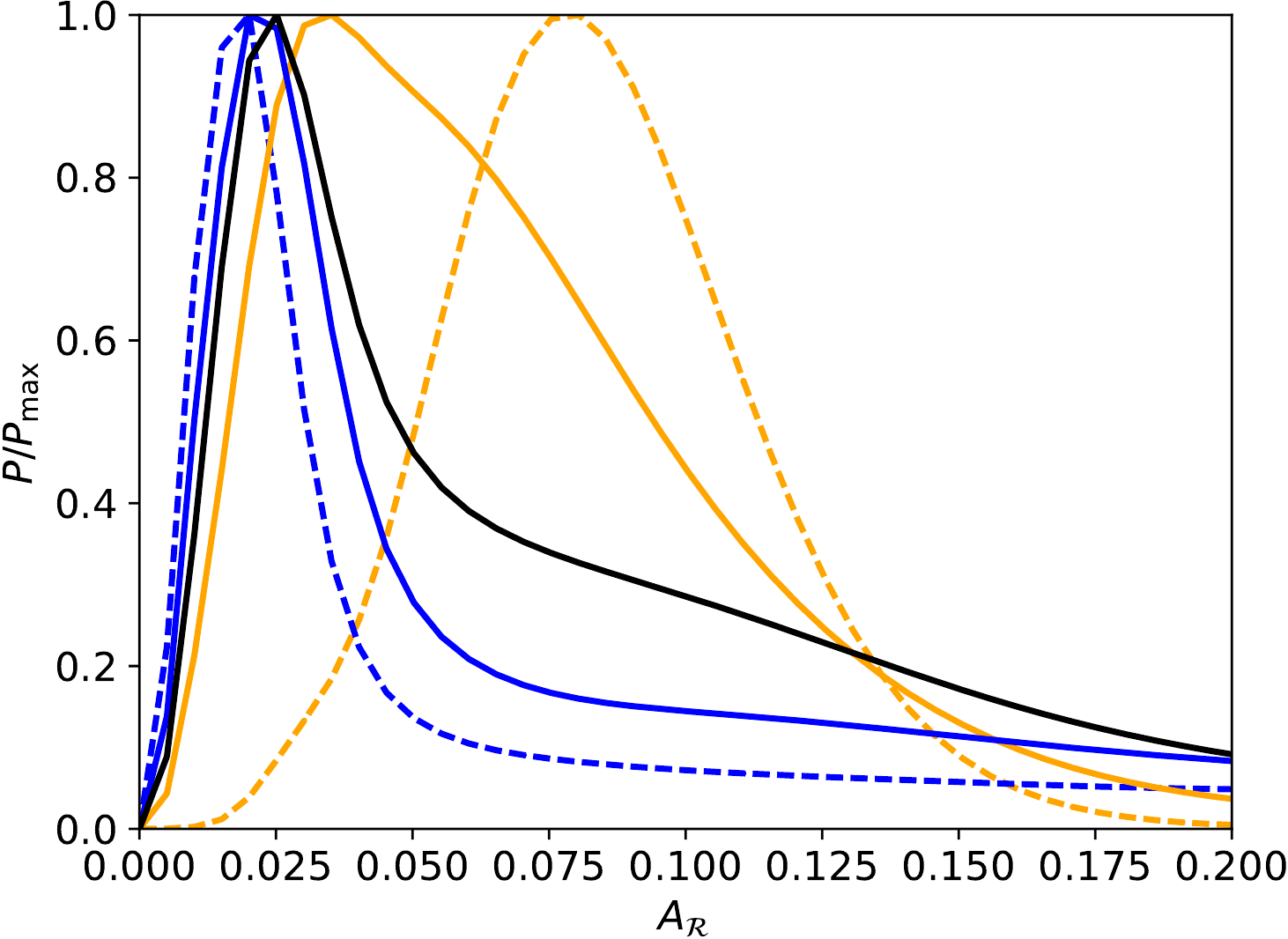}}
\caption{Posteriors for $k_{\rm c}$, $\Delta\ln k$, and $A_\mathcal{R}$ (top to
bottom panels) for the $\tanh$ model for temperature alone (black curves) and
temperature with isotropic (blue) and modulated (orange) polarization
simulations for the model parameters in Table~\ref{tab:bestfitparams}.  The
posteriors using polarization have been averaged over 500 polarization
realizations. Solid curves refer to \Planck\ polarization, while dashed curves
refer to cosmic-variance-limited polarization. The parameter $k_{\rm c}$ is
typically somewhat more constrained in the modulated than in the isotropic
polarization case, but cosmic variance is still significant.}
\label{fig:simposteriors} \end{figure}

\begin{figure}
\centerline{\includegraphics[width=\hsize]{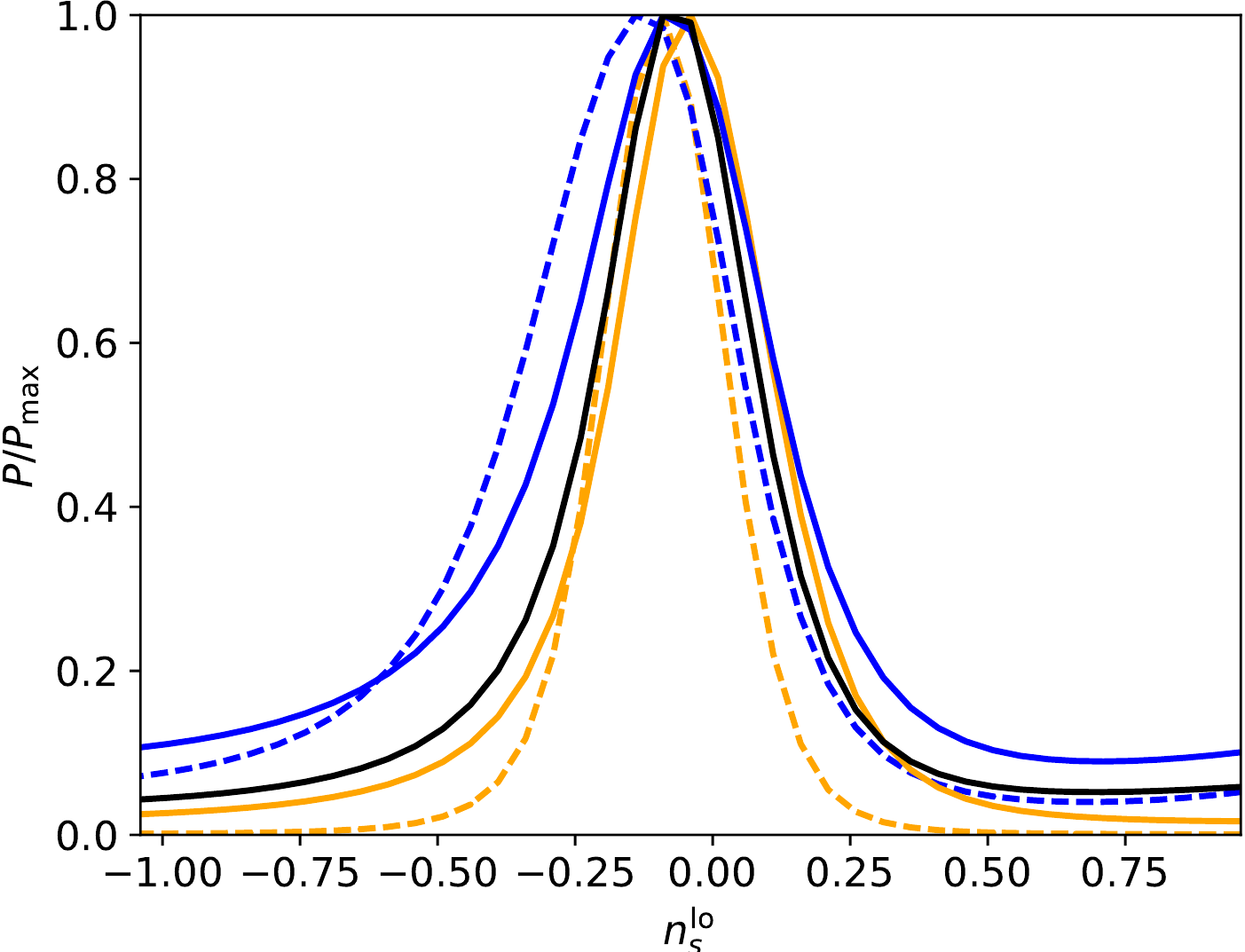}}
\centerline{\includegraphics[width=\hsize]{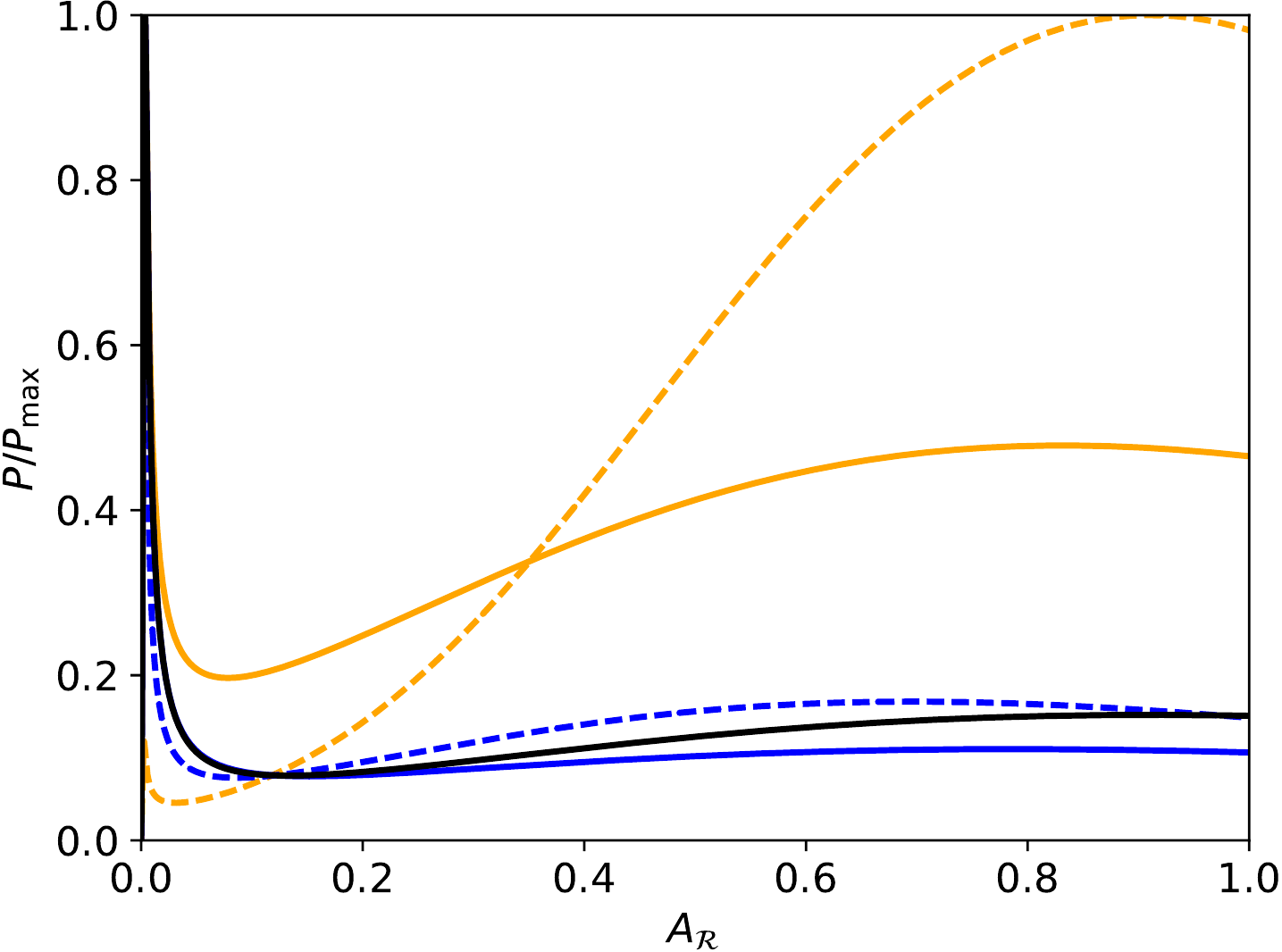}}
\caption{Posteriors for $n^{\rm lo}_s$ and $A_\mathcal{R}$ (top and bottom
panels, respectively) for the adiabatic power-law model for temperature alone
(black curves) and temperature with isotropic (blue) and modulated (orange)
polarization simulations for the parameters given in
Table~\ref{tab:bestfitparams}. The posteriors using polarization have been
averaged over 500 polarization realizations. Solid curves refer to \Planck\
polarization while dashed curves refer to cosmic-variance-limited
polarization.}
\label{fig:simposteriors_powerlaw}
\end{figure}

\begin{figure}
\centerline{\includegraphics[width=\hsize]{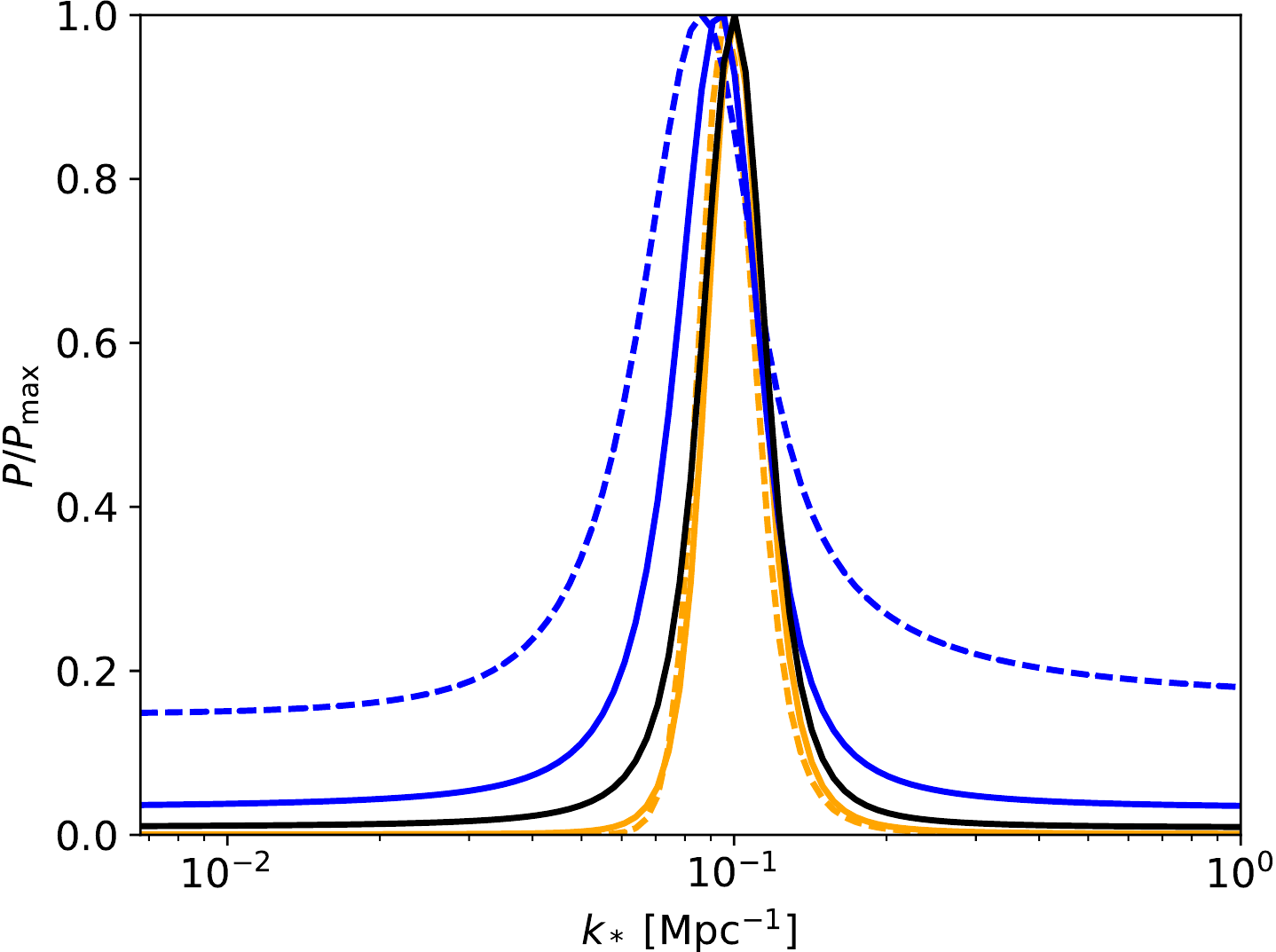}}
\centerline{\includegraphics[width=\hsize]{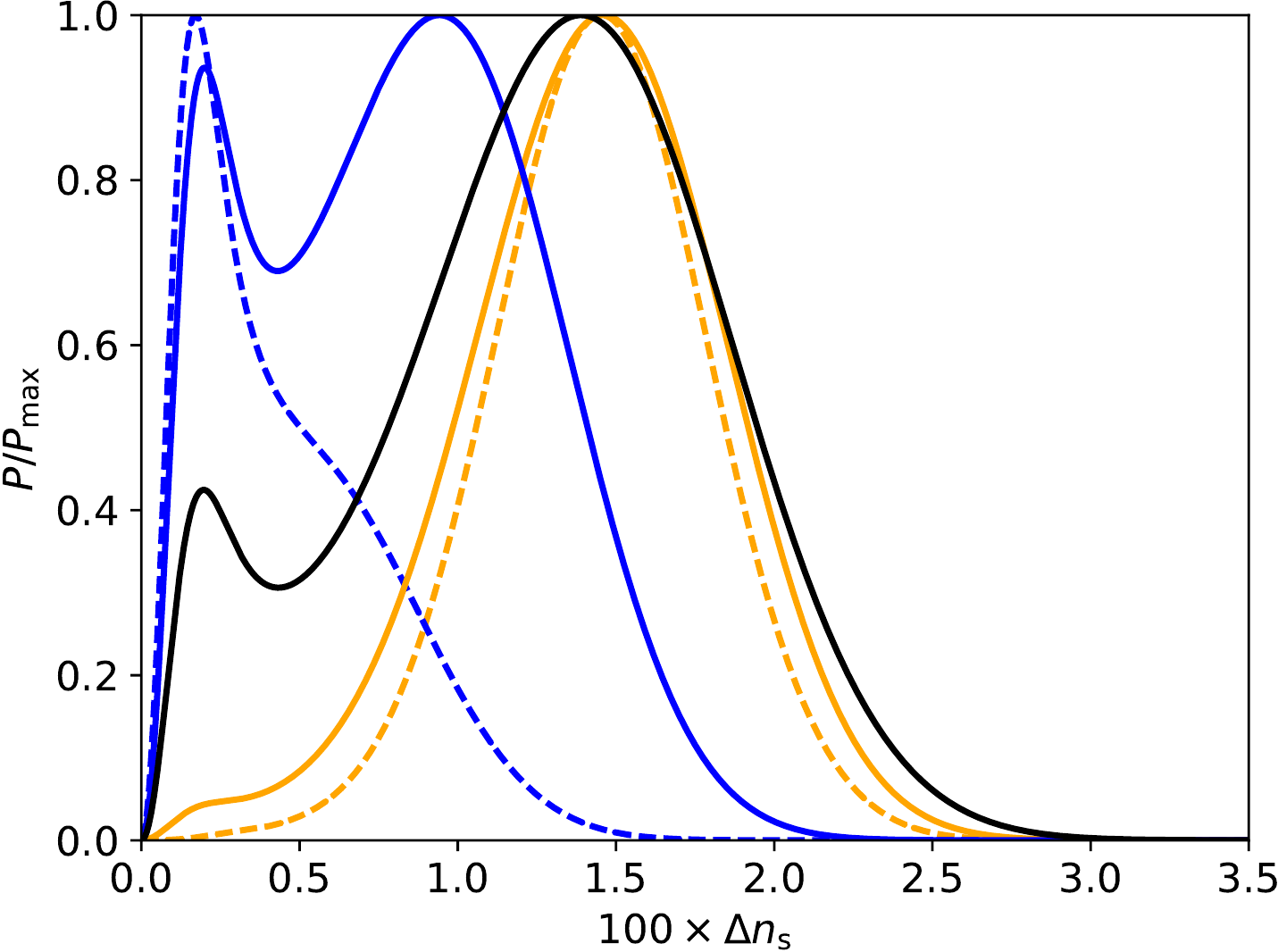}}
\caption{Posteriors for $k_*$ and $\Delta n_{\rm s}$ (top and bottom panels,
respectively) for the $n_{\rm s}$ gradient model for temperature alone (black
curves) and temperature with isotropic (blue) and modulated (orange)
polarization simulations, for the parameters given in
Table~\ref{tab:bestfitparams}. The posteriors using polarization have been
averaged over 500 polarization realizations. Solid curves refer to \Planck\
polarization while dashed curves refer to cosmic-variance-limited
polarization.}
\label{fig:simposteriors_ns}
\end{figure}

First we consider the case that the asymmetry does not have a physical origin
and is simply the result of fluctuations due to cosmic variance.  In this
case the polarization must be treated as statistically isotropic (apart from
the necessary $T$-$E$ correlation).  We apply Bayesian parameter fitting to the
combination of temperature data and polarization simulations (treated as if
they were data). In Figs.~\ref{fig:simposteriors}--\ref{fig:simposteriors_ns}
we show the marginalized posteriors of the modulation parameters for
temperature data alone (in black) and when adding statistically isotropic
polarization data averaged over 500 simulations and shown by the blue solid and
dashed curves for \Planck\ and cosmic-variance-limited polarization,
respectively. In general the addition of isotropic polarization data will act
to spread out the posteriors with respect to the temperature-only constraints.
However, we find that the addition of \emph{statistically isotropic}
polarization data increases the significance of a $\geq 3\sigma$ temperature
result (here we mean with respect to the amplitude parameter only) roughly 30\%
or 20\% of the time for \Planck\ or cosmic-variance-limited polarization,
respectively (for the $\tanh$ model).  This is mainly due to the initial
weakness of the temperature
signal.  Therefore, we urge caution when interpreting the addition of
polarization data to temperature.

Next we consider the case that the asymmetry is due to a real modulation, so
the polarization will be statistically anisotropic with a precise form
determined by the $k$-space modulation model. We generate modulated
polarization simulations (modulated with the temperature best-fit parameters
given in Table~\ref{tab:bestfitparams}) to combine with the temperature data to
forecast the type of constraints we expect to see in \Planck\ and
cosmic-variance-limited data if the modulation is real. We modify the existing
FFP8 simulations following the procedure in Appendix~\ref{sec:modsims}.

Results are also summarized in
Figs.~\ref{fig:simposteriors}--\ref{fig:simposteriors_ns}, where the addition
of modulated polarization is shown with the orange solid and dashed curves for
\Planck\ and cosmic-variance-limited polarization, respectively.  The curves
plotted are the mean posteriors averaged over 500 simulations. We see that in
general the addition of polarization makes the data more constraining. However,
these figures do not show how often we should expect to be able to distinguish
modulated polarization from statistically isotropic polarization, which will
necessarily be model dependent. We will be more quantitative about this in the
following subsection.

\subsection{Distinguishing modulated from isotropic polarization}
\label{sec:bayesfactor}

\begin{figure}
\centerline{\includegraphics[width=\hsize]{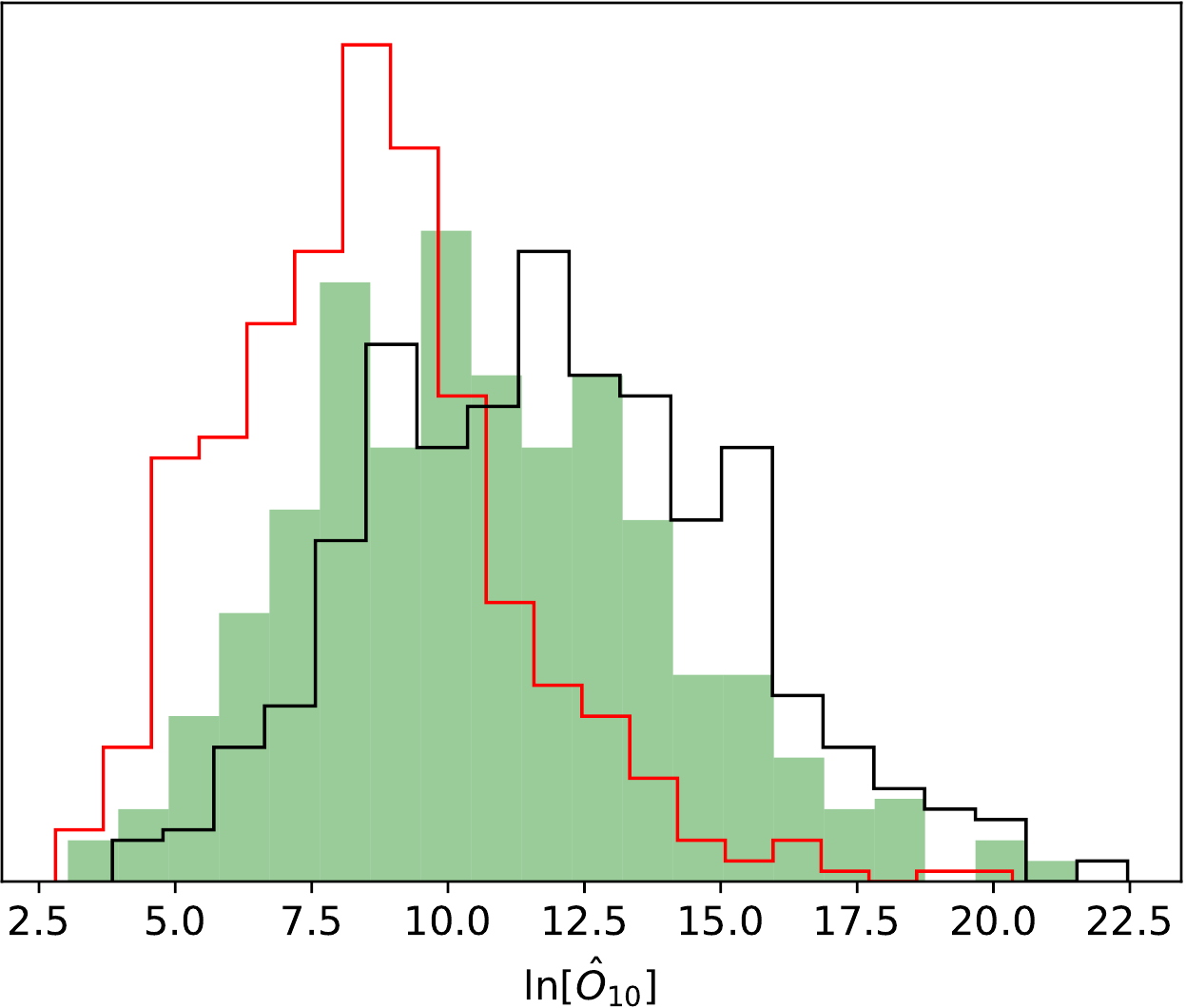}}
\centerline{\includegraphics[width=\hsize]{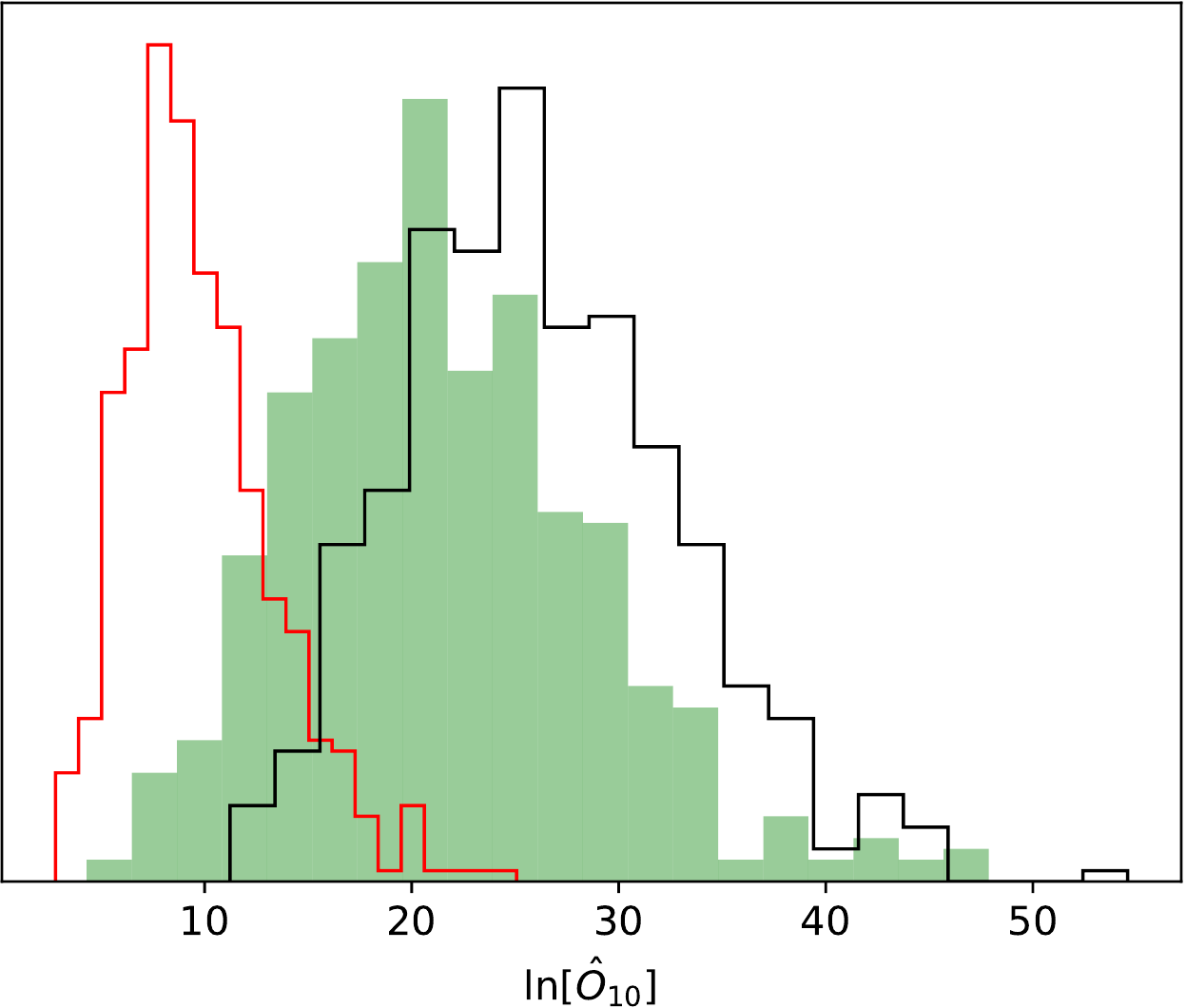}}
\caption{Histogram of the logarithm of $\hat{O}_{j0}$ [defined by
Eq.\eqref{eq:mod_odds}] for the $\tanh$ model using the \Planck\ temperature
data with 500 realizations of statistically isotropic (red outlines) or
modulated (black outlines and green filled) polarization as described in the text.  The top panel uses
\Planck\ polarization, while the bottom uses cosmic-variance-limited
polarization.  Large values of $\hat{O}_{j0}$ relative to the isotropic
histograms indicate that the modulation model should be preferred over \LCDM.}
\label{fig:tanh_oddsratio}
\end{figure}

\begin{figure}
\centerline{\includegraphics[width=\hsize]{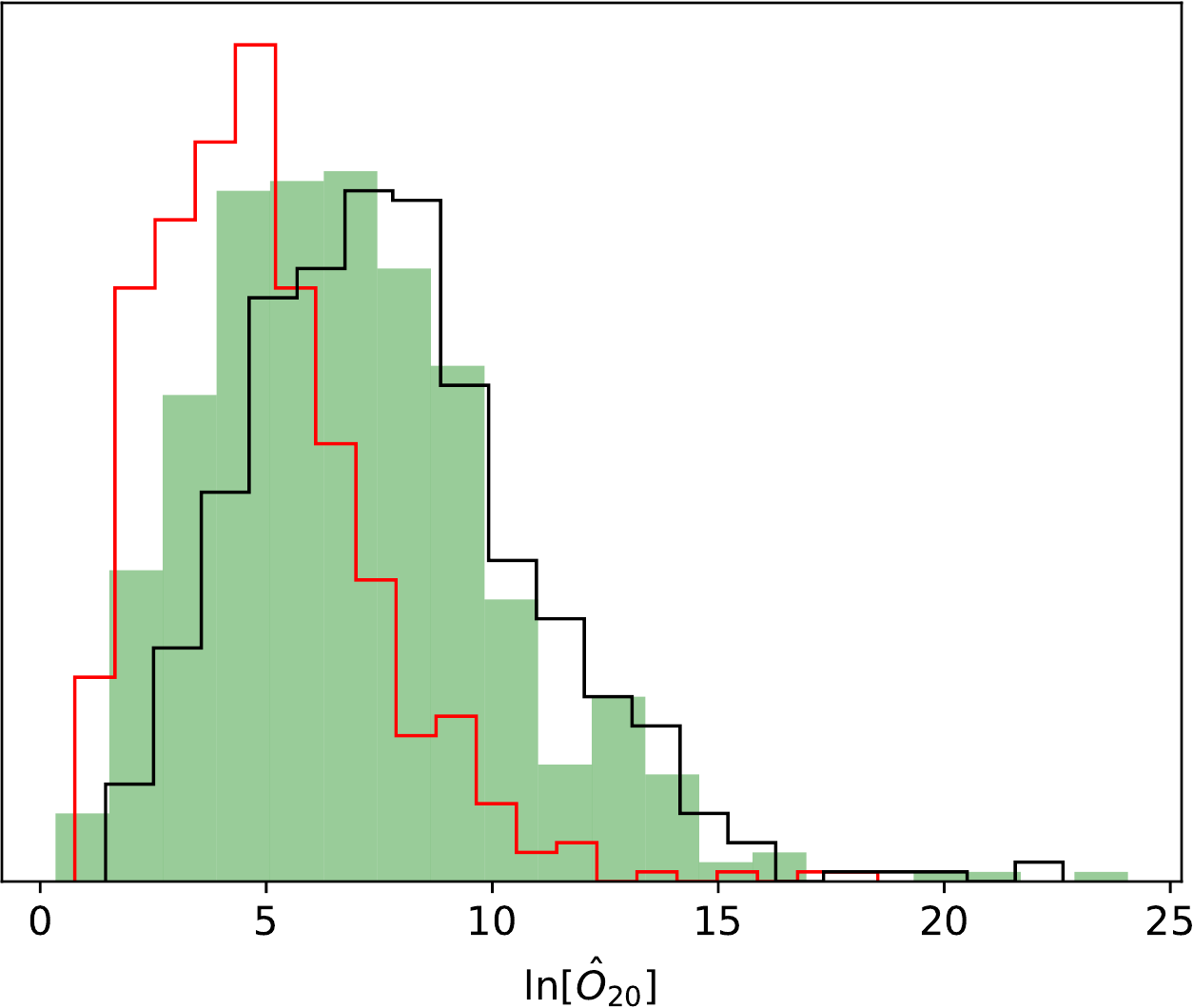}}
\centerline{\includegraphics[width=\hsize]{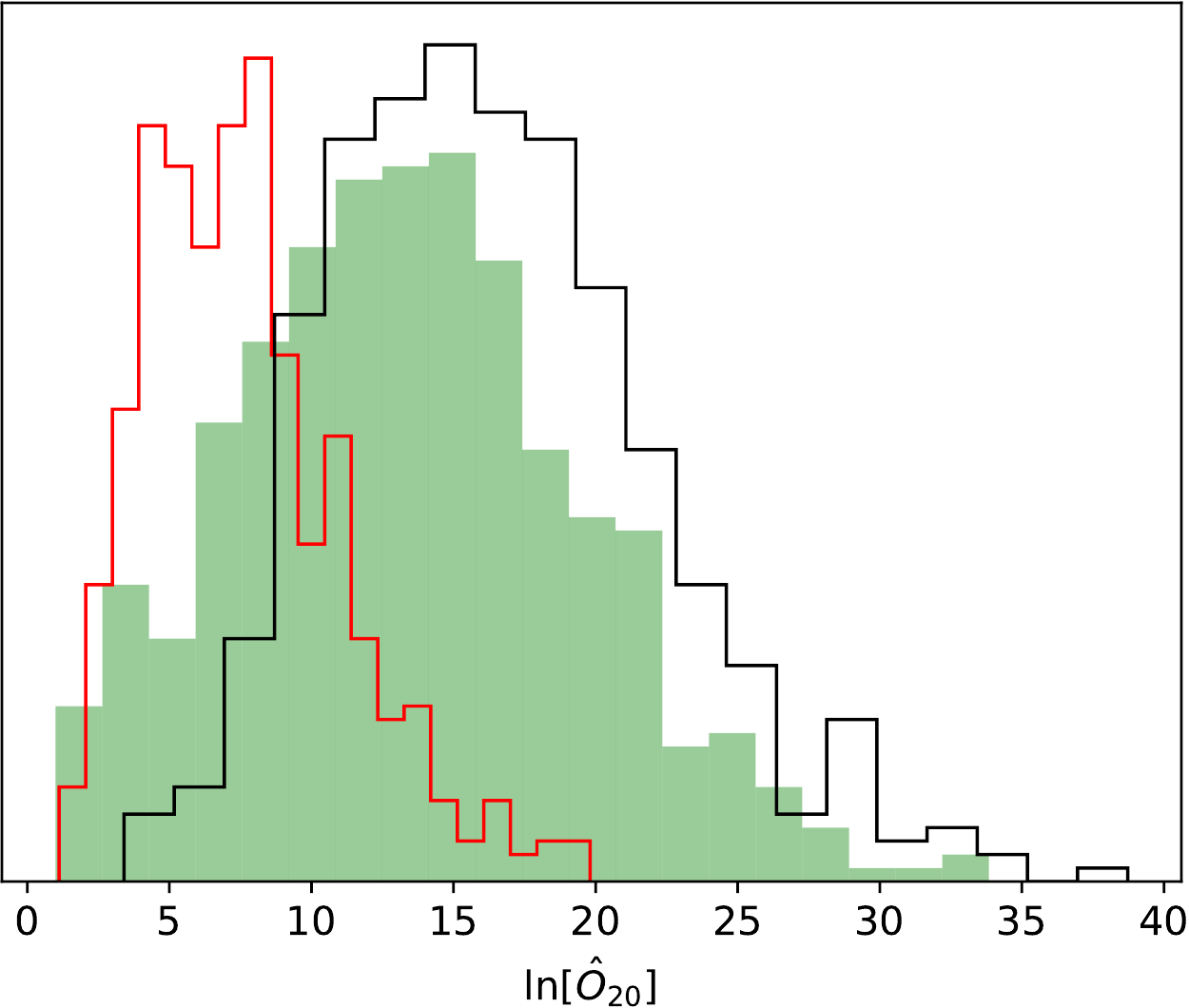}}
\caption{As in Fig.~\ref{fig:tanh_oddsratio} except for the adiabatic power-law
model.}
\label{fig:powerlaw_oddsratio}
\end{figure}

\begin{figure}
\centerline{\includegraphics[width=\hsize]{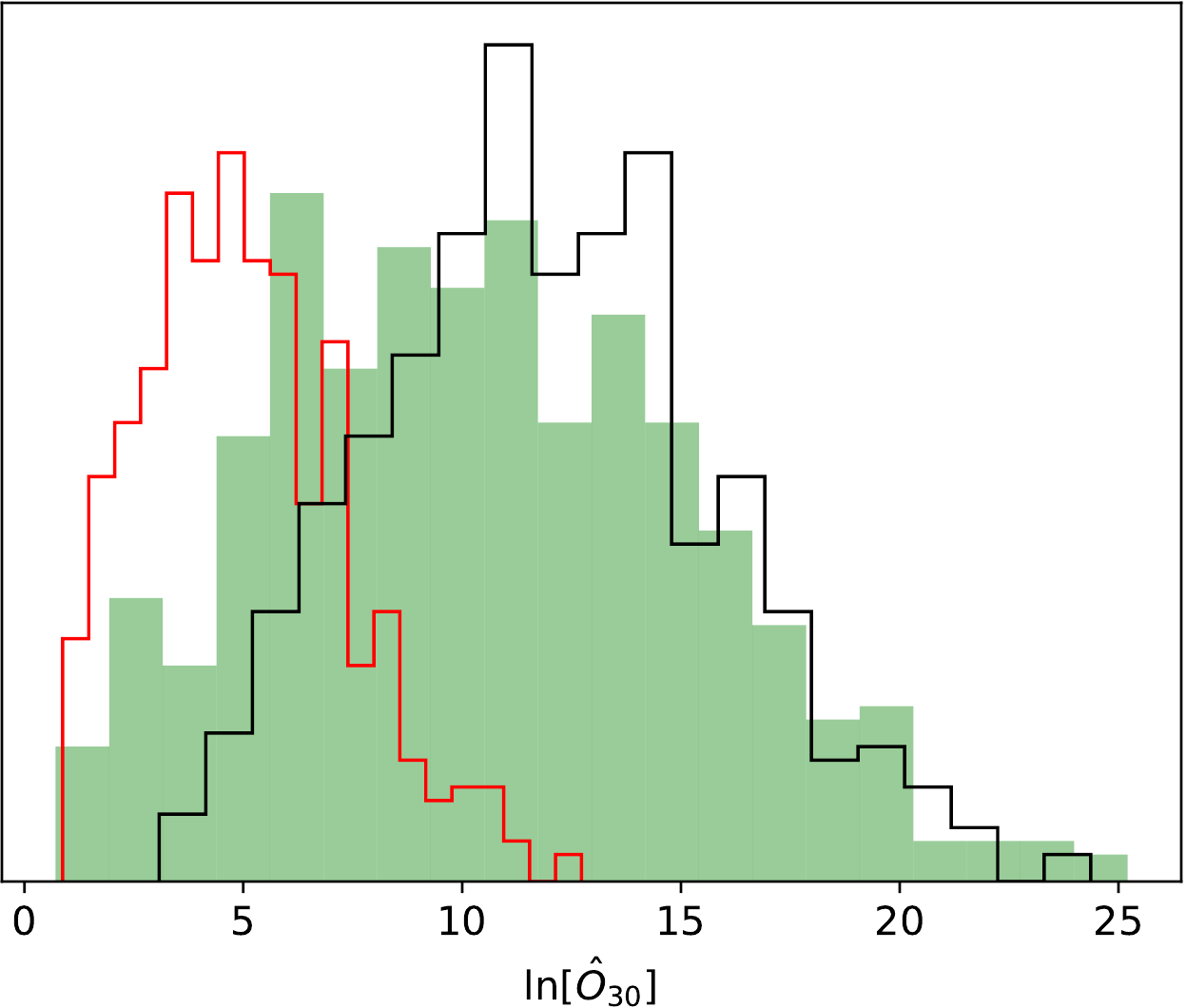}}
\centerline{\includegraphics[width=\hsize]{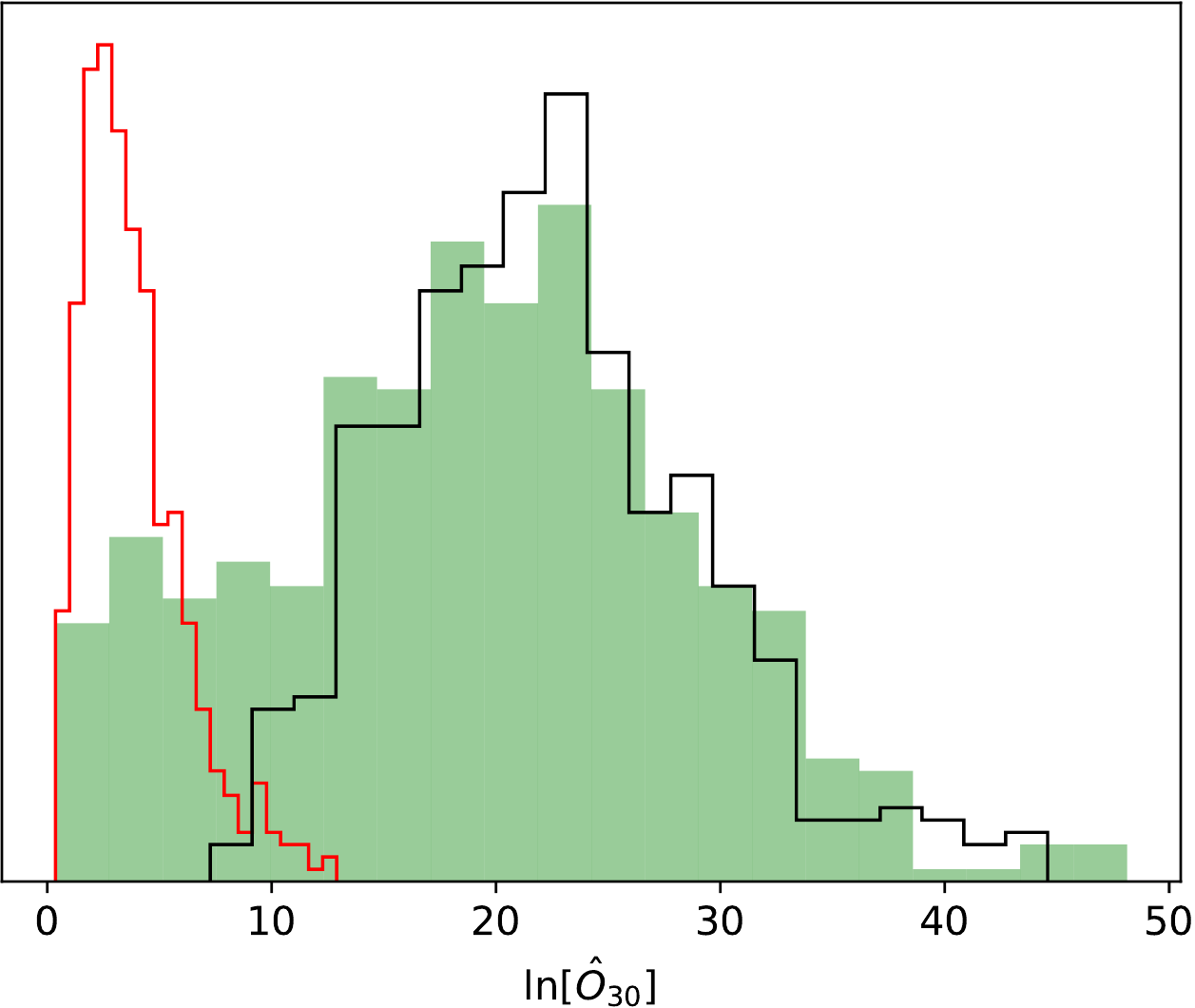}}
\caption{As in Fig.~\ref{fig:tanh_oddsratio} except for the $n_{\rm s}$
gradient model.}
\label{fig:ns_oddsratio}
\end{figure}

If the signal seen in temperature has a physical cause then it is clearly too
low in signal-to-noise to definitively distinguish from cosmic variance
fluctuations.  With the addition of polarization, however, we can assess how
well one could distinguish statistically anisotropic from statistically
isotropic data. With this in mind we define the quantity $\hat{O}_{jk}$ as the
ratio of the maximum likelihood of model $j$ to that of model $k$.  For
definiteness we will order the models in the following way:
\begin{enumerate}\addtocounter{enumi}{-1}
  \item \LCDM;
  \item $\tanh$ model;
  \item adiabatic power-law model;
  \item $n_{\rm s}$ gradient.
\end{enumerate}
For a modulation model $j$, the maximum likelihood is proportional to
$\exp[A^2/(2\sigma_X^2)]_{\max}$, whereas for \LCDM, it is proportional to
$\exp[-A^2/(2\sigma_X^2)]_{\max}$.  Therefore we can compute $\hat{O}_{j0}$ as
\begin{align}
  \hat{O}_{j0} &=
  \frac{\ld\{\exp\ld[A^2/\ld(2\sigma_X^2\rd)\rd]\rd\}_{\max}}
       {\ld\{\exp\ld[-A^2/\ld(2\sigma_X^2\rd)\rd]\rd\}_{\max}}.
  \label{eq:mod_odds}
\end{align}

Note that Eq.~\eqref{eq:mod_odds} is related to the often used odds ratio for
Bayesian model comparison, but without the Occam penalty factor
\cite{Gregory2005}.  For this reason the meaning of the \emph{absolute} value
of $\hat{O}_{j0}$ is irrelevant, though the \emph{relative} value between
different data (modulated or statistically isotropic) is valuable. We will thus
use $\hat{O}_{j0}$ as a proxy for distinguishing modulated data from
statistically isotropic data by comparing to statistically isotropic or
modulated simulations.

For the data we will choose to add either statistically isotropic or modulated
polarization to the existing \Planck\ temperature data.  The results are shown
in Figs.~\ref{fig:tanh_oddsratio}--\ref{fig:ns_oddsratio}, where the relation
between the colours and the type of simulations are as follows:
\begin{labeling}{simulations}
  \item [red:] statistically isotropic polarization added to temperature data;
  \item [green:] modulated polarization added to temperature data, where the
  modulation parameters are determined by randomly sampling the full likelihood
  of the temperature data;
  \item [black:] modulated polarization added to temperature data using the
  best-fit modulation parameters from Table~\ref{tab:bestfitparams}.
\end{labeling}
The figures show the histogram of the logarithm of $\hat{O}_{j0}$. Large values
of $\hat{O}_{j0}$ relative to the isotropic-polarization simulations indicate
that the modulation model should be preferred over \LCDM, and indeed the
modulated simulations are clearly shifted to the right of the statistically
isotropic simulations, with the black histograms further to the right than the
green. A large overlap in the distributions would indicate that it will be
difficult to distinguish modulated from statistically isotropic data using
polarization.

We would like to assess the probability of a detection of modulation assuming
that the polarization data are modulated as predicted by temperature.  In
Table~\ref{tab:2sigdetection} we indicate for each model the probability that
$\hat{O}_{j0}$ is greater than that of 95\% of isotropic (red) simulations---we
will loosely refer to this as a ``$2\sigma$''-detection. The ``best-fit''
values refer to polarization modulated with the temperature best-fit parameters
from Table~\ref{tab:bestfitparams}.  In this case the probability of a
$2\sigma$ detection with \Planck\ polarization ranges from $27\%$ for the
adiabatic power-law model to $75\%$ for the $n_{\rm s}$ gradient model.  For
cosmic-variance-limited polarization these probabilities increase
substantially, reaching $99\%$ for the $n_{\rm s}$ gradient model. However, as
the ``sampling'' columns in Table~\ref{tab:2sigdetection} show, when we sample
the modulation parameters from the full temperature posteriors, the
probabilities are reduced.

We see that even in the optimistic scenario that the true modulation parameters
are given by the best-fit values of the temperature likelihood, \Planck\ has a
low probability of distinguishing this from statistically isotropic data. The
exception is the case of the $n_{\rm s}$ gradient model, which has a large tail
out to very large $\hat{O}_{30}$ values even when simulated \Planck\
polarization data are used.  The situation improves considerably when
cosmic-variance-limited polarization is used, with the $\tanh$ and $n_{\rm
s}$-grad models being almost guaranteed to be detected in the scenario that the
true modulation model is given by the best-fit temperature parameters. These
large probabilities are diminished when we consider the case that polarization
is instead modulated with parameters randomly sampling the likelihood of the
temperature data, which is not constrained well.  However, these ``sampling''
probabilities are the best statements we can make about detectability of
modulation given the temperature signal, and should be considered as our most
conservative results.

\begin{table}
\begin{tabular}{lcccc}
\hline
\hline
\multirow{2}{*}{Model} & \multicolumn{2}{c}{\Planck} & \multicolumn{2}{c}{CV-limited} \\
                       & best fit & sampling         & best fit   & sampling \\
\hline
$\tanh$                & 37\%     & 23\%             & 94\%       & 75\% \\
ad.-PL                 & 27\%     & 20\%             & 63\%       & 45\% \\
$n_{\rm s}$-grad       & 75\%     & 58\%             & 99\%       & 84\% \\
\hline
\end{tabular}
\caption{Probability of a ``2$\sigma$''-detection (as defined in
Sec.~\ref{sec:bayesfactor}) of a real modulation as described by the model in
the first column given modulated \Planck\ or cosmic-variance-limited
polarization. The ``best fit'' columns refer to modulating polarization with
the best-fit values from the temperature data (see
Table~\ref{tab:bestfitparams}). The ``sampling'' columns refer to modulating
polarization using parameters chosen by randomly sampling the full likelihood
of the temperature data. The latter values are a more conservative approach to
how the polarization might be modulated and thus give smaller probabilities of
detection.}
\label{tab:2sigdetection}
\end{table}

\section{Discussion}
\label{sec:discussion}

In this paper we have applied to CMB polarization the formalism of
Ref.~\cite{Zibin2015}, which describes the effect of a spatially linear
modulation with arbitrary scale dependence. We have used the statistically
isotropic temperature and polarization simulations provided by the \Planck\
collaboration to estimate the decrease in uncertainty in the modulation
parameters when polarization is added. We have also generated asymmetric
polarization simulations to see how well we could test the possibility that the
modulation is a real, physical effect.  We have characterized the probability
of a ``2$\sigma$'' detection of our dipolar modulation models (introduced in
Sec.~\ref{sec:models}) when adding \Planck\ or cosmic-variance-limited
polarization data under the following important assumptions: 1) The modulation
model is correct (i.e., the primordial fluctuations are actually modulated
according to the model in question); 2) the polarization data are free of any
relevant systematic effects; and 3) the effective sky coverage is similar to
what is available for the \Planck\ temperature data (though our noiseless
polarization simulations use the full sky).  We found that, for the case of
\Planck\ polarization, we expect a probability of anywhere from 20\% to 75\%
for such a $2\sigma$ detection.  For cosmic-variance-limited polarization, the
probability increases to the range 45\% to 99\%.  We have shown that these
results are considerably stronger than a simple $\ell$-space analysis would
predict, due to the ability of polarization to resolve $k$-space detail more
sharply than temperature.

Our results are clearly strongly model dependent, with the $n_{\rm s}$ gradient
model being the most likely to be ruled out or confirmed. This is due to the
distinct scale dependence for this model, with substantial modulation at very
small scales.  This suggests that extending the polarization data to
$\ell_{\max} = 2000$ would provide a decisive test of this model~\footnote{In
addition, asymmetry constraints from large-scale structure surveys~\cite{gh12,
fvpbmb14,bafn14,sso17,zb17} should be important for $n_{\rm s}$-grad.  E.g.,
for our best-fit parameters, we find a modulation amplitude of about 1.6\%
at $k = 1\,\mathrm{Mpc}^{-1}$, which is already in strong tension with the
95\% upper limit based on quasar data in~\cite{Hirata09}.}.

Furthermore, we found that the probability that adding statistically isotropic
polarization spuriously \emph{increases} the significance of a $\geq 3\sigma$
signal (with respect to the amplitude) is 30\% or 20\% for \Planck\ or
cosmic-variance-limited polarization, respectively. This probability is large
due to the moderate strength of the modulation signal in temperature. Therefore
caution is warranted if polarization is found to increase the significance of
the temperature signal.

For a spatially linear modulation (as with our models) the addition of
polarization is our best short-term hope at detecting such a signal.  This is
because in practice the surface of last scattering is the furthest distance we
have access to and thus perturbations sourced there would be modulated with a
higher amplitude than observables such as lensing \cite{Zibin2015} or
integrated Sachs-Wolfe effect~\cite{short}, for example.  In spite of the poor
ability of \Planck\ polarization to address modulation, all is not lost, as a
CMB-S4 \cite{CMBS4} project or CORE \cite{CORE2016} should reach noise levels
such that $E$-mode polarization is essentially cosmic-variance limited. This
should provide a strong indication of the true nature of the dipole asymmetry
signal, at least for some models (recall the final two columns of
Table~\ref{tab:2sigdetection}). Farther into the future, 21\,cm measurements of
the dark ages ($z \gtrsim 30$) may be able to help in constraining the dipole
modulation models considered here due to the vastly larger number of modes
accessible to three-dimensional probes.

\section*{Acknowledgements}

   We thank Jens Chluba for comments on the draft.  This research was
supported by the Canadian Space Agency and the Natural Sciences and
Engineering Research Council of Canada.

\begin{appendix}

\section{Dipolar modulation of polarization}
\label{app:EBmod}

   In this Appendix we calculate the effect of a spatially linear modulation
of the primordial fluctuations on polarization $E$ and $B$ modes.  We will
only consider the ``lo'' component, i.e., the modulated fluctuations.  The
modulation of the $E$ and $B$ modes will differ in detail from that of
temperature [derived in \cite{Zibin2015}, and expressed in
Eqs.~(\ref{eq:totTanis}) and (\ref{eq:modulatedmodes})] due to the nonlocal
relation between $E$ and $B$ and the Stokes $Q$ and $U$ parameters, which, as
we will see, are modulated analogously to temperature.

   The polarization we see in direction $\nhat$ can be written as a line of
sight integral in terms of the temperature quadrupole, $a_{2m}^T(r\nhat)$,
seen at scatterer position $r\nhat$:
\beq
Q(\nhat) \pm iU(\nhat) = -\fr{\sqrt{6}}{10}\!\!\int\!\!dr\fr{d\tau}{dr}e^{-\tau(r)}
                          \!\sum_m a_{2m}^T(r\nhat){_{\pm2}Y_{2m}}(\nhat)
\label{QUfromquad}
\eeq
(see, e.g., \cite{hu99}).  Here $\2Ylm(\nhat)$ are the spin-2 spherical
harmonics and $\tau$ is the optical depth.  Importantly, for our purposes we
will not need to explicitly calculate the temperature quadrupole; instead, we
will only need to know how it is modulated across the sky in the presence of
a linear gradient in the primordial fluctuations.  The quadrupole
at the scatterer will be sourced by the
temperature anisotropies at the scatterer's last-scattering surface.
This means that different parts of that source will be modulated by
different amounts.  However, since the last-scattering surface is much
thinner than the distance to last scattering, to very good approximation we
can take the quadrupole $a_{2m}^T(r\nhat)$ to be modulated by the amplitude
given by the linear gradient evaluated at point $r\nhat$.  Errors due to
this approximation will be of order $\delta r/\rls \sim 10^{-3}$,
where $\delta r$ is a characteristic thickness of last scattering.
Again, due to the thinness of the last-scattering surface, this approximation
will hold independently of the radius $r$ of the quadrupole, and so for our
purposes we can place the scatterers at one radius ($\rls$) and write the
polarization in Eq.~(\ref{QUfromquad}) as
\beq
Q(\nhat) \pm iU(\nhat) \propto \sum_m a_{2m}^T(\rls\nhat){_{\pm2}Y_{2m}}(\nhat),
\label{QUfromquadLS}
\eeq
where the temperature quadrupole is modulated according to
\beq
a_{2m}^T(\rls\nhat) = a_{2m}^{T,i}(\rls\nhat)\ld(1 + A\nhat\cdot\dhat\rd).
\label{qaudmod}
\eeq
In this Appendix, superscript $i$ will indicate statistically isotropic and
homogeneous fields.

   Note that for the reionization bump at the very largest angular scales,
$\ell \lesssim 10$, we have $r_{\rm re}/\rls \simeq 0.7$.  Therefore for
polarization sourced at reionization, the error due to the variation of
modulation, $\delta r/\rls$, will be larger than for polarization from last
scattering.  Also, for a spatially linear modulation the amplitude of
modulation at reionization will be reduced by the factor $r_{\rm re}/\rls$
compared with that in Eq.~(\ref{qaudmod}).  However, given the low
statistical weight for these few very-largest-scale modes, we ignore this
effect here.  Therefore our approach will slightly overestimate the
modulation of the reionization bump.

   Combining Eqs.~(\ref{QUfromquadLS}) and (\ref{qaudmod}) gives
\bea
\!\!\!\!\!\!\!\!\!Q(\nhat) \pm iU(\nhat)
   &=& \ld[Q^i(\nhat) \pm iU^i(\nhat)\rd]\ld(1 + A\nhat\cdot\dhat\rd)\label{QUmod}\\
   &=& \ld[Q^i(\nhat) \pm iU^i(\nhat)\rd]\nn\\
 &\times& \!\!\ld(1 + \sqrt{\fr{4\pi}{3}}\sum_M\Delta X_M Y_{1M}(\nhat)\rd).
\eea
In words, the Stokes parameters are simply dipole modulated, just as
temperature was in Eq.~(\ref{eq:totTanis}).  Again, this approximation will
be good {\em for our purposes}, i.e., for the sake of quantifying the effect
of the modulation on polarization.  Note that Eq.~(\ref{QUmod}) was taken as
the starting point for polarization modulation in \cite{gkjr16}.

   The $E$- and $B$-mode multipole moments are defined by
\beq
Q(\nhat) \pm iU(\nhat) = -\sum_{\ell m}\ld(\alme \pm i\almb\rd)\2Ylm(\nhat),
\eeq
which implies
\beq
\alme \pm i\almb = -\int\ld[Q(\nhat) \pm iU(\nhat)\rd]
                   {}_{\pm2}^{}Y_{\ell m}^*(\nhat)d\Omega_\nhat.
\eeq
Combining the previous three expressions gives, for the modulated $E$ and $B$
multipoles,\begin{widetext}
\beq
\alme \pm i\almb = \almei \pm i\almbi + \sum_M\Delta X_M\sum_{\ell'm'}
                   \ld(\almpei \pm i\almpbi\rd){}_{\pm2}^{}\xi_{\ell m\ell'm'}^M,
\label{aEBmod}
\eeq
where
\beq
{}_{\pm2}^{}\xi_{\ell m\ell'm'}^M \equiv \sqrt{\fr{4\pi}{3}}
   \int\p2Ylm(\nhat)Y_{1M}(\nhat){}_{\pm2}^{}Y_{\ell m}^*(\nhat)d\Omega_\nhat
\label{2xidef}
\eeq
generalizes the usual $\xi_{\ell m\ell'm'}^M$ matrix in Eq.~(\ref{eq:xidef})
to spin-2 fields.

   To evaluate the ${}_{\pm2}^{}\xi_{\ell m\ell'm'}^M$ coefficients we can
write the spherical harmonics in terms of the rotation
matrices~\cite{edmonds74}, with the result
\bea
{}_{\pm2}^{}\xi_{\ell m\ell'm'}^M
  &=& (-1)^m\sqrt{(2\ell + 1)(2\ell' + 1)}
      \ld(\begin{array}{ccc} \ell' & \,1\, & \ell\\
                               m'  & \,M\, &  -m
          \end{array}\rd)
      \ld(\begin{array}{ccc} \ell' & \,1\, & \ell\\
                             \mp2  & \,0\, & \pm2
          \end{array}\rd).
\eea
For the case $M = 0$ (i.e., a polar modulation) the non-zero coefficients can
therefore be evaluated to be
\beq
{}_{\pm2}^{}\xi_{\ell m\ell + 1m}^0
   = \oo{\ell + 1}\sqrt{\fr{(\ell + m + 1)(\ell - m + 1)(\ell + 3)(\ell - 1)}
                           {(2\ell + 3)(2\ell + 1)}},
\eeq
\beq
{}_{\pm2}^{}\xi_{\ell m\ell m}^0 = \mp\fr{2m}{(\ell + 1)\ell},
\eeq
\beq
{}_{\pm2}^{}\xi_{\ell m\ell - 1m}^0
   = \oo{\ell}\sqrt{\fr{(\ell + m)(\ell - m)(\ell + 2)(\ell - 2)}
                           {(2\ell + 1)(2\ell - 1)}}.
\eeq
These expressions agree with those in \cite{gkjr16}.  Note that
${}_{\pm2}^{}\xi_{\ell m\ell m}^0 \ne 0$ (for $m \ne 0$), so we expect
$\ell$, $\ell$ coupling in $E$ and $B$ modes.  Also note that
${}_{\pm2}^{}\xi_{\ell m\ell \pm 1m}^0$ are symmetric and
${}_{\pm2}^{}\xi_{\ell m\ell m}^0$ is antisymmetric with respect to a sign
change of the spin index.

   Using these symmetry properties we can now evaluate Eq.~(\ref{aEBmod}) for
the case of modulation along the polar axis, taking sums and differences to
isolate the $E$ and $B$ modes.  The result is
\beq
\alme = \almei + \Delta X_0\ld(a_{\ell + 1m}^{E,i}\,{}_2^{}\xi_{\ell m\ell + 1m}^0
                             + a_{\ell - 1m}^{E,i}\,{}_2^{}\xi_{\ell m\ell - 1m}^0
                            + ia_{\ell m}^{B,i}\,{}_2^{}\xi_{\ell m\ell m}^0\rd),
\eeq
\beq
\almb = \almbi + \Delta X_0\ld(a_{\ell + 1m}^{B,i}\,{}_2^{}\xi_{\ell m\ell + 1m}^0
                             + a_{\ell - 1m}^{B,i}\,{}_2^{}\xi_{\ell m\ell - 1m}^0
                            - ia_{\ell m}^{E,i}\,{}_2^{}\xi_{\ell m\ell m}^0\rd).
\eeq
These imply
\beq
\ld\bra\alme a_{\ell'm'}^{E*}\rd\ket = C_\ell^E\del_{\ell'\ell}\del_{m'm}
   + \Delta X_0\ld(C_\ell^E + C_{\ell'}^E\rd){}_2^{}\xi_{\ell m\ell'm}^0
     \ld(\del_{\ell'\ell - 1} + \del_{\ell'\ell + 1}\rd)\del_{m'm},
\label{almecov}
\eeq
\beq
\ld\bra\almb a_{\ell'm'}^{B*}\rd\ket = C_\ell^B\del_{\ell'\ell}\del_{m'm}
   + \Delta X_0\ld(C_\ell^B + C_{\ell'}^B\rd){}_2^{}\xi_{\ell m\ell'm}^0
     \ld(\del_{\ell'\ell - 1} + \del_{\ell'\ell + 1}\rd)\del_{m'm},
\label{almbcov}
\eeq
\beq
\ld\bra\alme a_{\ell'm'}^{B*}\rd\ket = i\Delta X_0\ld(C_\ell^E + C_{\ell}^B\rd)
   {}_2^{}\xi_{\ell m\ell m}^0\del_{\ell'\ell}\del_{m'm},
\label{ebcoup}
\eeq
to first order in $\Delta X_0$.  The power spectra here are for the isotropic
(unmodulated) fields, i.e., $C_\ell^E \equiv
\ld\bra a_{\ell m}^{E,i}a_{\ell m}^{E,i*}\rd\ket$ etc., but agree with those of
the modulated fields to first order in $\Delta X_0$.
Using Eq.~(\ref{eq:modulatedmodes}) for the modulated temperature modes,
we find
\beq
\bra a_{\ell m}^Ta_{\ell'm'}^{E*}\ket = C_\ell^{TE}\delta_{\ell'\ell}\delta_{m'm}
   + \Delta X_0\ld[C_\ell^{TE}\ld({}_2^{}\xi_{\ell m\ell - 1m}^0\del_{\ell'\ell - 1}
   + {}_2^{}\xi_{\ell m\ell + 1m}^0\del_{\ell'\ell + 1}\rd)\del_{m'm}
   + C_{\ell'}^{TE}\xi_{\ell m\ell'm'}^0\rd],
\eeq
\beq
\bra a_{\ell m}^Ta_{\ell'm'}^{B*}\ket = i\Delta X_0C_\ell^{TE}
   {}_2^{}\xi_{\ell m\ell m}^0\del_{\ell'\ell}\del_{m'm},
\eeq
\end{widetext}also to lowest order in $\Delta X_0$.  Note crucially that we
find coupling between $B$ modes and $E$ or $T$ modes, which of course
vanishes in the statistically isotropic case.  However, we have
\beq
\sum_m \ld\bra\alme a_{\ell m}^{B*}\rd\ket =
\sum_m \bra a_{\ell m}^Ta_{\ell m}^{B*}\ket = 0,
\eeq
since ${}_{\pm2}^{}\xi_{\ell m\ell m}^0$ is antisymmetric with respect to $m$.

\section{Filtering}
\label{sec:filter}

In Ref.~\cite{Zibin2015} a simplified noise model was used when treating the
temperature data.  The model did not account for variations in the noise level
due to the \Planck\ scanning strategy, or scale-dependence of the noise power,
or foreground signal.  Not accounting for these effects was deemed adequate due
to the noise power being subdominant on the scales probed. However, for
polarization this is not the case.  Here we describe our approach to account
for the scale dependence of the noise and foreground power (as used similarly
in Ref.~\cite{planck2014-a17}).  The corrections below are applied after the
filtering process and are the cause of the small differences between the
temperature results here and those in Ref.~\cite{Zibin2015}.

Our starting point will be filtered data as in \cite{Zibin2015}, denoted here
as $s_{X,\, \ell m}$, where $X = T$ or $E$.  We multiply these by a quality
factor $Q^X_\ell$ to obtain filtered data that are closer to optimal, defining
\begin{align}
  X_{\ell m} &= Q^X_\ell s_{X,\, \ell m}.
  \label{eq:qualityfilter}
\end{align}
The choice of $Q^X_\ell$ is determined by the following two requirements:
\begin{align}
  F^X_\ell &= \frac{Q^X_\ell}{C^{XX}_\ell + N^{XX}_\ell},  \\
  F^X_\ell &= \frac{f^{-1}_{\rm sky}}{2\ell + 1} \sum_m |X_{\ell m}|^2.
  \label{eq:requirements}
\end{align}
Here $f_{\rm sky} = \sum_p M_p/N_{\rm pix}$, where $M_p$ is the map in pixel
space and $N_{\rm pix}$ is the total number of pixels. Further details can be
found in Appendix A.1 of Ref.~\cite{planck2014-a17}.

\section{Simulating modulation parameters}

\subsection{Isotropic estimates}
\label{sec:isoestimates}

The FFP8 simulations require modification in order to combine isotropic
polarization data with temperature data.  This is because the polarization
simulations are not correlated with temperature data in the way that the true
polarization data are. While this is a small correction, we describe below how
we perform it.

For each polarization simulation to be included with the temperature data we
modify the modulation estimator ($\tilde{X}^{WZ}_M$) in the following way:
\begin{align}
  \tilde{X}_M^{WZ\,\rm cor} &= \tilde{X}_M^{WZ} + \tilde{X}^{TT,\,\rm data}_M
  \frac{\operatorname{Cor}\ld(\tilde{X}_M^{WZ},
  \tilde{X}^{TT}_M\rd)}{\operatorname{Var}\ld(\tilde{X}^{TT}_M\rd)}.
  \label{eq:correlated}
\end{align}
The correlation and variance are estimated with the statistically isotropic
FFP8 simulations. Note that this procedure only modifies the values of the
estimators (the $\tilde{X}_M$'s) and \emph{not} the CMB simulations themselves.
This provides a significant computational speed-up when analyzing a large
number of simulations.

If the $\tilde{X}_M$'s are Gaussian (which we verify to be true to high
accuracy with our
simulations) then this approach is exact and amounts to simply shifting the
mean of $\tilde{X}_M^{TE}$ and $\tilde{X}_M^{EE}$ by an amount given by the
\emph{fixed} temperature data \cite{Bunn2016}.

\subsection{Anisotropic estimates}
\label{sec:modsims}

In the appendix of Ref.~\cite{Hanson2009} it was demonstrated how to generate
anisotropic maps from isotropic ones. Such an algorithm is convenient, but can
be computationally expensive when scanning over many different anisotropic
models. In this Appendix we will demonstrate our strategy for quickly
generating modulated {\it estimates\/} ($\tilde{X}_M$'s) by using isotropic
estimates (thus skipping the step of generating maps, filtering them, and
computing estimates from them).

For simplicity we will assume that we want to generate modulation in the
$+\hat{z}$-direction; however, a general direction can be implemented by simply
breaking the direction into components. The following will make use of binned
versions of the estimators of Eqs.~\eqref{eq:est0}--\eqref{eq:est1}. These can
be written as
\begin{align}
  \tilde{X}^{WZ}_{0,\ell} &= \frac{6\sum_{m} A_{\ell m}
  S^{(WZ)}_{\ell m\,\ell+1\, m}}
  {\delta C^{WZ}_{\ell \ell +1} (\ell + 1)F^{(W}_{\ell}F^{Z)}_{\ell+1}}.
  \label{eq:ttbinned_estimators}
\end{align}
Thus we see that at each multipole an estimate of the amplitude (and direction)
can be made. If we want to generate an estimate of an anisotropic simulation we
can modify an estimate of an isotropic simulation in the following way. First
we compute Eq.~\eqref{eq:ttbinned_estimators} for an isotropic simulation at
the desired modulation parameters (e.g., $\tilde{p}_i = \{\tilde{k}_{\rm
c},\,\Delta\ln\tilde{k}\}$ for the $\tanh$ model), which implies a particular
anisotropic power spectrum $\widetilde{\delta C}^{WZ}_{\ell\ell+1}$.  Then an
anisotropic binned estimator can be obtained as
\begin{align}
  \left.\tilde{X}^{WZ\,{\rm ani}}_{0,\ell}\right|_{\tilde{p}_i} &=
  \left.\tilde{X}^{WZ\,{\rm iso}}_{0,\ell}\right|_{\tilde{p}_i} +
  \tilde{A},
  \label{eq:ani_binned}
\end{align}
where $\tilde{A}$ is the desired amplitude of modulation. The full estimator,
Eqs.~\eqref{eq:est0}--\eqref{eq:est1}, for a general modulation model ($\delta
C_{\ell\ell+1}$) can be recovered by
\begin{align}
  \tilde{X}^{WZ\,{\rm ani}}_{0} &= \frac{\sum_\ell\left.\tilde{X}^{WZ\,{\rm
  ani}}_{0,\ell}\right|_{\tilde{p}_i}\!\!\!\widetilde{\delta C}^{WZ}_{\ell\ell+1}
  \delta C^{WZ}_{\ell\ell+1} (\ell+1) F^{(W}_{\ell} F^{Z)}_{\ell+1}}
  {\sum_\ell\ld(\delta C^{WZ}_{\ell\ell+1}\rd)^2 (\ell+1) F^{(W}_{\ell}
  F^{Z)}_{\ell+1}}.
  \label{eq:ani_fullest}
\end{align}

\section{Detection and removal of aberration}
\label{sec:aberration}

Aberration due to our velocity relative to the CMB frame adds a term to
the CMB temperature multipole covariance given by~\cite{Challinor:2002zh}
\begin{align}
  \left< a_{\ell m} a^*_{\ell+1 m} \right> &= -\beta A_{\ell m} \left[(\ell +
  2) C_{\ell+1} - \ell C_{\ell}\right].
  \label{eq:abcovariance}
\end{align}
Here the $a_{\ell m}$'s are defined in a coordinate system where the dipole
direction, $(l, b) = (264^\circ, 48^\circ)$, is aligned with the polar
direction, and $\beta = 1.23\times10^{-3}$ is the magnitude of the temperature
dipole~\cite{Adam:2015vua}.  In our notation this implies an asymmetry spectrum
of the form
\begin{align}
  \delta C_{\ell\ell+1} &= -2\left[(\ell + 2)C_{\ell+1} - \ell C_{\ell}\right].
  \label{eq:aberration}
\end{align}
Using this in our estimator gives us a constraint on $\beta$. With $\ell_{\max}
= 2000$ we obtain $\beta = (1.5\pm0.5)\times10^{-3}$ in direction $(l, b) =
(281^\circ, 57^\circ)\pm22^\circ$, i.e., a roughly $3\sigma$ detection of
aberration, consistent with the observed CMB dipole and the results of
\cite{Aghanim:2013suk}.

We are then able to remove this signal from the temperature data using the
method outlined in Appendix \ref{sec:modsims}.  Specifically we use
Eqs.~\eqref{eq:ani_binned} and \eqref{eq:ani_fullest} with $\tilde{A} = -\beta$
and $\widetilde{\delta C}^{WZ}_{\ell\ell+1}$ given by
Eq.~(\ref{eq:aberration}).  Note that the high-$\ell$ and oscillatory nature of
Eq.~(\ref{eq:aberration}) means that this procedure only noticeably affects the
results for the $n_{\rm s}$ gradient model.

\end{appendix}

\bibliography{Planck_bib,modulation}

\begin{thebibliography}{56}
\expandafter\ifx\csname natexlab\endcsname\relax\def\natexlab#1{#1}\fi
\expandafter\ifx\csname bibnamefont\endcsname\relax
  \def\bibnamefont#1{#1}\fi
\expandafter\ifx\csname bibfnamefont\endcsname\relax
  \def\bibfnamefont#1{#1}\fi
\expandafter\ifx\csname citenamefont\endcsname\relax
  \def\citenamefont#1{#1}\fi
\expandafter\ifx\csname url\endcsname\relax
  \def\url#1{\texttt{#1}}\fi
\expandafter\ifx\csname urlprefix\endcsname\relax\def\urlprefix{URL }\fi
\providecommand{\bibinfo}[2]{#2}
\renewcommand{\eprint}[1]{arXiv:\href{http://arxiv.org/abs/#1}{#1}}
\providecommand{\doi}[2]{\href{http://dx.doi.org/#1}{#2}}

\bibitem[{\citenamefont{Eriksen et~al.}(2004)\citenamefont{Eriksen, Hansen,
  Banday, Gorski, and Lilje}}]{Eriksen2004}
\bibinfo{author}{\bibfnamefont{H.~K.} \bibnamefont{Eriksen}},
  \bibinfo{author}{\bibfnamefont{F.~K.} \bibnamefont{Hansen}},
  \bibinfo{author}{\bibfnamefont{A.~J.} \bibnamefont{Banday}},
  \bibinfo{author}{\bibfnamefont{K.~M.} \bibnamefont{Gorski}},
  \bibnamefont{and} \bibinfo{author}{\bibfnamefont{P.~B.} \bibnamefont{Lilje}},
  \bibinfo{journal}{Astrophys. J.}
  \textbf{\doi{10.1086/382267}{\bibinfo{volume}{605}}},
  \doi{10.1086/382267}{\bibinfo{pages}{14}} (\bibinfo{year}{2004}),
  \bibinfo{note}{[Erratum: Astrophys. J.609,1198(2004)]},
  \eprint{astro-ph/0307507} [astro-ph].

\bibitem[{\citenamefont{Hanson and Lewis}(2009)}]{Hanson2009}
\bibinfo{author}{\bibfnamefont{D.}~\bibnamefont{Hanson}} \bibnamefont{and}
  \bibinfo{author}{\bibfnamefont{A.}~\bibnamefont{Lewis}},
  \bibinfo{journal}{Phys. Rev. D}
  \textbf{\doi{10.1103/PhysRevD.80.063004}{\bibinfo{volume}{80}}},
  \bibinfo{eid}{063004} (\bibinfo{year}{2009}), \eprint{0908.0963}
  [astro-ph.CO].

\bibitem[{\citenamefont{Bennett et~al.}(2011)}]{Bennett2011}
\bibinfo{author}{\bibfnamefont{C.~L.} \bibnamefont{Bennett}}
  \bibnamefont{et~al.}, \bibinfo{journal}{Astrophys. J. Suppl.}
  \textbf{\doi{10.1088/0067-0049/192/2/17}{\bibinfo{volume}{192}}},
  \doi{10.1088/0067-0049/192/2/17}{\bibinfo{pages}{17}} (\bibinfo{year}{2011}),
  \eprint{1001.4758} [astro-ph.CO].

\bibitem[{\citenamefont{Ade et~al.}(2014{\natexlab{a}})}]{planck2013-p09}
\bibinfo{collaboration}{Planck} Collaboration, \bibinfo{journal}{Astron.
  Astrophys.}
  \textbf{\doi{10.1051/0004-6361/201321534}{\bibinfo{volume}{571}}},
  \doi{10.1051/0004-6361/201321534}{\bibinfo{pages}{A23}}
  (\bibinfo{year}{2014}{\natexlab{a}}), \eprint{1303.5083} [astro-ph.CO].

\bibitem[{\citenamefont{Ade et~al.}(2016{\natexlab{a}})}]{planck2014-a18}
\bibinfo{collaboration}{Planck} Collaboration, \bibinfo{journal}{Astron.
  Astrophys.}
  \textbf{\doi{10.1051/0004-6361/201526681}{\bibinfo{volume}{594}}},
  \doi{10.1051/0004-6361/201526681}{\bibinfo{pages}{A16}}
  (\bibinfo{year}{2016}{\natexlab{a}}), \eprint{1506.07135} [astro-ph.CO].

\bibitem[{\citenamefont{Scott et~al.}(2016)\citenamefont{Scott, Contreras,
  Narimani, and Ma}}]{Scott2016}
\bibinfo{author}{\bibfnamefont{D.}~\bibnamefont{Scott}},
  \bibinfo{author}{\bibfnamefont{D.}~\bibnamefont{Contreras}},
  \bibinfo{author}{\bibfnamefont{A.}~\bibnamefont{Narimani}}, \bibnamefont{and}
  \bibinfo{author}{\bibfnamefont{Y.-Z.} \bibnamefont{Ma}},
  \bibinfo{journal}{JCAP}
  \textbf{\doi{10.1088/1475-7516/2016/06/046}{\bibinfo{volume}{1606}}},
  \doi{10.1088/1475-7516/2016/06/046}{\bibinfo{pages}{046}}
  (\bibinfo{year}{2016}), \eprint{1603.03550} [astro-ph.CO].

\bibitem[{\citenamefont{Hirata}(2009)}]{Hirata09}
\bibinfo{author}{\bibfnamefont{C.~M.} \bibnamefont{Hirata}},
  \bibinfo{journal}{JCAP}
  \textbf{\doi{10.1088/1475-7516/2009/09/011}{\bibinfo{volume}{0909}}},
  \doi{10.1088/1475-7516/2009/09/011}{\bibinfo{pages}{011}}
  (\bibinfo{year}{2009}), \eprint{0907.0703} [astro-ph.CO].

\bibitem[{\citenamefont{Fernández-Cobos
  et~al.}(2014)\citenamefont{Fernández-Cobos, Vielva, Pietrobon, Balbi,
  Martínez-González, and Barreiro}}]{fvpbmb14}
\bibinfo{author}{\bibfnamefont{R.}~\bibnamefont{Fernández-Cobos}},
  \bibinfo{author}{\bibfnamefont{P.}~\bibnamefont{Vielva}},
  \bibinfo{author}{\bibfnamefont{D.}~\bibnamefont{Pietrobon}},
  \bibinfo{author}{\bibfnamefont{A.}~\bibnamefont{Balbi}},
  \bibinfo{author}{\bibfnamefont{E.}~\bibnamefont{Martínez-González}},
  \bibnamefont{and} \bibinfo{author}{\bibfnamefont{R.~B.}
  \bibnamefont{Barreiro}}, \bibinfo{journal}{Mon. Not. Roy. Astron. Soc.}
  \textbf{\doi{10.1093/mnras/stu749}{\bibinfo{volume}{441}}},
  \doi{10.1093/mnras/stu749}{\bibinfo{pages}{2392}} (\bibinfo{year}{2014}),
  \eprint{1312.0275} [astro-ph.CO].

\bibitem[{\citenamefont{Yoon et~al.}(2014)\citenamefont{Yoon, Huterer,
  Gibelyou, Kov\'acs, and Szapudi}}]{yhgks14}
\bibinfo{author}{\bibfnamefont{M.}~\bibnamefont{Yoon}},
  \bibinfo{author}{\bibfnamefont{D.}~\bibnamefont{Huterer}},
  \bibinfo{author}{\bibfnamefont{C.}~\bibnamefont{Gibelyou}},
  \bibinfo{author}{\bibfnamefont{A.}~\bibnamefont{Kov\'acs}}, \bibnamefont{and}
  \bibinfo{author}{\bibfnamefont{I.}~\bibnamefont{Szapudi}},
  \bibinfo{journal}{Mon. Not. Roy. Astron. Soc.}
  \textbf{\doi{10.1093/mnrasl/slu133}{\bibinfo{volume}{445}}},
  \doi{10.1093/mnrasl/slu133}{\bibinfo{pages}{L60}} (\bibinfo{year}{2014}),
  \eprint{1406.1187} [astro-ph.CO].

\bibitem[{\citenamefont{Baghram et~al.}(2014)\citenamefont{Baghram, Abolhasani,
  Firouzjahi, and Namjoo}}]{bafn14}
\bibinfo{author}{\bibfnamefont{S.}~\bibnamefont{Baghram}},
  \bibinfo{author}{\bibfnamefont{A.~A.} \bibnamefont{Abolhasani}},
  \bibinfo{author}{\bibfnamefont{H.}~\bibnamefont{Firouzjahi}},
  \bibnamefont{and} \bibinfo{author}{\bibfnamefont{M.~H.}
  \bibnamefont{Namjoo}}, \bibinfo{journal}{JCAP}
  \textbf{\doi{10.1088/1475-7516/2014/12/036}{\bibinfo{volume}{1412}}},
  \doi{10.1088/1475-7516/2014/12/036}{\bibinfo{pages}{036}}
  (\bibinfo{year}{2014}), \eprint{1406.7277} [astro-ph.CO].

\bibitem[{\citenamefont{Zibin and Contreras}(2017)}]{Zibin2015}
\bibinfo{author}{\bibfnamefont{J.~P.} \bibnamefont{Zibin}} \bibnamefont{and}
  \bibinfo{author}{\bibfnamefont{D.}~\bibnamefont{Contreras}},
  \bibinfo{journal}{Phys. Rev.}
  \textbf{\doi{10.1103/PhysRevD.95.063011}{\bibinfo{volume}{D95}}},
  \doi{10.1103/PhysRevD.95.063011}{\bibinfo{pages}{063011}}
  (\bibinfo{year}{2017}), \eprint{1512.02618} [astro-ph.CO].

\bibitem[{\citenamefont{Hassani et~al.}(2016)\citenamefont{Hassani, Baghram,
  and Firouzjahi}}]{hbf16}
\bibinfo{author}{\bibfnamefont{F.}~\bibnamefont{Hassani}},
  \bibinfo{author}{\bibfnamefont{S.}~\bibnamefont{Baghram}}, \bibnamefont{and}
  \bibinfo{author}{\bibfnamefont{H.}~\bibnamefont{Firouzjahi}},
  \bibinfo{journal}{JCAP}
  \textbf{\doi{10.1088/1475-7516/2016/05/044}{\bibinfo{volume}{1605}}},
  \doi{10.1088/1475-7516/2016/05/044}{\bibinfo{pages}{044}}
  (\bibinfo{year}{2016}), \eprint{1511.05534} [astro-ph.CO].

\bibitem[{\citenamefont{Shiraishi et~al.}(2016)\citenamefont{Shiraishi,
  Mu\~noz, Kamionkowski, and Raccanelli}}]{smkr16}
\bibinfo{author}{\bibfnamefont{M.}~\bibnamefont{Shiraishi}},
  \bibinfo{author}{\bibfnamefont{J.~B.} \bibnamefont{Mu\~noz}},
  \bibinfo{author}{\bibfnamefont{M.}~\bibnamefont{Kamionkowski}},
  \bibnamefont{and}
  \bibinfo{author}{\bibfnamefont{A.}~\bibnamefont{Raccanelli}},
  \bibinfo{journal}{Phys. Rev.}
  \textbf{\doi{10.1103/PhysRevD.93.103506}{\bibinfo{volume}{D93}}},
  \doi{10.1103/PhysRevD.93.103506}{\bibinfo{pages}{103506}}
  (\bibinfo{year}{2016}), \eprint{1603.01206} [astro-ph.CO].

\bibitem[{\citenamefont{Dai et~al.}(2013)\citenamefont{Dai, Jeong,
  Kamionkowski, and Chluba}}]{pesky}
\bibinfo{author}{\bibfnamefont{L.}~\bibnamefont{Dai}},
  \bibinfo{author}{\bibfnamefont{D.}~\bibnamefont{Jeong}},
  \bibinfo{author}{\bibfnamefont{M.}~\bibnamefont{Kamionkowski}},
  \bibnamefont{and} \bibinfo{author}{\bibfnamefont{J.}~\bibnamefont{Chluba}},
  \bibinfo{journal}{Phys. Rev.}
  \textbf{\doi{10.1103/PhysRevD.87.123005}{\bibinfo{volume}{D87}}},
  \doi{10.1103/PhysRevD.87.123005}{\bibinfo{pages}{123005}}
  (\bibinfo{year}{2013}), \eprint{1303.6949} [astro-ph.CO].

\bibitem[{\citenamefont{Dvorkin et~al.}(2008)\citenamefont{Dvorkin, Peiris, and
  Hu}}]{dph08}
\bibinfo{author}{\bibfnamefont{C.}~\bibnamefont{Dvorkin}},
  \bibinfo{author}{\bibfnamefont{H.~V.} \bibnamefont{Peiris}},
  \bibnamefont{and} \bibinfo{author}{\bibfnamefont{W.}~\bibnamefont{Hu}},
  \bibinfo{journal}{Phys. Rev.}
  \textbf{\doi{10.1103/PhysRevD.77.063008}{\bibinfo{volume}{D77}}},
  \doi{10.1103/PhysRevD.77.063008}{\bibinfo{pages}{063008}}
  (\bibinfo{year}{2008}), \eprint{0711.2321} [astro-ph].

\bibitem[{\citenamefont{Paci et~al.}(2010)\citenamefont{Paci, Gruppuso,
  Finelli, Cabella, de~Rosa, Mandolesi, and Natoli}}]{Paci2010}
\bibinfo{author}{\bibfnamefont{F.}~\bibnamefont{Paci}},
  \bibinfo{author}{\bibfnamefont{A.}~\bibnamefont{Gruppuso}},
  \bibinfo{author}{\bibfnamefont{F.}~\bibnamefont{Finelli}},
  \bibinfo{author}{\bibfnamefont{P.}~\bibnamefont{Cabella}},
  \bibinfo{author}{\bibfnamefont{A.}~\bibnamefont{de~Rosa}},
  \bibinfo{author}{\bibfnamefont{N.}~\bibnamefont{Mandolesi}},
  \bibnamefont{and} \bibinfo{author}{\bibfnamefont{P.}~\bibnamefont{Natoli}},
  \bibinfo{journal}{Mon. Not. Roy. Astron. Soc.}
  \textbf{\doi{10.1111/j.1365-2966.2010.16905.x}{\bibinfo{volume}{407}}},
  \doi{10.1111/j.1365-2966.2010.16905.x}{\bibinfo{pages}{399}}
  (\bibinfo{year}{2010}), \eprint{1002.4745}.

\bibitem[{\citenamefont{Chang and Wang}(2013)}]{Chang2013}
\bibinfo{author}{\bibfnamefont{Z.}~\bibnamefont{Chang}} \bibnamefont{and}
  \bibinfo{author}{\bibfnamefont{S.}~\bibnamefont{Wang}}
  (\bibinfo{year}{2013}), \eprint{1312.6575} [astro-ph.CO].

\bibitem[{\citenamefont{Paci et~al.}(2013)\citenamefont{Paci, Gruppuso,
  Finelli, De~Rosa, Mandolesi, and Natoli}}]{Paci2013}
\bibinfo{author}{\bibfnamefont{F.}~\bibnamefont{Paci}},
  \bibinfo{author}{\bibfnamefont{A.}~\bibnamefont{Gruppuso}},
  \bibinfo{author}{\bibfnamefont{F.}~\bibnamefont{Finelli}},
  \bibinfo{author}{\bibfnamefont{A.}~\bibnamefont{De~Rosa}},
  \bibinfo{author}{\bibfnamefont{N.}~\bibnamefont{Mandolesi}},
  \bibnamefont{and} \bibinfo{author}{\bibfnamefont{P.}~\bibnamefont{Natoli}},
  \bibinfo{journal}{Mon. Not. Roy. Astron. Soc.}
  \textbf{\doi{10.1093/mnras/stt1219}{\bibinfo{volume}{434}}},
  \doi{10.1093/mnras/stt1219}{\bibinfo{pages}{3071}} (\bibinfo{year}{2013}),
  \eprint{1301.5195}.

\bibitem[{\citenamefont{Ghosh et~al.}(2016{\natexlab{a}})\citenamefont{Ghosh,
  Kothari, Jain, and Rath}}]{Ghosh2015}
\bibinfo{author}{\bibfnamefont{S.}~\bibnamefont{Ghosh}},
  \bibinfo{author}{\bibfnamefont{R.}~\bibnamefont{Kothari}},
  \bibinfo{author}{\bibfnamefont{P.}~\bibnamefont{Jain}}, \bibnamefont{and}
  \bibinfo{author}{\bibfnamefont{P.~K.} \bibnamefont{Rath}},
  \bibinfo{journal}{JCAP}
  \textbf{\doi{10.1088/1475-7516/2016/01/046}{\bibinfo{volume}{1601}}},
  \doi{10.1088/1475-7516/2016/01/046}{\bibinfo{pages}{046}}
  (\bibinfo{year}{2016}{\natexlab{a}}), \eprint{1507.04078} [astro-ph.CO].

\bibitem[{\citenamefont{Kothari et~al.}(2016)\citenamefont{Kothari, Ghosh,
  Rath, Kashyap, and Jain}}]{Kothari2015}
\bibinfo{author}{\bibfnamefont{R.}~\bibnamefont{Kothari}},
  \bibinfo{author}{\bibfnamefont{S.}~\bibnamefont{Ghosh}},
  \bibinfo{author}{\bibfnamefont{P.~K.} \bibnamefont{Rath}},
  \bibinfo{author}{\bibfnamefont{G.}~\bibnamefont{Kashyap}}, \bibnamefont{and}
  \bibinfo{author}{\bibfnamefont{P.}~\bibnamefont{Jain}},
  \bibinfo{journal}{Mon. Not. Roy. Astron. Soc.}
  \textbf{\doi{10.1093/mnras/stw1039}{\bibinfo{volume}{460}}},
  \doi{10.1093/mnras/stw1039}{\bibinfo{pages}{1577}} (\bibinfo{year}{2016}),
  \eprint{1503.08997} [astro-ph.CO].

\bibitem[{\citenamefont{Kothari}(2015)}]{Kothari2015a}
\bibinfo{author}{\bibfnamefont{R.}~\bibnamefont{Kothari}}
  (\bibinfo{year}{2015}), \eprint{1508.03547} [astro-ph.CO].

\bibitem[{\citenamefont{Namjoo et~al.}(2015)\citenamefont{Namjoo, Abolhasani,
  Assadullahi, Baghram, Firouzjahi, and Wands}}]{Namjoo2015}
\bibinfo{author}{\bibfnamefont{M.~H.} \bibnamefont{Namjoo}},
  \bibinfo{author}{\bibfnamefont{A.~A.} \bibnamefont{Abolhasani}},
  \bibinfo{author}{\bibfnamefont{H.}~\bibnamefont{Assadullahi}},
  \bibinfo{author}{\bibfnamefont{S.}~\bibnamefont{Baghram}},
  \bibinfo{author}{\bibfnamefont{H.}~\bibnamefont{Firouzjahi}},
  \bibnamefont{and} \bibinfo{author}{\bibfnamefont{D.}~\bibnamefont{Wands}},
  \bibinfo{journal}{JCAP}
  \textbf{\doi{10.1088/1475-7516/2015/05/015}{\bibinfo{volume}{1505}}},
  \doi{10.1088/1475-7516/2015/05/015}{\bibinfo{pages}{015}}
  (\bibinfo{year}{2015}), \eprint{1411.5312} [astro-ph.CO].

\bibitem[{\citenamefont{Bunn et~al.}(2016)\citenamefont{Bunn, Xue, and
  Zheng}}]{Bunn2016}
\bibinfo{author}{\bibfnamefont{E.~F.} \bibnamefont{Bunn}},
  \bibinfo{author}{\bibfnamefont{Q.}~\bibnamefont{Xue}}, \bibnamefont{and}
  \bibinfo{author}{\bibfnamefont{H.}~\bibnamefont{Zheng}},
  \bibinfo{journal}{Phys. Rev.}
  \textbf{\doi{10.1103/PhysRevD.94.103512}{\bibinfo{volume}{D94}}},
  \doi{10.1103/PhysRevD.94.103512}{\bibinfo{pages}{103512}}
  (\bibinfo{year}{2016}), \eprint{1608.05070} [astro-ph.CO].

\bibitem[{\citenamefont{Ade et~al.}(2016{\natexlab{b}})}]{planck2014-a14}
\bibinfo{collaboration}{Planck} Collaboration, \bibinfo{journal}{Astron.
  Astrophys.}
  \textbf{\doi{10.1051/0004-6361/201527103}{\bibinfo{volume}{594}}},
  \doi{10.1051/0004-6361/201527103}{\bibinfo{pages}{A12}}
  (\bibinfo{year}{2016}{\natexlab{b}}), \eprint{1509.06348} [astro-ph.CO].

\bibitem[{\citenamefont{Aghanim et~al.}(2014)}]{Aghanim:2013suk}
\bibinfo{collaboration}{Planck} Collaboration, \bibinfo{journal}{Astron.
  Astrophys.}
  \textbf{\doi{10.1051/0004-6361/201321556}{\bibinfo{volume}{571}}},
  \doi{10.1051/0004-6361/201321556}{\bibinfo{pages}{A27}}
  (\bibinfo{year}{2014}), \eprint{1303.5087} [astro-ph.CO].

\bibitem[{\citenamefont{Prunet et~al.}(2005)\citenamefont{Prunet, Uzan,
  Bernardeau, and Brunier}}]{pubb05}
\bibinfo{author}{\bibfnamefont{S.}~\bibnamefont{Prunet}},
  \bibinfo{author}{\bibfnamefont{J.-P.} \bibnamefont{Uzan}},
  \bibinfo{author}{\bibfnamefont{F.}~\bibnamefont{Bernardeau}},
  \bibnamefont{and} \bibinfo{author}{\bibfnamefont{T.}~\bibnamefont{Brunier}},
  \bibinfo{journal}{Phys. Rev.}
  \textbf{\doi{10.1103/PhysRevD.71.083508}{\bibinfo{volume}{D71}}},
  \doi{10.1103/PhysRevD.71.083508}{\bibinfo{pages}{083508}}
  (\bibinfo{year}{2005}), \eprint{astro-ph/0406364} [astro-ph].

\bibitem[{\citenamefont{Ade et~al.}(2016{\natexlab{c}})}]{planck2014-a24}
\bibinfo{collaboration}{Planck} Collaboration, \bibinfo{journal}{Astron.
  Astrophys.}
  \textbf{\doi{10.1051/0004-6361/201525898}{\bibinfo{volume}{594}}},
  \doi{10.1051/0004-6361/201525898}{\bibinfo{pages}{A20}}
  (\bibinfo{year}{2016}{\natexlab{c}}), \eprint{1502.02114} [astro-ph.CO].

\bibitem[{\citenamefont{Ade et~al.}(2015)}]{Ade:2015tva}
\bibinfo{collaboration}{BICEP2/Keck/Planck} Collaboration,
  \bibinfo{journal}{Phys. Rev. Lett.}
  \textbf{\doi{10.1103/PhysRevLett.114.101301}{\bibinfo{volume}{114}}},
  \doi{10.1103/PhysRevLett.114.101301}{\bibinfo{pages}{101301}}
  (\bibinfo{year}{2015}), \eprint{1502.00612} [astro-ph.CO].

\bibitem[{\citenamefont{Ade et~al.}(2016{\natexlab{d}})}]{BKP16}
\bibinfo{collaboration}{Keck Array/BICEP2} Collaboration,
  \bibinfo{journal}{Phys. Rev. Lett.}
  \textbf{\doi{10.1103/PhysRevLett.116.031302}{\bibinfo{volume}{116}}},
  \doi{10.1103/PhysRevLett.116.031302}{\bibinfo{pages}{031302}}
  (\bibinfo{year}{2016}{\natexlab{d}}), \eprint{1510.09217} [astro-ph.CO].

\bibitem[{\citenamefont{Contreras et~al.}(2017)\citenamefont{Contreras,
  Hutchinson, Moss, Scott, and Zibin}}]{short}
\bibinfo{author}{\bibfnamefont{D.}~\bibnamefont{Contreras}},
  \bibinfo{author}{\bibfnamefont{J.}~\bibnamefont{Hutchinson}},
  \bibinfo{author}{\bibfnamefont{A.}~\bibnamefont{Moss}},
  \bibinfo{author}{\bibfnamefont{D.}~\bibnamefont{Scott}}, \bibnamefont{and}
  \bibinfo{author}{\bibfnamefont{J.~P.} \bibnamefont{Zibin}}
  (\bibinfo{year}{2017}), \eprint{1709.10134} [astro-ph.CO].

\bibitem[{\citenamefont{{Moss} et~al.}(2011)\citenamefont{{Moss}, {Scott},
  {Zibin}, and {Battye}}}]{Moss2011}
\bibinfo{author}{\bibfnamefont{A.}~\bibnamefont{{Moss}}},
  \bibinfo{author}{\bibfnamefont{D.}~\bibnamefont{{Scott}}},
  \bibinfo{author}{\bibfnamefont{J.~P.} \bibnamefont{{Zibin}}},
  \bibnamefont{and} \bibinfo{author}{\bibfnamefont{R.}~\bibnamefont{{Battye}}},
  \bibinfo{journal}{\prd}
  \textbf{\doi{10.1103/PhysRevD.84.023014}{\bibinfo{volume}{84}}},
  \bibinfo{eid}{023014} (\bibinfo{year}{2011}), \eprint{1011.2990}
  [astro-ph.CO].

\bibitem[{\citenamefont{Scott and Frolop}(2014)}]{Scott:2014qea}
\bibinfo{author}{\bibfnamefont{D.}~\bibnamefont{Scott}} \bibnamefont{and}
  \bibinfo{author}{\bibfnamefont{A.}~\bibnamefont{Frolop}}
  (\bibinfo{year}{2014}), \eprint{1403.8145} [astro-ph.CO].

\bibitem[{\citenamefont{Chluba et~al.}(2014)\citenamefont{Chluba, Dai, Jeong,
  Kamionkowski, and Yoho}}]{Chluba2014}
\bibinfo{author}{\bibfnamefont{J.}~\bibnamefont{Chluba}},
  \bibinfo{author}{\bibfnamefont{L.}~\bibnamefont{Dai}},
  \bibinfo{author}{\bibfnamefont{D.}~\bibnamefont{Jeong}},
  \bibinfo{author}{\bibfnamefont{M.}~\bibnamefont{Kamionkowski}},
  \bibnamefont{and} \bibinfo{author}{\bibfnamefont{A.}~\bibnamefont{Yoho}},
  \bibinfo{journal}{Mon. Not. Roy. Astron. Soc.}
  \textbf{\doi{10.1093/mnras/stu921}{\bibinfo{volume}{442}}},
  \doi{10.1093/mnras/stu921}{\bibinfo{pages}{670}} (\bibinfo{year}{2014}),
  \eprint{1404.2798} [astro-ph.CO].

\bibitem[{\citenamefont{Zibin}(2014)}]{zibin14}
\bibinfo{author}{\bibfnamefont{J.~P.} \bibnamefont{Zibin}},
  \bibinfo{journal}{Phys. Rev.}
  \textbf{\doi{10.1103/PhysRevD.89.121301}{\bibinfo{volume}{D89}}},
  \doi{10.1103/PhysRevD.89.121301}{\bibinfo{pages}{121301}}
  (\bibinfo{year}{2014}), \eprint{1404.4866} [astro-ph.CO].

\bibitem[{\citenamefont{Erickcek et~al.}(2009)\citenamefont{Erickcek, Hirata,
  and Kamionkowski}}]{ehk09}
\bibinfo{author}{\bibfnamefont{A.~L.} \bibnamefont{Erickcek}},
  \bibinfo{author}{\bibfnamefont{C.~M.} \bibnamefont{Hirata}},
  \bibnamefont{and}
  \bibinfo{author}{\bibfnamefont{M.}~\bibnamefont{Kamionkowski}},
  \bibinfo{journal}{Phys. Rev.}
  \textbf{\doi{10.1103/PhysRevD.80.083507}{\bibinfo{volume}{D80}}},
  \doi{10.1103/PhysRevD.80.083507}{\bibinfo{pages}{083507}}
  (\bibinfo{year}{2009}), \eprint{0907.0705} [astro-ph.CO].

\bibitem[{\citenamefont{Hajian and Souradeep}(2003)}]{Hajian2003}
\bibinfo{author}{\bibfnamefont{A.}~\bibnamefont{Hajian}} \bibnamefont{and}
  \bibinfo{author}{\bibfnamefont{T.}~\bibnamefont{Souradeep}},
  \bibinfo{journal}{Astrophys. J.}
  \textbf{\doi{10.1086/379757}{\bibinfo{volume}{597}}},
  \doi{10.1086/379757}{\bibinfo{pages}{L5}} (\bibinfo{year}{2003}),
  \eprint{astro-ph/0308001} [astro-ph].

\bibitem[{\citenamefont{Hajian and Souradeep}(2006)}]{Hajian2006}
\bibinfo{author}{\bibfnamefont{A.}~\bibnamefont{Hajian}} \bibnamefont{and}
  \bibinfo{author}{\bibfnamefont{T.}~\bibnamefont{Souradeep}},
  \bibinfo{journal}{Phys. Rev.}
  \textbf{\doi{10.1103/PhysRevD.74.123521}{\bibinfo{volume}{D74}}},
  \doi{10.1103/PhysRevD.74.123521}{\bibinfo{pages}{123521}}
  (\bibinfo{year}{2006}), \eprint{astro-ph/0607153} [astro-ph].

\bibitem[{\citenamefont{Ade et~al.}(2014{\natexlab{b}})}]{planck2013-p12}
\bibinfo{collaboration}{Planck} Collaboration, \bibinfo{journal}{Astron.
  Astrophys.}
  \textbf{\doi{10.1051/0004-6361/201321543}{\bibinfo{volume}{571}}},
  \doi{10.1051/0004-6361/201321543}{\bibinfo{pages}{A17}}
  (\bibinfo{year}{2014}{\natexlab{b}}), \eprint{1303.5077} [astro-ph.CO].

\bibitem[{\citenamefont{Ade et~al.}(2016{\natexlab{e}})}]{planck2014-a17}
\bibinfo{collaboration}{Planck} Collaboration, \bibinfo{journal}{Astron.
  Astrophys.}
  \textbf{\doi{10.1051/0004-6361/201525941}{\bibinfo{volume}{594}}},
  \doi{10.1051/0004-6361/201525941}{\bibinfo{pages}{A15}}
  (\bibinfo{year}{2016}{\natexlab{e}}), \eprint{1502.01591} [astro-ph.CO].

\bibitem[{\citenamefont{Jeong et~al.}(2014)\citenamefont{Jeong, Chluba, Dai,
  Kamionkowski, and Wang}}]{jcdkw14}
\bibinfo{author}{\bibfnamefont{D.}~\bibnamefont{Jeong}},
  \bibinfo{author}{\bibfnamefont{J.}~\bibnamefont{Chluba}},
  \bibinfo{author}{\bibfnamefont{L.}~\bibnamefont{Dai}},
  \bibinfo{author}{\bibfnamefont{M.}~\bibnamefont{Kamionkowski}},
  \bibnamefont{and} \bibinfo{author}{\bibfnamefont{X.}~\bibnamefont{Wang}},
  \bibinfo{journal}{Phys. Rev.}
  \textbf{\doi{10.1103/PhysRevD.89.023003}{\bibinfo{volume}{D89}}},
  \doi{10.1103/PhysRevD.89.023003}{\bibinfo{pages}{023003}}
  (\bibinfo{year}{2014}), \eprint{1309.2285} [astro-ph.CO].

\bibitem[{\citenamefont{Adam et~al.}(2016{\natexlab{a}})}]{planck2014-a11}
\bibinfo{collaboration}{Planck} Collaboration, \bibinfo{journal}{Astron.
  Astrophys.}
  \textbf{\doi{10.1051/0004-6361/201525936}{\bibinfo{volume}{594}}},
  \doi{10.1051/0004-6361/201525936}{\bibinfo{pages}{A9}}
  (\bibinfo{year}{2016}{\natexlab{a}}), \eprint{1502.05956} [astro-ph.CO].

\bibitem[{\citenamefont{Gregory}(2005)}]{Gregory2005}
\bibinfo{author}{\bibfnamefont{P.~C.} \bibnamefont{Gregory}},
  \emph{\bibinfo{title}{Bayesian Logical Data Analysis for the Physical
  Sciences}} (\bibinfo{publisher}{Cambridge University Press},
  \bibinfo{year}{2005}).

\bibitem[{\citenamefont{Abazajian et~al.}(2016)}]{CMBS4}
\bibinfo{collaboration}{CMB-S4} Collaboration (\bibinfo{year}{2016}),
  \eprint{1610.02743} [astro-ph.CO].

\bibitem[{\citenamefont{{Delabrouille}
  et~al.}(2016)\citenamefont{{Delabrouille}, {de Bernardis}, {Bouchet}, and
  {the CORE Collaboration}}}]{CORE2016}
\bibinfo{author}{\bibfnamefont{J.}~\bibnamefont{{Delabrouille}}},
  \bibinfo{author}{\bibfnamefont{P.}~\bibnamefont{{de Bernardis}}},
  \bibinfo{author}{\bibfnamefont{F.~R.} \bibnamefont{{Bouchet}}},
  \bibnamefont{and} \bibinfo{author}{\bibnamefont{{the CORE Collaboration}}},
  \bibinfo{journal}{A proposal in response to the ESA call for a Medium Size
  space mission for launch in 2029-2030}  (\bibinfo{year}{2016}).

\bibitem[{\citenamefont{Hu}(2000)}]{hu99}
\bibinfo{author}{\bibfnamefont{W.}~\bibnamefont{Hu}},
  \bibinfo{journal}{Astrophys. J.}
  \textbf{\doi{10.1086/308279}{\bibinfo{volume}{529}}},
  \doi{10.1086/308279}{\bibinfo{pages}{12}} (\bibinfo{year}{2000}),
  \eprint{astro-ph/9907103} [astro-ph].

\bibitem[{\citenamefont{Ghosh et~al.}(2016{\natexlab{b}})\citenamefont{Ghosh,
  Kothari, Jain, and Rath}}]{gkjr16}
\bibinfo{author}{\bibfnamefont{S.}~\bibnamefont{Ghosh}},
  \bibinfo{author}{\bibfnamefont{R.}~\bibnamefont{Kothari}},
  \bibinfo{author}{\bibfnamefont{P.}~\bibnamefont{Jain}}, \bibnamefont{and}
  \bibinfo{author}{\bibfnamefont{P.~K.} \bibnamefont{Rath}},
  \bibinfo{journal}{JCAP}
  \textbf{\doi{10.1088/1475-7516/2016/01/046}{\bibinfo{volume}{1601}}},
  \doi{10.1088/1475-7516/2016/01/046}{\bibinfo{pages}{046}}
  (\bibinfo{year}{2016}{\natexlab{b}}), \eprint{1507.04078} [astro-ph.CO].

\bibitem[{\citenamefont{Edmonds}(1974)}]{edmonds74}
\bibinfo{author}{\bibfnamefont{A.~R.} \bibnamefont{Edmonds}},
  \emph{\bibinfo{title}{Angular Momentum in Quantum Mechanics}}
  (\bibinfo{publisher}{Princeton University Press},
  \bibinfo{address}{Princeton}, \bibinfo{year}{1974}).

\bibitem[{\citenamefont{Challinor and van Leeuwen}(2002)}]{Challinor:2002zh}
\bibinfo{author}{\bibfnamefont{A.}~\bibnamefont{Challinor}} \bibnamefont{and}
  \bibinfo{author}{\bibfnamefont{F.}~\bibnamefont{van Leeuwen}},
  \bibinfo{journal}{Phys. Rev.}
  \textbf{\doi{10.1103/PhysRevD.65.103001}{\bibinfo{volume}{D65}}},
  \doi{10.1103/PhysRevD.65.103001}{\bibinfo{pages}{103001}}
  (\bibinfo{year}{2002}), \eprint{astro-ph/0112457} [astro-ph].

\bibitem[{\citenamefont{Adam et~al.}(2016{\natexlab{b}})}]{Adam:2015vua}
\bibinfo{collaboration}{Planck} Collaboration, \bibinfo{journal}{Astron.
  Astrophys.}
  \textbf{\doi{10.1051/0004-6361/201525820}{\bibinfo{volume}{594}}},
  \doi{10.1051/0004-6361/201525820}{\bibinfo{pages}{A8}}
  (\bibinfo{year}{2016}{\natexlab{b}}), \eprint{1502.01587} [astro-ph.CO].

\bibitem[{\citenamefont{Hansen et~al.}(2009)\citenamefont{Hansen, Banday,
  Gorski, Eriksen, and Lilje}}]{hansenetal09}
\bibinfo{author}{\bibfnamefont{F.}~\bibnamefont{Hansen}},
  \bibinfo{author}{\bibfnamefont{A.}~\bibnamefont{Banday}},
  \bibinfo{author}{\bibfnamefont{K.}~\bibnamefont{Gorski}},
  \bibinfo{author}{\bibfnamefont{H.}~\bibnamefont{Eriksen}}, \bibnamefont{and}
  \bibinfo{author}{\bibfnamefont{P.}~\bibnamefont{Lilje}},
  \bibinfo{journal}{Astrophys. J.}
  \textbf{\doi{10.1088/0004-637X/704/2/1448}{\bibinfo{volume}{704}}},
  \doi{10.1088/0004-637X/704/2/1448}{\bibinfo{pages}{1448}}
  (\bibinfo{year}{2009}), \eprint{0812.3795} [astro-ph].

\bibitem[{\citenamefont{Flender and Hotchkiss}(2013)}]{fh13}
\bibinfo{author}{\bibfnamefont{S.}~\bibnamefont{Flender}} \bibnamefont{and}
  \bibinfo{author}{\bibfnamefont{S.}~\bibnamefont{Hotchkiss}},
  \bibinfo{journal}{JCAP}
  \textbf{\doi{10.1088/1475-7516/2013/09/033}{\bibinfo{volume}{1309}}},
  \doi{10.1088/1475-7516/2013/09/033}{\bibinfo{pages}{033}}
  (\bibinfo{year}{2013}), \eprint{1307.6069} [astro-ph.CO].

\bibitem[{\citenamefont{Quartin and Notari}(2015)}]{qn15}
\bibinfo{author}{\bibfnamefont{M.}~\bibnamefont{Quartin}} \bibnamefont{and}
  \bibinfo{author}{\bibfnamefont{A.}~\bibnamefont{Notari}},
  \bibinfo{journal}{JCAP}
  \textbf{\doi{10.1088/1475-7516/2015/01/008}{\bibinfo{volume}{1501}}},
  \doi{10.1088/1475-7516/2015/01/008}{\bibinfo{pages}{008}}
  (\bibinfo{year}{2015}), \eprint{1408.5792} [astro-ph.CO].

\bibitem[{\citenamefont{Aiola et~al.}(2015)\citenamefont{Aiola, Wang, Kosowsky,
  Kahniashvili, and Firouzjahi}}]{awkkf15}
\bibinfo{author}{\bibfnamefont{S.}~\bibnamefont{Aiola}},
  \bibinfo{author}{\bibfnamefont{B.}~\bibnamefont{Wang}},
  \bibinfo{author}{\bibfnamefont{A.}~\bibnamefont{Kosowsky}},
  \bibinfo{author}{\bibfnamefont{T.}~\bibnamefont{Kahniashvili}},
  \bibnamefont{and}
  \bibinfo{author}{\bibfnamefont{H.}~\bibnamefont{Firouzjahi}},
  \bibinfo{journal}{Phys. Rev.}
  \textbf{\doi{10.1103/PhysRevD.92.063008}{\bibinfo{volume}{D92}}},
  \doi{10.1103/PhysRevD.92.063008}{\bibinfo{pages}{063008}}
  (\bibinfo{year}{2015}), \eprint{1506.04405} [astro-ph.CO].

\bibitem[{\citenamefont{Gibelyou and Huterer}(2012)}]{gh12}
\bibinfo{author}{\bibfnamefont{C.}~\bibnamefont{Gibelyou}} \bibnamefont{and}
  \bibinfo{author}{\bibfnamefont{D.}~\bibnamefont{Huterer}},
  \bibinfo{journal}{Mon. Not. Roy. Astron. Soc.}
  \textbf{\doi{10.1111/j.1365-2966.2012.22032.x}{\bibinfo{volume}{427}}},
  \doi{10.1111/j.1365-2966.2012.22032.x}{\bibinfo{pages}{1994}}
  (\bibinfo{year}{2012}), \eprint{1205.6476} [astro-ph.CO].

\bibitem[{\citenamefont{Shiraishi et~al.}(2017)\citenamefont{Shiraishi,
  Sugiyama, and Okumura}}]{sso17}
\bibinfo{author}{\bibfnamefont{M.}~\bibnamefont{Shiraishi}},
  \bibinfo{author}{\bibfnamefont{N.~S.} \bibnamefont{Sugiyama}},
  \bibnamefont{and} \bibinfo{author}{\bibfnamefont{T.}~\bibnamefont{Okumura}},
  \bibinfo{journal}{Phys. Rev.}
  \textbf{\doi{10.1103/PhysRevD.95.063508}{\bibinfo{volume}{D95}}},
  \doi{10.1103/PhysRevD.95.063508}{\bibinfo{pages}{063508}}
  (\bibinfo{year}{2017}), \eprint{1612.02645} [astro-ph.CO].

\bibitem[{\citenamefont{Zhai and Blanton}(2017)}]{zb17}
\bibinfo{author}{\bibfnamefont{Z.}~\bibnamefont{Zhai}} \bibnamefont{and}
  \bibinfo{author}{\bibfnamefont{M.}~\bibnamefont{Blanton}}
  (\bibinfo{year}{2017}), \eprint{1707.06555} [astro-ph.CO].

\end{thebibliography}

\end{document}